\documentclass[12pt,a4paper]{article}

\usepackage{ifthen} 
\usepackage{multirow}
\usepackage{booktabs} 
\newboolean{pdflatex}
\setboolean{pdflatex}{true} 

\newboolean{articletitles}
\setboolean{articletitles}{true} 

\newboolean{uprightparticles}
\setboolean{uprightparticles}{false} 

\def\paperauthors{LHCb collaboration} 
\def\paperasciititle{Search for CP violation in D+->phipi+ decays } 
 \def\papertitle{Search for $C\!P$ violation in $D^+ \to \phi \pi^+$ decays} 
\def\paperkeywords{{High Energy Physics}, {LHCb}} 
\def\papercopyright{\the\year\ CERN for the benefit of the LHCb collaboration} 
\def\paperlicence{CC BY 4.0 licence}
\def\paperlicenceurl{https://creativecommons.org/licenses/by/4.0/}

\newif\ifEnableSectionTOCLinks
\EnableSectionTOCLinksfalse 

\usepackage[top=1in, bottom=1.25in, left=1in, right=1in]{geometry}

\usepackage{microtype}
\usepackage{lineno}  
\usepackage{xspace} 
\usepackage{caption} 

\usepackage{graphicx}  
\usepackage{color}
\usepackage{colortbl}
\graphicspath{{./figs/}} 

\usepackage{amsmath} 
\usepackage{amssymb}
\usepackage{amsfonts}
\usepackage{upgreek} 

\newcommand*\patchAmsMathEnvironmentForLineno[1]{%
\expandafter\let\csname old#1\expandafter\endcsname\csname #1\endcsname
\expandafter\let\csname oldend#1\expandafter\endcsname\csname
end#1\endcsname
 \renewenvironment{#1}%
   {\linenomath\csname old#1\endcsname}%
   {\csname oldend#1\endcsname\endlinenomath}%
}
\newcommand*\patchBothAmsMathEnvironmentsForLineno[1]{%
  \patchAmsMathEnvironmentForLineno{#1}%
  \patchAmsMathEnvironmentForLineno{#1*}%
}
\AtBeginDocument{%
\patchBothAmsMathEnvironmentsForLineno{equation}%
\patchBothAmsMathEnvironmentsForLineno{align}%
\patchBothAmsMathEnvironmentsForLineno{flalign}%
\patchBothAmsMathEnvironmentsForLineno{alignat}%
\patchBothAmsMathEnvironmentsForLineno{gather}%
\patchBothAmsMathEnvironmentsForLineno{multline}%
\patchBothAmsMathEnvironmentsForLineno{eqnarray}%
}

\usepackage[pdftex,
            pdfauthor={\paperauthors},
            pdftitle={\paperasciititle},
            pdfkeywords={\paperkeywords}]{hyperref}
\usepackage{hyperxmp}
\hypersetup{
    pdfcopyright={Copyright (C) \papercopyright},
    pdflicenseurl={\paperlicenceurl}
}

\usepackage[colorinlistoftodos,textsize=scriptsize]{todonotes}

\usepackage[bottom,flushmargin,hang,multiple]{footmisc}

\usepackage[all]{hypcap} 

\usepackage{xspace}
\usepackage{upgreek}

\def\lhcb   {\mbox{LHCb}\xspace}

\def\lhc    {\mbox{LHC}\xspace}

\def\velo   {VELO\xspace}
\def\rich   {RICH\xspace}

\def\MagUp {\mbox{\em Mag\kern -0.05em Up}\xspace}
\def\MagDown {\mbox{\em MagDown}\xspace}

\ifthenelse{\boolean{uprightparticles}}%
{

 \def\Ppi         {\ensuremath{\uppi}\xspace}

 \def\PDelta      {\ensuremath{\Delta}\xspace}
 \def\PXi         {\ensuremath{\Xi}\xspace}
 \def\PLambda     {\ensuremath{\Lambda}\xspace}
 \def\PSigma      {\ensuremath{\Sigma}\xspace}
 \def\POmega      {\ensuremath{\Omega}\xspace}
 \def\PUpsilon    {\ensuremath{\Upsilon}\xspace}
 \let\oldPi\Pi
 \def\PPi         {\ensuremath{\oldPi}\xspace}

 \def\PB      {\ensuremath{\mathrm{B}}\xspace}
 \def\PD      {\ensuremath{\mathrm{D}}\xspace}

 \def\PK      {\ensuremath{\mathrm{K}}\xspace}
 \def\Pb      {\ensuremath{\mathrm{b}}\xspace}
 \def\Pc      {\ensuremath{\mathrm{c}}\xspace}
 \def\Pd      {\ensuremath{\mathrm{d}}\xspace}

 \def\Pp      {\ensuremath{\mathrm{p}}\xspace}

 \def\Ps      {\ensuremath{\mathrm{s}}\xspace}
 
 \def\Pu      {\ensuremath{\mathrm{u}}\xspace}
 \def\thebaroffset{0.0em}
}
{

 \def\Ppi         {\ensuremath{\pi}\xspace}

 \mathchardef\PDelta="7101
 \mathchardef\PXi="7104
 \mathchardef\PLambda="7103
 \mathchardef\PSigma="7106
 \mathchardef\POmega="710A
 \mathchardef\PUpsilon="7107
 \mathchardef\PPi="7105
 \def\PB      {\ensuremath{B}\xspace}
 \def\PD      {\ensuremath{D}\xspace}

 \def\PK      {\ensuremath{K}\xspace}
 \def\Pb      {\ensuremath{b}\xspace}
 \def\Pc      {\ensuremath{c}\xspace}
 \def\Pd      {\ensuremath{d}\xspace}

 \def\Pp      {\ensuremath{p}\xspace}

 \def\Ps      {\ensuremath{s}\xspace}
 
 \def\Pu      {\ensuremath{u}\xspace}
 \def\thebaroffset{0.18em}
}
\newcommand{\offsetoverline}[2][\thebaroffset]{\kern #1\overline{\kern -#1 #2}}%

\makeatletter
\ifcase \@ptsize \relax
  \newcommand{\miniscule}{\@setfontsize\miniscule{4}{5}}
\or
  \newcommand{\miniscule}{\@setfontsize\miniscule{5}{6}}
\or
  \newcommand{\miniscule}{\@setfontsize\miniscule{5}{6}}
\fi
\makeatother

\DeclareRobustCommand{\optbar}[1]{\shortstack{{\miniscule (\rule[.5ex]{1.25em}{.18mm})}
  \\ [-.7ex] $#1$}}

\def\uquark    {{\ensuremath{\Pu}}\xspace}

\def\dquark    {{\ensuremath{\Pd}}\xspace}
\def\dquarkbar {{\ensuremath{\overline \dquark}}\xspace}

\def\squark    {{\ensuremath{\Ps}}\xspace}
\def\squarkbar {{\ensuremath{\overline \squark}}\xspace}

\def\cquark    {{\ensuremath{\Pc}}\xspace}
\def\cquarkbar {{\ensuremath{\overline \cquark}}\xspace}

\def\bquark    {{\ensuremath{\Pb}}\xspace}

\def\pion   {{\ensuremath{\Ppi}}\xspace}
\def\piz    {{\ensuremath{\pion^0}}\xspace}
\def\pip    {{\ensuremath{\pion^+}}\xspace}
\def\pim    {{\ensuremath{\pion^-}}\xspace}
\def\pipm   {{\ensuremath{\pion^\pm}}\xspace}

\def\kaon    {{\ensuremath{\PK}}\xspace}
\def\Kbar    {{\ensuremath{\offsetoverline{\PK}}}\xspace}

\def\KorKbar {\kern \thebaroffset\optbar{\kern -\thebaroffset \PK}{}\xspace}
\def\Kz      {{\ensuremath{\kaon^0}}\xspace}
\def\Kzb     {{\ensuremath{\Kbar{}^0}}\xspace}
\def\Kp      {{\ensuremath{\kaon^+}}\xspace}
\def\Km      {{\ensuremath{\kaon^-}}\xspace}

\def\KS      {{\ensuremath{\kaon^0_{\mathrm{S}}}}\xspace}

\def\KL      {{\ensuremath{\kaon^0_{\mathrm{L}}}}\xspace}

\def\Dbar    {{\ensuremath{\offsetoverline{\PD}}}\xspace}
\def\D       {{\ensuremath{\PD}}\xspace}

\def\DorDbar {\kern \thebaroffset\optbar{\kern -\thebaroffset \PD}\xspace}
\def\Dz      {{\ensuremath{\D^0}}\xspace}

\def\Dp      {{\ensuremath{\D^+}}\xspace}
\def\Dm      {{\ensuremath{\D^-}}\xspace}
\def\Dpm     {{\ensuremath{\D^\pm}}\xspace}

\def\DpDm    {\ensuremath{\Dp {\kern -0.16em \Dm}}\xspace}

\def\Dsp     {{\ensuremath{\D^+_\squark}}\xspace}

\def\B       {{\ensuremath{\PB}}\xspace}

\def\BorBbar {\kern \thebaroffset\optbar{\kern -\thebaroffset \PB}\xspace}

\def\Bd      {{\ensuremath{\B^0}}\xspace}

\def\BdorBdbar {\kern \thebaroffset\optbar{\kern -\thebaroffset \Bd}\xspace}

\def\Bs      {{\ensuremath{\B^0_\squark}}\xspace}

\def\BsorBsbar {\kern \thebaroffset\optbar{\kern -\thebaroffset \Bs}\xspace}

\def\Y#1S{\ensuremath{\PUpsilon{(#1S)}}\xspace}

\def\proton      {{\ensuremath{\Pp}}\xspace}

\def\Lz          {{\ensuremath{\PLambda}}\xspace}

\def\LorLbar     {\kern \thebaroffset\optbar{\kern -\thebaroffset \PLambda}\xspace}

\def\Lc          {{\ensuremath{\Lz^+_\cquark}}\xspace}

\newcommand{\decay}[2]{\mbox{\ensuremath{#1\!\to #2}}\xspace}

\def\to                 {\ensuremath{\rightarrow}\xspace}

\def\CP                {{\ensuremath{C\!P}}\xspace}

\def\Vud  {{\ensuremath{V_{\uquark\dquark}^{\phantom{\ast}}}}\xspace}

\def\Vus  {{\ensuremath{V_{\uquark\squark}^{\phantom{\ast}}}}\xspace}

\def\Vcds  {{\ensuremath{V_{\cquark\dquark}^\ast}}\xspace}

\def\Vcss  {{\ensuremath{V_{\cquark\squark}^\ast}}\xspace}

\def\AT#1     {\ensuremath{A_{\mathrm{T}}^{#1}}\xspace}           

\def\C#1      {\ensuremath{\mathcal{C}_{#1}}\xspace}                       
\def\Cp#1     {\ensuremath{\mathcal{C}_{#1}^{'}}\xspace}                    
\def\Ceff#1   {\ensuremath{\mathcal{C}_{#1}^{\mathrm{(eff)}}}\xspace}        
\def\Cpeff#1  {\ensuremath{\mathcal{C}_{#1}^{'\mathrm{(eff)}}}\xspace}       
\def\Ope#1    {\ensuremath{\mathcal{O}_{#1}}\xspace}                       
\def\Opep#1   {\ensuremath{\mathcal{O}_{#1}^{'}}\xspace}                    

\newcommand{\ket}[1]{\ensuremath{|#1\rangle}}              

\newcommand{\nospaceunit}[1]{\ensuremath{\text{#1}}}
\newcommand{\aunit}[1]{\ensuremath{\text{\,#1}}}

\newcommand{\tev}{\aunit{Te\kern -0.1em V}\xspace}
\newcommand{\gev}{\aunit{Ge\kern -0.1em V}\xspace}
\newcommand{\mev}{\aunit{Me\kern -0.1em V}\xspace}
\newcommand{\kev}{\aunit{ke\kern -0.1em V}\xspace}
\newcommand{\ev}{\aunit{e\kern -0.1em V}\xspace}

\newcommand{\mevc}{\ensuremath{\aunit{Me\kern -0.1em V\!/}c}\xspace}
\newcommand{\gevc}{\ensuremath{\aunit{Ge\kern -0.1em V\!/}c}\xspace}
\newcommand{\mevcc}{\ensuremath{\aunit{Me\kern -0.1em V\!/}c^2}\xspace}
\newcommand{\gevcc}{\ensuremath{\aunit{Ge\kern -0.1em V\!/}c^2}\xspace}

\def\mm   {\aunit{mm}\xspace}

\def\mum  {\ensuremath{\,\upmu\nospaceunit{m}}\xspace}

\def\fb   {\ensuremath{\aunit{fb}}\xspace}
\def\invfb   {\ensuremath{\fb^{-1}}\xspace}

\def\sec  {\ensuremath{\aunit{s}}\xspace}

\def\ps   {\ensuremath{\aunit{ps}}\xspace}

\newcommand{\stat}{\aunit{(stat)}\xspace}
\newcommand{\syst}{\aunit{(syst)}\xspace}

\newcommand{\chisq}{\ensuremath{\chi^2}\xspace}
\newcommand{\chisqndf}{\ensuremath{\chi^2/\mathrm{ndf}}\xspace}

\def\deriv {\ensuremath{\mathrm{d}}}

\def\gsim{{~\raise.15em\hbox{$>$}\kern-.85em
          \lower.35em\hbox{$\sim$}~}\xspace}
\def\lsim{{~\raise.15em\hbox{$<$}\kern-.85em
          \lower.35em\hbox{$\sim$}~}\xspace}

\def\PDF {PDF\xspace}

\def\pt         {\ensuremath{p_{\mathrm{T}}}\xspace}

\def\ptot       {\ensuremath{p}\xspace}

\def\mrad{\aunit{mrad}\xspace}

\def\evtgen     {\mbox{\textsc{EvtGen}}\xspace}

\def\geant      {\mbox{\textsc{Geant4}}\xspace}

\def\photos     {\mbox{\textsc{Photos}}\xspace}

\def\pythia     {\mbox{\textsc{Pythia}}\xspace}

\def\tell1  {TELL1\xspace}
\def\ukl1   {UKL1\xspace}

\newcommand{\ie}{\mbox{\itshape i.e.}\xspace}

\newcommand{\lhcborcid}[1]{\href{https://orcid.org/#1}{\hspace*{0.1em}\raisebox{-0.45ex}{\includegraphics[width=1em]{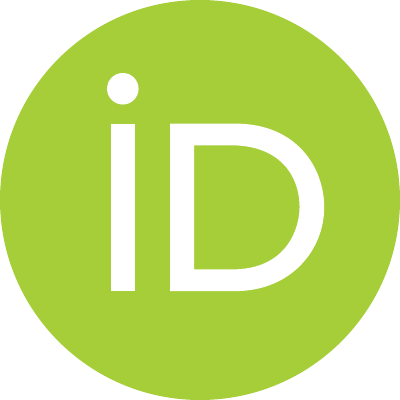}}}}

\hypersetup{
  colorlinks   = true, 
  urlcolor     = blue, 
  linkcolor    = blue, 
  citecolor    = red   
}

\ifEnableSectionTOCLinks
    \usepackage[explicit]{titlesec} 
    
    \let\oldcontentsline\contentsline
    \renewcommand

    \titleformat{\section}{\normalfont\Large\bf}{\hyperlink{tocsection.\thesection}{{\thesection} \parbox[t]{\dimexpr\textwidth-1pc}{#1}}}{1pc}{}

    \titleformat{\subsection}{\normalfont\bf}{\hyperlink{tocsubsection.\thesubsection}{{\thesubsection} \parbox[t]{\dimexpr\textwidth-1pc}{#1}}}{1pc}{}

    \titleformat{name=\section,numberless}[display]{}{}{0pt}{\normalfont\Huge\bfseries #1}
\fi

\usepackage{cite} 
\usepackage{mciteplus}

\newcommand{\Araw}{{\ensuremath{A}\xspace}}
\newcommand{\Araws}{{\ensuremath{A^s}}\xspace}
\newcommand{\ACPdir}{{\ensuremath{a_\CP} \xspace}}
\newcommand{\DeltaACP}{{\ensuremath{{\Delta\ACPdir}\xspace}}}

\newcommand{\Aprod}{{\ensuremath{A_{\mathrm{ P}}}}\xspace}
\newcommand{\Aprods}{{\ensuremath{A^s_{\mathrm{ P}}}}\xspace}
\newcommand{\Adet}{{\ensuremath{A_{\mathrm{ D}}}}\xspace}
\newcommand{\Adets}{{\ensuremath{A^s_{\mathrm{ D}}}}\xspace}
\newcommand{\DeltaAs}{\ensuremath{\Delta A^s}\xspace} 
\newcommand{\DeltaAsTilde}{\ensuremath{\Delta \tilde{A}^s}\xspace} 
\newcommand{\ADTilde}{\ensuremath{\tilde{A}^s_\mathrm{D}(\KS)}\xspace} 
\newcommand{\ADTilderdelta}{\ensuremath{\tilde{A}^s_\mathrm{D}(\KS;\rpi,\deltapi)}\xspace} 
\newcommand{\ADTilderdeltarho}{\ensuremath{\tilde{A}^s_\mathrm{D}(\KS;\rpi,\deltapi,\varepsilon_\rho)}\xspace}

\def\IP {\ensuremath{\textrm{IP}}\xspace}

\def\ipDp {\ensuremath{\textrm{IP}}(\Dp)\xspace}
\def\ipDpsm {\ensuremath{\textrm{IP}(\Dp)_{\text{smeared}}}\xspace}
\def\jsu {\ensuremath{\textrm{Johnson } \text{S}_\text{U}}\xspace}
\def\pvalue {\ensuremath{\textit{p}\textrm{-value}}\xspace}

\def\Kone{{\ensuremath{\kaon^0_{\mathrm{1}}}}\xspace}
\def\Ktwo{{\ensuremath{\kaon^0_{\mathrm{2}}}}\xspace}
\def\alphaS{{\ensuremath{\alpha_{\mathrm{S}}}}\xspace}
\def\alphaL{{\ensuremath{\alpha_{\mathrm{L}}}}\xspace}
\def\alphaSL{{\ensuremath{\alpha_{\mathrm{S,L}}}}\xspace}
\def\alphaSpm{{\ensuremath{\alpha_{\mathrm{S}}^{(\pm)}}}\xspace}
\def\alphaLpm{{\ensuremath{\alpha_{\mathrm{L}}^{(\pm)}}}\xspace}
\def\alphaSLpm{{\ensuremath{\alpha_{\mathrm{S,L}}^{(\pm)}}}\xspace}

\newcommand{\KSLL}{\ensuremath{\KS(\rm{LL})}\xspace} 
\newcommand{\KSDD}{\ensuremath{\KS(\rm{DD})}\xspace}

\newcommand{\DpToKmKppi}{\mbox{\ensuremath{\decay{\Dp}{\Km\Kp\pip}}}\xspace}
\newcommand{\DpTophipi}{\mbox{\ensuremath{\decay{\Dp}{\phi\pip}}}\xspace}
\newcommand{\DpToKSpi}{\mbox{\ensuremath{\decay{\Dp}{\KS\pip}}}\xspace}
\newcommand{\DmToKSpi}{\mbox{\ensuremath{\decay{\Dm}{\KS\pim}}}\xspace}
\newcommand{\DpmToKSpi}{\mbox{\ensuremath{\decay{\Dpm}{\KS\pipm}}}\xspace}
\newcommand{\DpToKzpi}{\mbox{\ensuremath{\decay{\Dp}{\Kz\pip}}}\xspace}
\newcommand{\DpToKzbpi}{\mbox{\ensuremath{\decay{\Dp}{\Kzb\pip}}}\xspace}
\newcommand{\DpToKSLLpi}{\mbox{\ensuremath{\decay{\Dp}{\KSLL\pip}}}\xspace}
\newcommand{\DpToKSDDpi}{\mbox{\ensuremath{\decay{\Dp}{\KSDD\pip}}}\xspace}

\newcommand{\ACPDPhiPi}{{\ensuremath{\ACPdir(\DpTophipi)}}\xspace}

\newcommand{\rpi}{\ensuremath{r_\pi}\xspace} 
\newcommand{\deltapi}{\ensuremath{\delta_\pi}\xspace} 
\newcommand{\rpideltapi}{\ensuremath{(\rpi\, , \, \deltapi)}\xspace} 
 
\newcommand{\qrpideltapi}{\ensuremath{(q \, , \, \rpi\, , \, \deltapi)}\xspace} 
\newcommand{\qrpideltapiDeltarho}{\ensuremath{(q \, , \, \rpi\, , \, \deltapi\, , \, \varepsilon_\rho)}\xspace} 
\newcommand{\BigRpi}{\ensuremath{R_\pi}\xspace} 

\usepackage[capitalize]{cleveref}
\usepackage{siunitx}
\begin{document}

\renewcommand{\thefootnote}{\fnsymbol{footnote}}
\setcounter{footnote}{1}


\begin{titlepage}
\pagenumbering{roman}

\vspace*{-1.5cm}
\centerline{\large EUROPEAN ORGANIZATION FOR NUCLEAR RESEARCH (CERN)}
\vspace*{1.5cm}
\noindent
\begin{tabular*}{\linewidth}{lc@{\extracolsep{\fill}}r@{\extracolsep{0pt}}}
\ifthenelse{\boolean{pdflatex}}
{\vspace*{-1.5cm}\mbox{\!\!\!\includegraphics[width=.14\textwidth]{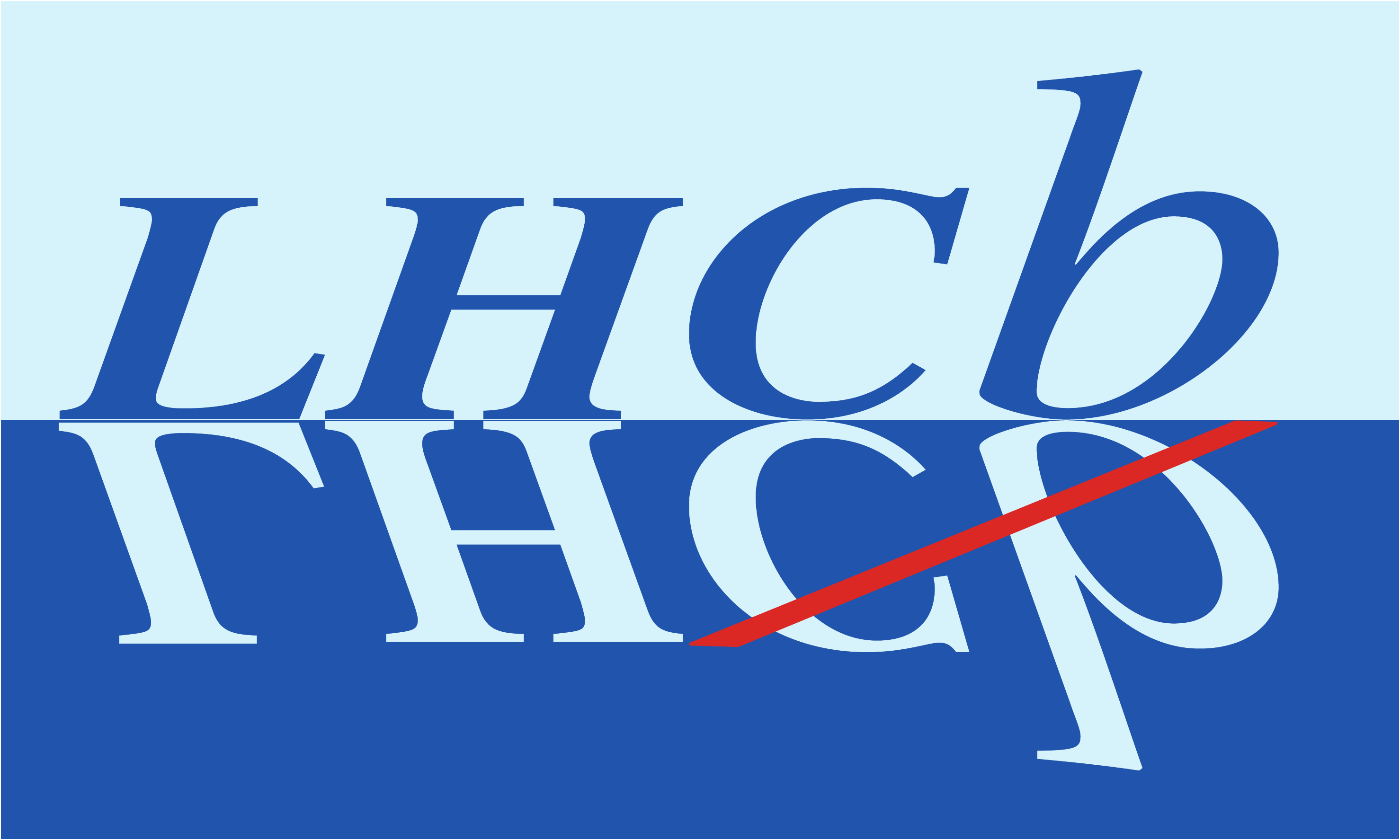}} & &}%
{\vspace*{-1.2cm}\mbox{\!\!\!\includegraphics[width=.12\textwidth]{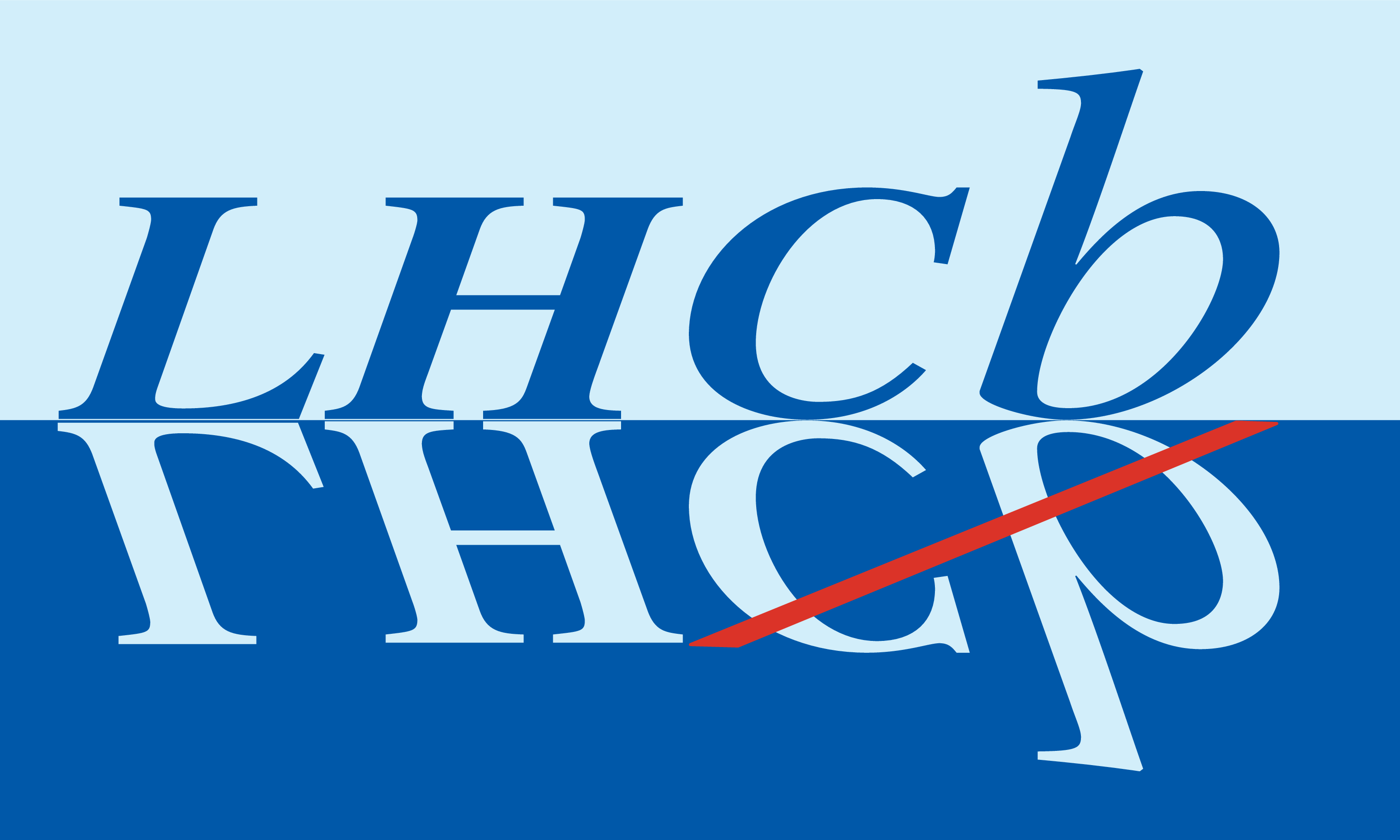}} & &}%
\\
 & & CERN-EP-2026-158 \\  
 & & LHCb-PAPER-2026-011 \\  
 & & November 6, 2026 \\
 & & \\
\end{tabular*}

\vspace*{2.0cm}

{\normalfont\bfseries\boldmath\huge
\begin{center}
  \papertitle 
\end{center}
}

\vspace*{2.0cm}

\begin{center}
\paperauthors\footnote{Authors are listed at the end of this paper.}
\end{center}

\vspace{\fill}

\begin{abstract}
    \noindent
A search for charge-parity ($C\!P$) violation in the Cabibbo-suppressed \mbox{$D^+ \to \phi \pi^+$} decay is presented, using proton-proton collision data corresponding to an integrated luminosity of $6\text{\,fb}^{-1}$, collected at a centre-of-mass energy of $13\text{\,Te\kern -0.1em V}$ with the LHCb detector during Run~2.
An abundant sample of \mbox{$D^+ \to K_{\rm S}^0 \pi^+$} decays is employed to correct for asymmetries arising from the production of the charmed meson and from detection effects associated with the charged pion accompanying the $\phi$ meson. Asymmetries induced by interference of multiple processes, such as neutral kaon mixing, regeneration of neutral kaons, and interference between the Cabibbo-favoured \mbox{$D^+ \to \kern 0.18em\overline{\kern -0.18em K}^0 \pi^+$} and the doubly Cabibbo-suppressed \mbox{$D^+ \to {K}^0 \pi^+$} decay, are subtracted.
The direct $C\!P$ asymmetry in the \mbox{$D^+ \to \phi \pi^+$} decay is measured to be
\begin{equation*}
a_{C\!P}(D^+ \to \phi \pi^+) = \left(0.1 \pm 4.9\text{ (stat)} \pm 2.4\text{ (syst)} \right)\times 10^{-4}.
\end{equation*}
For the first time at a hadron collider, the modulus of the ratio and the relative strong phase of the \mbox{$D^+ \to {K}^0 \pi^+$} to \mbox{$D^+ \to \kern 0.18em\overline{\kern -0.18em K}^0 \pi^+$} decay amplitudes are also investigated. Two-dimensional confidence intervals are reported for these parameters.    
\end{abstract}


\begin{center}
  Published in Journal of High Energy Physics, JHEP 10 (2026) 256
\end{center}

\vspace{\fill}

{\footnotesize 
\centerline{\copyright~\papercopyright. \href{\paperlicenceurl}{\paperlicence}.}}
\vspace*{2mm}

\end{titlepage}


\newpage
\setcounter{page}{2}
\mbox{~}
%
%
%
%


\renewcommand{\thefootnote}{\arabic{footnote}}
\setcounter{footnote}{0}


\cleardoublepage


\pagestyle{plain} 
\setcounter{page}{1}
\pagenumbering{arabic}



\section{Introduction}\label{sec:introduction}

The violation of the invariance of fundamental interactions under the combined action of charge conjugation ($C$) and parity ($P$) transformations, known as \CP violation, is a necessary ingredient to explain the observed dominance of matter over antimatter in the Universe~\cite{Sakharov:1967dj}. Within the Standard Model (SM) of particle physics, \CP violation originates from a single complex phase in the Cabibbo--Kobayashi--Maskawa (CKM) quark-mixing matrix~\cite{Cabibbo:1963yz,Kobayashi:1973fv}. Although this mechanism has been firmly established experimentally in the strange- and beauty-meson sectors, the amount of \CP violation it generates is insufficient to account for the observed cosmological matter--antimatter asymmetry~\cite{Dine:2003ax}. This discrepancy motivates precision searches for additional sources of \CP violation, particularly in systems where the SM contributions are expected to be strongly suppressed. Charmed hadrons provide a unique laboratory for such searches. Unlike $K$ and $B$ mesons, in which the decaying quark is down-type, charm decays involve up-type quarks and therefore probe a complementary flavour sector of the SM. As a result, they are sensitive to different classes of potential new interactions. Within the SM, 
\CP-violating effects in charm decays are expected to be very small, typically of the order of $10^{-4}$ to $10^{-3}$, owing to the strong suppression of flavour-changing neutral currents by the Glashow--Iliopoulos--Maiani (GIM) mechanism and the smallness of the relevant CKM matrix elements~\cite{Grossman:2006jg,Lenz:2020awd}. 

Singly Cabibbo-suppressed \decay{\D}{f}(\decay{\Dbar}{\offsetoverline{f}}) decays, where the initial \D state can be \mbox{$\Dp,  \Dz, \Dsp$}, provide one of the most sensitive tests of \CP violation through the measurement of the direct asymmetry, defined as 
\begin{equation}
\label{eq:acp}
\ACPdir(\decay{\D}{f}) \equiv
\frac{ \Gamma(\decay{\D}{f}) - \Gamma(\decay{\Dbar}{\offsetoverline{f}})}
{ \Gamma(\decay{\D}{f}) + \Gamma(\decay{\Dbar}{\offsetoverline{f}})},
\end{equation}
where \mbox{$\Gamma(\decay{\D}{f})$} \ (\mbox{$\Gamma(\decay{\Dbar}{\offsetoverline{f}})$}) denotes the partial decay width of the initial $D\ (\Dbar)$ meson into the final state $f\ (\offsetoverline{f})$. 
The most prominent processes of this type are the \decay{\Dz}{\Kp\Km} and \decay{\Dz}{\pip\pim} decays, in which \CP violation was observed for the first time in a charm-quark transition. In 2019, the LHCb collaboration measured the difference of the direct \CP asymmetries, ${\DeltaACP = \ACPdir(\Dz\to\Kp\Km) - \ACPdir(\Dz\to\pip\pim) = (-15.7\pm2.9)\times10^{-4}}$~\cite{LHCb-PAPER-2019-006},
which represents a significant deviation from zero. Theoretical uncertainties related to nonperturbative strong-interaction effects prevent a rigorous assessment of its compatibility with the SM~\cite{Grossman:2006jg,Li:2012cfa,Cheng:2012wr,Khodjamirian:2017zdu,Chala:2019fdb,Grossman:2019xcj}. This result has triggered renewed theoretical interest in the charm sector~\cite{Buccella:2019kpn,Li:2019hho,Schacht:2021jaz,Cheng:2019ggx,Dery:2019ysp,Wang:2020gmn,Bause:2020obd,Dery:2021mll,Cheng:2021yrn,Lenz:2023rlq,Pich:2023kim,Gavrilova:2023fzy}, 
highlighting the importance of searching
for \CP-violating effects in other charm decays.

Among the singly Cabibbo-suppressed decays of the \Dp meson, the $\decay{\Dp}{\Kp\Km\pip}$ channel is particularly well suited for high-precision measurements of direct \CP violation, owing to its large yield.\footnote{The inclusion of charge-conjugate processes is implied throughout the text except in the discussion of asymmetries.}
In particular, the $\phi$ intermediate state in the \DpTophipi decay provides a favourable topology for \CP-violation studies, since the corresponding tree-level amplitude is colour-suppressed, potentially enhancing the relative impact of a new-physics penguin contribution~\cite{Atwood:2012ac}.
Consequently, precise measurements of the direct \CP asymmetry in the \DpTophipi decay, denoted as \ACPDPhiPi, provide a complementary probe of \CP violation with respect to the singly Cabibbo-suppressed $\Dz\to\Kp\Km$ and $\Dz\to\pip\pim$ channels.

Experimentally, the precision of the \ACPDPhiPi measurement is limited by the accuracy in controlling detection and production asymmetries. These effects are largely cancelled by measuring the \CP-violating asymmetry of interest relative to that of one or more abundant calibration modes with similar kinematic properties, for which the physics asymmetry is expected to be much smaller than the overall experimental uncertainty. In this case, the \DpToKSpi decay mode is employed to measure \ACPDPhiPi, with the neutral kaon reconstructed from its decay into two charged pions. It receives contributions from the Cabibbo-favoured (CF) \DpToKzbpi and the doubly Cabibbo-suppressed (DCS) \DpToKzpi amplitudes, which are dominated by tree-level \decay{\cquark}{\squark \dquarkbar \uquark} and \decay{\cquark}{\dquark \squarkbar \uquark}  transitions, respectively.  
Since these transitions are unaffected by QCD penguin and chromomagnetic dipole operators, these channels are not expected to exhibit any measurable direct \CP violation~\cite{Pajero:2022vev}.

Using the \DpToKSpi decay, however, requires accurate modelling of the 
asymmetry induced by the presence of a long-lived neutral kaon in the final state, which traverses multiple detector layers before decaying. This neutral-kaon asymmetry, which depends on the decay time of the neutral kaon, originates from several interfering physical processes: the different strong interactions of $\Kz$ and $\Kzb$ mesons with nuclei of the detector material, mixing and \CP violation in the $\Kz$--$\Kzb$ system,
and the interference of the neutral kaon time evolution with the DCS and CF amplitudes of the \Dp decay.
The first two are well understood and can be accurately predicted given a detailed description of the detector material~\cite{Good:1957zza,Sozzi:1087897,Pais:1955sm}. The third, instead, arises when two or more charm decay amplitudes contribute to the same final state. In the specific case of the \DpToKSpi decay, the interference between the CF and DCS decay amplitudes \mbox{$\mathcal{A}(\DpToKzbpi)$} and \mbox{$\mathcal{A}(\DpToKzpi )$}, combined with neutral-kaon mixing and regeneration, induces an additional time-dependent asymmetry. 
This additional asymmetry source must be properly accounted for in precision measurements~\cite{Yu:2017oky}, but has been neglected in all measurements in these channels prior to this work~\cite{LHCb-PAPER-2019-002,LHCb-PAPER-2022-024,Belle:2012ygt}.
This effect can be parametrised in terms of three observables: the relative magnitude of the two amplitudes, \rpi, their strong-phase difference, \deltapi, and their relative weak phase, $\varphi$, defined as 
\begin{equation}\label{eq:DCS-CF_ratio}
    \rpi \, e^{i(\deltapi + \varphi)} \equiv \frac{\mathcal{A}(\Dp \to \Kz \pip)}{\mathcal{A}(\Dp \to \Kzb \pip)}.
\end{equation}
The weak phase $\varphi\equiv \arg[-(\Vcds \Vus)/(\Vcss \Vud)]$ is fixed by CKM matrix elements and  its current experimental measurement is $(-6.2 \pm 0.4)\times10^{-4}$
in the SM~\cite{CKMfitter2015,UTfit-UT}.
On the other hand, \rpi and \deltapi are governed by strong-interaction dynamics. Experimental constraints on the combination of these parameters can be obtained by comparing the rates of $\Dp\to K^0_{\rm S/L} \pip$ decays~\cite{CLEO:2007rhw}, while model-dependent estimates were obtained within the factorisation-assisted topological amplitudes (FAT) theoretical framework~\cite{Wang:2017ksn,Yu:2017oky}. Within this approach, the strong parameters are predicted to be
$\rpi = -0.073 \pm 0.004$, and
$\deltapi = +1.39 \pm 0.05$~rad or $-1.39 \pm 0.05$~rad. The adopted phase convention assigns a negative value to \rpi, following Ref.~\cite{Wang:2017ksn}, while the strong phase $\deltapi$ is predicted up to a sign.

This article presents a new measurement of the direct \CP asymmetry in the \DpTophipi 
decay using the Run~2 dataset collected by the LHCb experiment between 2015 and 2018.
The result supersedes the previous LHCb measurement based on a partial dataset~\cite{LHCb-PAPER-2019-002}, and  
utilises \DpToKSpi decays as calibration, to precisely remove detection asymmetries.
For the first time, the phenomenological model of the neutral-kaon asymmetry in matter has been extended to accurately account for the contribution from incoherent \KS regeneration and CF-DCS interference. 
The measurement of \ACPDPhiPi is performed by applying different constraints to the strong parameters governing these effects, \rpi and \deltapi, using both experimental and theoretical inputs.
The inclusion of these new effects in the model 
enables the usage of the full available data sample, including for the first time the \DpToKSpi decays where the \KS mesons decay downstream of the vertex detector acceptance.  
These decays correspond to longer kaon decay times, where interference and regeneration effects are amplified and can be controlled more precisely. Their inclusion also increases the yield of \DpToKSpi decays by more than a factor of two.
In addition to the \CP asymmetry measurement, this article also reports the first determination at a hadron collider of the values of the strong parameters \rpi and \deltapi of \DpToKSpi decays.

\section{Measurement overview}
\label{sec:overview}
The analysis described in this article is based on the data sample of \proton\proton collisions at a centre-of-mass energy of $13\,\tev$ collected with the \lhcb detector during Run~2 of the \lhc, corresponding to an integrated luminosity of $6\,\invfb$. The \Dp mesons used in the analysis have been promptly produced in \proton\proton collisions.
The strategy employed to measure the direct \CP asymmetry \ACPDPhiPi has been revised and improved with respect to the approach presented in Ref.~\cite{LHCb-PAPER-2019-002}, which used only a subset of the Run~2 dataset collected between 2015 and 2017.
The present analysis therefore extends the measurement to the full Run~2 data sample by incorporating data collected in 2018 and by reanalysing the 2015--2017 sample with a unified selection and analysis procedure. Moreover, \DpToKSpi candidates in which the neutral kaon decays downstream of the vertex detector acceptance, previously excluded in Ref.~\cite{LHCb-PAPER-2019-002}, are added to the data sample. 
The online and offline requirements used to select candidates are significantly more stringent than those of Ref.~\cite{LHCb-PAPER-2019-002}, in order to robustly control detector-induced asymmetries. This is 
particularly important given the inclusion of longer-lived neutral kaons, which are particularly useful to constrain the hadronic parameters \rpi and \deltapi. 
Overall, the final dataset used in this measurement is only about a factor of 1.5 larger than that of Ref.~\cite{LHCb-PAPER-2019-002}. 
The new analysis strategy developed in this article, however, will allow to extend this measurement to much more abundant data samples, where the effect of DCS-CF interference will no longer be subdominant and will need to be accurately modelled in order not to bias the results.

The raw asymmetry for a decay mode \decay{\Dp}{f} is defined as
\begin{equation}\label{eq:raw_asy_def}
\Araw(\decay{\Dp}{f}) \equiv
\frac{N(\decay{\Dp}{f}) - N(\decay{\Dm}{\bar{f}})}
     {N(\decay{\Dp}{f}) + N(\decay{\Dm}{\bar{f}})},
\end{equation}
where $N(\decay{\Dp}{f})$ and $N(\decay{\Dm}{\bar{f}})$ denote the signal yields. In general, the raw asymmetry receives contributions not only from the \CP asymmetry of interest but also from production and detection asymmetries, which will be referred to as nuisance asymmetries.
Since the initial \proton\proton state is not \CP invariant, and the detector covers a limited solid angle, the production cross-sections of prompt \Dp and \Dm mesons can differ, giving rise to a production asymmetry $\Aprod(\Dp)$.
Furthermore, detection asymmetries $\Adet(f)$ arise from differences in the reconstruction and identification efficiencies for particles and antiparticles traversing the detector material.
To first order approximation, the raw asymmetry is given by
\begin{equation}\label{eq:Araw_linear}
\Araw(\decay{\Dp}{f}) \approx \Aprod(\Dp) + \ACPdir(\decay{\Dp}{f}) + \Adet(f),
\end{equation}
where $ \ACPdir(\decay{\Dp}{f})$ denotes the direct \CP asymmetry in the generic \decay{\Dp}{f} decay and higher-order terms, involving products of at least three asymmetries, are negligible at the current level of precision.
For typical nuisance asymmetries of order $1\%$, \cref{eq:Araw_linear} is accurate up to corrections of $\mathcal{O}(10^{-6})$. 

In the specific case of \DpTophipi decays, for any sample $s$, the raw asymmetry receives contributions from the production asymmetry of the prompt \Dp meson, $\Aprods(\Dp)$ and from the detection asymmetry of the accompanying pion, $\Adets(\pip)$, 
\begin{equation}
\Araws(\DpTophipi) =
\Aprods(\Dp) + \Adets(\pip) + \ACPDPhiPi,
\end{equation}
where \ACPDPhiPi is the direct \CP asymmetry to be determined and is expected to be independent of the chosen sample. 
The detection asymmetry associated with the charged kaons from the $\phi$ decay is assumed to vanish, since the $\phi$ is a self-conjugate state, that decays into the symmetric $\Kp\Km$ final state.
The detection asymmetries can be removed by exploiting the \DpToKSpi decays, whose raw asymmetry can be expressed in a similar way as
\begin{equation}
\Araws(\Dp \to \KS \pip) =
\Aprods(\Dp) + \Adets(\pip) + \Adets(\KS),
\end{equation}
where the term $\Adets(\KS)$ is the asymmetry associated with the experimental detection of the neutral kaon, in the sample $s$, and includes the contribution of the direct \CP violation arising in the interference between the CF and DCS decays.
Subtracting these two raw asymmetries defines the observable
\begin{equation}\label{eq:DeltaAKS0}
\begin{split}
\DeltaAs &\equiv
\Araws(\decay{\Dp}{\KS\pip}) - \Araws(\DpTophipi) \\
&\approx \Adets(\KS) - \ACPDPhiPi.
\end{split}
\end{equation}
An accurate modelling of the neutral-kaon detection asymmetry is required to measure the direct \CP asymmetry in the \DpTophipi decay. This contribution depends primarily on the momentum vector of the neutral kaon and on the amount of detector material it traverses before decaying into two charged pions, and therefore the decay time of the neutral kaon is used as the primary observable sensitive to this effect.
To study the time dependence of 
$\Adet(\KS)$, the \DpToKSpi candidates are partitioned into 13 independent subsamples, which will be referred to as bins, according to the measured \KS decay time. 
To ensure the cancellation of nuisance asymmetries, the offline selections for the \DpToKSpi and \DpTophipi channels must be strictly aligned. Furthermore, to maintain this cancellation within each specific decay-time bin $s$, the \DpTophipi candidates are split into 13 independent subsamples as described in~\cref{sec:nuisance_asys}, and each is assigned to a decay-time bin of the \DpToKSpi sample. 
Within each bin, the kinematic distributions of the \DpTophipi candidates are weighted to match those of the corresponding \DpToKSpi candidates. This procedure yields a smaller final statistical uncertainty on \ACPDPhiPi than would be obtained by weighting the kinematics of the \DpToKSpi candidates to those of the signal \DpTophipi candidates.

The observable $\DeltaAs$ of \cref{eq:DeltaAKS0} is measured independently in each decay-time bin $s$ for all subsamples, defined by data-taking year and \lhcb magnet polarity. The compatibility of the resulting measurements is verified within each bin, after which they are combined. The combined values of $\DeltaAs$ are then compared with the corresponding predictions for the average neutral-kaon detection asymmetry, \ADTilde, in order to extract \ACPDPhiPi.
The predicted values \ADTilde are computed by means of an analytical model of the neutral-kaon time evolution as it traverses the detector material. Such a model~\cite{Pais:1955sm,Good:1957zza,Fetscher:1996fa} has been used in several measurements~\cite{LHCB-PAPER-2014-013,LHCb-PAPER-2019-002,LHCb-PAPER-2022-024}. Given the detector material map, as well as the \KS momentum and decay time, the model accurately describes the interaction of a neutral kaon with the detector medium, including both the coherent forward scattering off all nuclei (transmission regeneration) and neutral-kaon mixing. For the first time, the model is extended to include the interference between CF and DCS charm decay amplitudes in the presence of \Kz--\Kzb mixing, parametrised by the hadronic parameters \rpi and \deltapi and the weak phase $\varphi$. Additionally, this model now also includes the contribution from the coherent forward scattering of the neutral kaon with all nucleons within individual nuclei, where different nuclei contribute incoherently (diffraction regeneration)~\cite{Charpak:275831,Baldini:1996ss,Kleinknecht:1973ny,PhysRev.124.1223}.

The measurement of \ACPDPhiPi is obtained by exploiting the fact that the time-dependent neutral-kaon asymmetry vanishes in the limit of zero \KS decay time, up to a small correction due to direct \CP violation in the \DpToKSpi decay arising from the interference of CF and DCS amplitudes~\cite{Wang:2017ksn}. This correction is known and incorporated into the model of \ADTilde. The direct \CP asymmetry in the CF \DpToKzbpi and DCS \DpToKzpi decays is neglected, in the assumption that both processes occur only through tree-level amplitudes. The overall constant term, measured with a template fit to data, provides a direct determination of \ACPDPhiPi.
The best-fit value of \ACPDPhiPi was examined only after the analysis strategy was finalised, to avoid experimenter bias.
In addition, this paper presents a frequentist template-based fit to the \DeltaAs values, allowing for the direct determination of confidence intervals for the strong parameters \rpi and \deltapi from the data. This represents the first measurement of these parameters in a hadronic environment.

\section{Detector and simulation}
\label{sec:detector}

The \lhcb detector~\cite{LHCb-DP-2008-001,LHCb-DP-2014-002} is a single-arm forward spectrometer covering the pseudorapidity range $2<\eta <5$, designed for the study of particles containing \bquark or \cquark quarks. The detector used to collect the data analysed in this paper includes a high-precision tracking system consisting of a silicon-strip vertex detector surrounding the $pp$ interaction region (Vertex Locator, \velo~\cite{LHCb-TDR-005}), a large-area silicon-strip detector located upstream of a dipole magnet with a bending power of about $4{\mathrm{\,T\,m}}$, and three stations of silicon-strip detectors and straw drift tubes placed downstream of the magnet.
The tracking system provides a measurement of the momentum, \ptot, of charged particles with a relative uncertainty that varies from 0.5\% at low momentum to 1.0\% at 200\gevc.
The minimum distance of a track to a primary vertex (PV), the impact parameter (\IP), is measured with a resolution of $(15+29/\pt)\mum$, where \pt is the component of the momentum transverse to the beam, in\,\gevc.
The magnetic field deflects oppositely charged particles in opposite directions, which can lead to detection asymmetries.
Therefore, its polarity is reversed around every two weeks throughout the data taking to reduce such effects.
The \lhcb coordinate system is a right-handed system centred at the nominal $pp$ interaction point, with the $z$ axis aligned along the beam direction and pointing towards the downstream detectors, the $y$ axis pointing vertically upwards, and the $x$ axis oriented horizontally. The magnetic field is aligned with the $y$ axis; when it points upwards the configuration is referred to as \MagUp, while the reversed configuration is denoted as \MagDown.
Different types of charged hadrons are distinguished using information from two ring-imaging Cherenkov (\rich) detectors.
Photons, electrons and hadrons are identified by a calorimeter system consisting of scintillating-pad and preshower detectors, an electromagnetic and a hadronic calorimeter. Muons are identified by a system composed of alternating layers of iron and multiwire proportional chambers.

The online event selection is performed by a trigger, which consists of a hardware stage followed by a two-level software stage, which applies a full event reconstruction.
At the hardware-trigger stage, events are required to contain a muon with high \pt or a hadron, photon, or electron with high transverse energy deposited in the calorimeters.
For hadrons, the transverse energy threshold is $3.5\gev$.
In between the two software stages, an alignment and calibration of the detector are performed in near real-time~\cite{LHCb-PROC-2015-011} and updated constants are made available for the trigger, ensuring high-quality tracking and particle identification (PID) information.
The excellent performance of the online reconstruction
offers the opportunity to perform physics analyses directly using candidates reconstructed at the trigger level~\cite{LHCb-DP-2012-004,LHCb-DP-2016-001}, which the present analysis exploits.
The storage of only the triggered candidates enables a reduction in the event size by an order of magnitude.

The measurement described in this paper is primarily based on data-driven techniques. However, simulation is used to estimate the neutral-kaon asymmetry, $\Adets(\KS)$, in each subsample $s$, as it encodes the \lhcb material distribution. It is also employed to assess systematic uncertainties associated with the residual contamination of $\Dp$ mesons originating from $b$-hadron decays. In the simulation, $pp$ collisions are generated using \pythia~\cite{Sjostrand:2007gs,*Sjostrand:2006za} with a specific \lhcb configuration~\cite{LHCb-PROC-2010-056}.
Decays of unstable particles are described by \evtgen~\cite{Lange:2001uf}, in which final-state radiation is generated using \photos~\cite{davidson2015photos}.
The interaction of the generated particles with the detector, and its response, are implemented using the \geant toolkit~\cite{Allison:2006ve, *Agostinelli:2002hh} as described in Ref.~\cite{LHCb-PROC-2011-006}. In order to increase the speed of producing simulated decays and allow much larger samples to be saved on disk, the simulation of a single specific decay process can be enabled, ignoring all other particles emerging from the $pp$ collisions.

\section{Candidate selection}\label{sec:selection}

\begin{figure}
    \centering
    \includegraphics[page = 1, width = 0.45\textwidth]{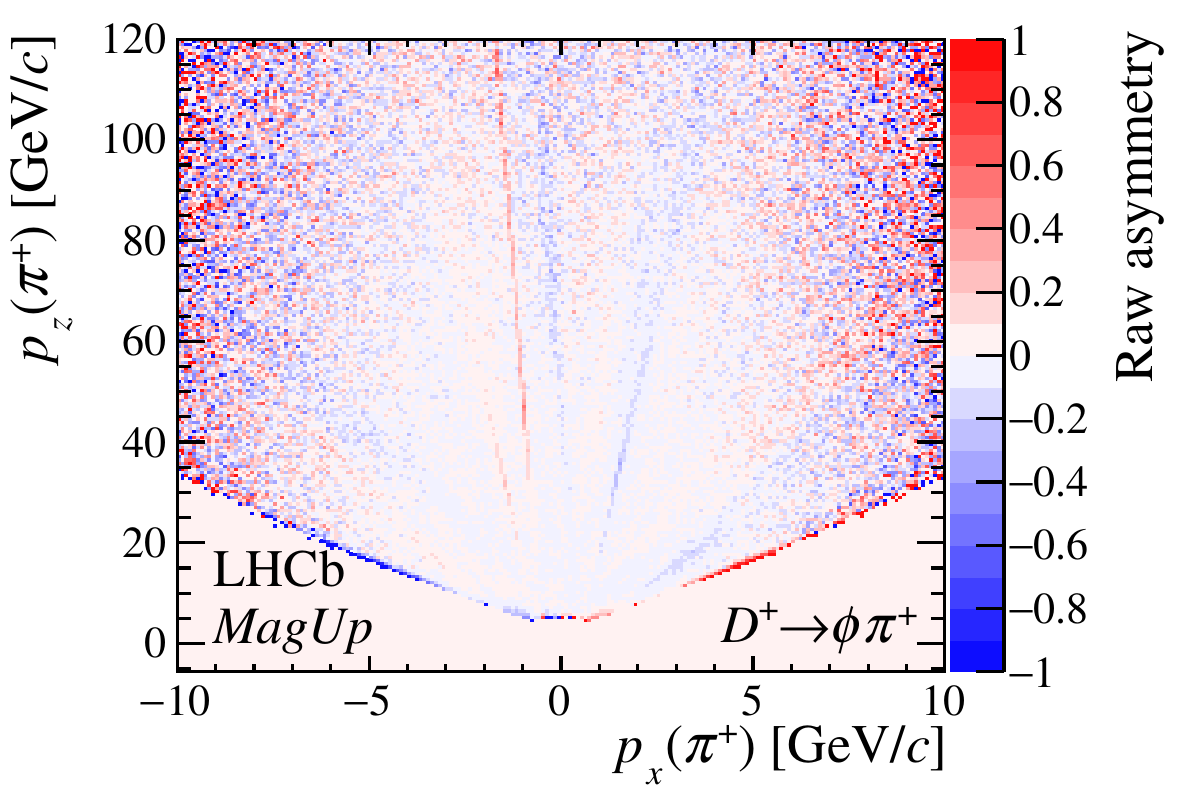}
    \includegraphics[page = 1, width = 0.45\textwidth]{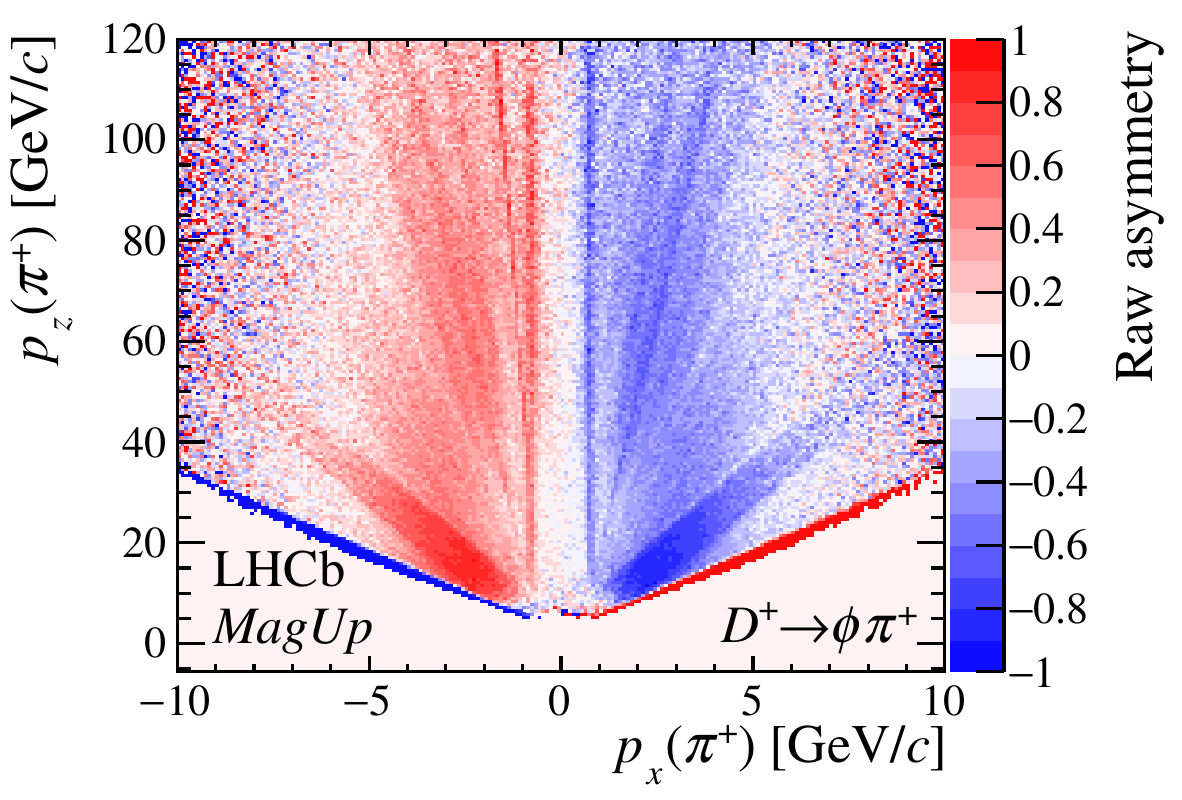}
    \caption{Distribution of the raw asymmetry in the \mbox{$D^+ \to \phi \pi^+$} sample in the $p_x(\pi^+)$--$p_z(\pi^+)$ plane for Run~2 data collected with positive magnet polarity, shown for different hardware-trigger categories.
(Left) Candidates selected independently of the $D^+$ decay products, or selected via a large transverse-energy deposit in the calorimeters from both decay products of the $\phi$ meson; these candidates are used in the analysis.
(Right) Discarded candidates selected by the hardware trigger based on a large transverse-energy deposit in the calorimeters from the $\pi^+$.}
    \label{fig:Fig1}
\end{figure}

The \DpToKSpi and \DpTophipi decay candidates, with $\KS \to \pip\pim$ and $\phi \to \Kp\Km$, are fully reconstructed online and selected by a dedicated trigger. 
Additional selection criteria are then applied offline to further suppress background contamination.
The hardware-trigger decisions used in this analysis are either completely independent of the \Dp decay products or, alternatively, require a large transverse energy deposit in the calorimeters from both decay products of the \KS or $\phi$ meson.
This requirement removes approximately 25\% of signal candidates in both decay modes, which would otherwise be selected exclusively through the energy deposited by the pion from the \Dp decay or by a single decay product of the \KS or the $\phi$ meson.
This requirement removes regions of phase space, mainly close to the edges of the calorimeter acceptance, which exhibit large detection asymmetries.
The impact of the hardware-trigger requirement on the charge asymmetries is illustrated in \cref{fig:Fig1}, which shows that the rejected sample displays raw asymmetries from 10 to 100\% in extended regions in the $p_x(\pip)$--$p_z(\pip)$ plane.

The first stage of the software trigger requires at least one track, reconstructed using information from all three tracking detectors, with high \pt and significant IP with respect to all PVs. In the offline selection, the \pip meson from the \Dp decay is required to fulfil these criteria; it is therefore referred to hereafter as the \textit{trigger pion}.
Candidates in which the \pip meson from the \Dp decay is not selected at the first-stage software trigger are excluded from this analysis, even if they would otherwise be selected by the trigger using information from the decay products of the \KS or $\phi$ meson.
As a result, approximately 57\% (90\%) of \DpTophipi (\DpToKSpi) candidates are retained. 
This selection aims to equalise the first-stage software-trigger conditions between the two decay modes,
and to ensure a robust cancellation of detection charge asymmetries at this level.

The second-stage software trigger selects events compatible with the \DpToKSpi and \DpToKmKppi decay topologies.
The \DpToKSpi decay is reconstructed by combining a \KS meson candidate with a well-reconstructed charged track, forming a \Dp candidate with a good-quality vertex fit. Each charged track is required to be compatible with the pion mass hypothesis using information from the \rich detectors. The neutral kaon candidate is required to be displaced from the PV and is reconstructed from two oppositely charged pion tracks, originating from a common vertex with a good fit quality, an IP significance with respect to its associated PV larger than six Gaussian-equivalent standard deviations, and an invariant mass compatible with the known \KS mass~\cite{PDG2024}.
Based on the tracking information used to reconstruct the charged pions, neutral kaons are classified into two exclusive categories:
those in which the decay occurred early enough for the pions to be reconstructed with at least three 
hits in different silicon layers of the \velo (\textit{long}, indicated
in the following by \KSLL); and those decaying later, such that segments of pion tracks can only be formed in the silicon-strip tracker upstream of the magnet, and the tracking stations after the magnet (\textit{downstream}, indicated in the following by \KSDD). 
The \KSLL candidates have flight distances up to 600\mm, corresponding to decay times of [0.0, 0.4]$\tau_\KS$, where $\tau_\KS$ is the known \KS lifetime~\cite{PDG2024}.
The \KSDD have flight distances between 350 and 2240\mm, corresponding to decay times of [0.1, 3.0]$\tau_\KS$.
Due to the absence of \velo information, the momentum and flight-distance resolutions of \KSDD candidates are generally worse than those of \KSLL candidates. In particular, the $\pip\pim$ invariant-mass resolution is approximately twice that of \KSLL candidates. In addition, the uncertainty on the longitudinal position of the \Dp decay vertex, which is typically 0.5\mm for \DpToKSLLpi candidates, is about 10\mm for \DpToKSDDpi candidates.
The \DpToKmKppi decay is reconstructed by combining three well-reconstructed charged tracks into a common vertex, forming a \Dp candidate with good vertex-fit quality. All charged tracks are required to have \pt greater than 0.25\gevc. Particle-identification requirements, based on information from the \rich detectors, are applied to distinguish kaons from pions.
This selection helps suppress residual combinatorial background, as well as partially reconstructed and misidentified decays.

For both the \DpToKSpi and \DpToKmKppi decay modes, the \Dp candidate is required to have an invariant mass compatible with the known value~\cite{PDG2024}, and a decay time with respect to the PV that fits best its flight direction larger than 0.25\ps and 0.4\ps, respectively.
The angle between the \Dp momentum and the vector connecting the PV to the \Dp decay vertex is required to be smaller than 10\mrad (17\mrad) for the \DpToKSpi (\DpToKmKppi) channel. All charged tracks are required to be displaced from the PV, with an IP significance larger than six (two) Gaussian-equivalent standard deviations for the \DpToKSpi (\DpToKmKppi) channel. 
The \pt of the \Dp candidate is required to be sufficiently large to suppress random combinations of low-momentum tracks. Specifically, the sum of the transverse momenta of the \Dp decay products must exceed $2\gevc$ for \DpToKSpi candidates and $3\gevc$ for \DpToKmKppi candidates.

The offline selection is designed to equalise the kinematics of the different \Dp decay samples and to minimise background contributions as well as spurious asymmetries. 
For the \DpToKSpi samples, stricter mass requirements with respect to the software trigger are imposed on the pion pair to suppress combinatorial background from random track pairs. In particular, the two pion candidates are required to have an invariant mass within $\pm 35\mevcc \ (\pm 64\mevcc)$ of the known \KS mass for \KSLL (\KSDD) candidates. 
For the \DpTophipi decay mode, the two kaon candidates are required to form an invariant mass within $\pm 10\ \mevcc$ of the known $\phi$ mass~\cite{PDG2024}, ensuring a clean selection of the resonant decay. 
In particular, to match the tight software-trigger selection on the \Dp \pt in the \DpTophipi sample, a requirement of $\pt(\Dp) > 2.8\ \gevc$ is applied to all candidates, thereby reducing kinematic differences between the \DpToKSpi and \DpTophipi samples. 
The trigger pion, with momentum components $p_{x,y,z}$, is required to satisfy four fiducial conditions on its angular variables $\theta_{x,y} = \arctan(p_{x,y}/p_z)$: 
a momentum-dependent acceptance cut $ |\theta_x| <  0.3 - (1\gevc)/\sqrt{p_x^2 + p_z^2}$; a cut excluding the elliptical region, $(\theta_x/0.027)^2 + (\theta_y/0.017)^2 > 1$; a minimum vertical angle requirement \mbox{$|\theta_y| > 0.001$}; a veto rejecting candidates inside the regions given by $|\theta_y| < 0.005$ and \mbox{$0.06 < |\theta_x| < 0.1$}.
These requirements remove regions with very large asymmetries due to the detector acceptance, and reject about 15\% of the signal candidates that are produced close to the beam pipe or that travel close to the edges of the PID and calorimeter systems. This approach closely follows that used in other \lhcb precision measurements~\cite{LHCb-PAPER-2019-006,LHCb-PAPER-2022-024,LHCb-PAPER-2024-008}, and ensures precise and robust cancellation of instrumental charge asymmetries.

A fraction of reconstructed \Dp candidates originates from the decays of long-lived $b$ hadrons (secondary decays), thereby contaminating both signal samples. Since the production asymmetries of $b$ hadrons generally differ from the \Dp production asymmetry, any difference in the residual secondary fraction between the \DpToKSpi and \DpTophipi samples could bias the determination of the \DeltaAs observables.
Unlike promptly produced \Dp mesons, secondary \Dp decays do not originate directly from the PV. Their contribution can therefore be suppressed by requiring the \Dp candidate to have a small impact parameter with respect to the PV. The exact threshold values depend on the decay mode and on the \KS reconstruction category (\KSLL or \KSDD). In particular, the resolution of the \ipDp variable is significantly worse in \DpToKSDDpi decays, since the \Dp decay vertex is reconstructed from the trigger pion and a displaced \KS candidate. The long \KS flight distance amplifies the uncertainty on the \Dp direction and decay-vertex position.
\begin{figure}
    \centering
    \includegraphics[page = 1, width = 0.45\textwidth]{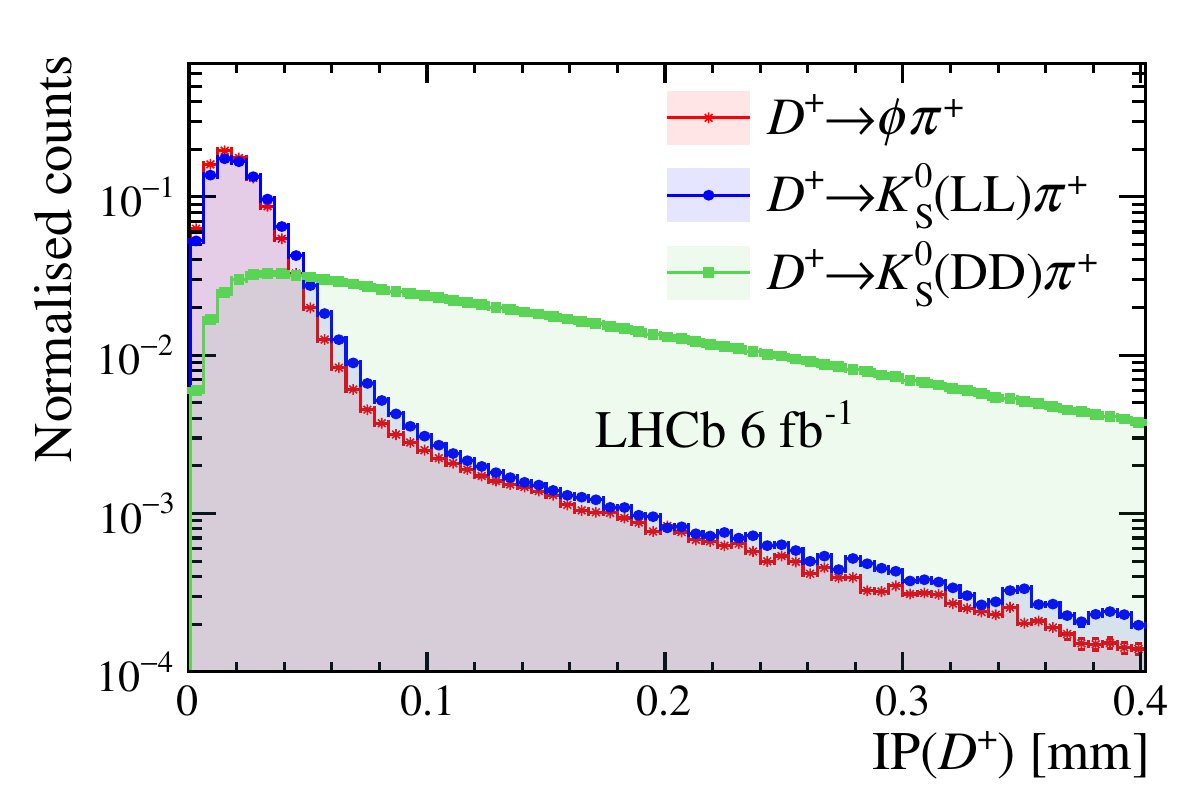}
    \includegraphics[page = 1, width = 0.45\textwidth]{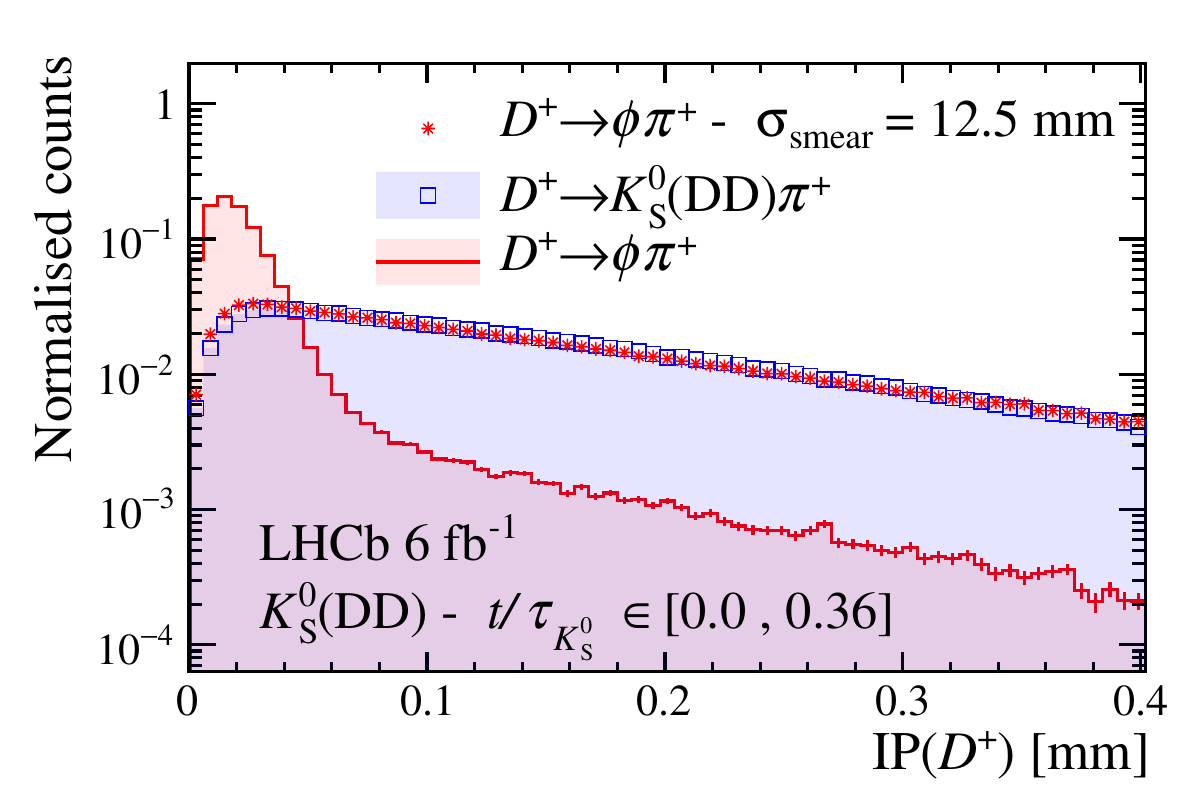}

    \caption{(Left) Distribution of $\mathrm{IP}(D^+)$ in the \mbox{$D^+ \to K_{\mathrm{S}}^0(\mathrm{LL}) \pi^+$}, \mbox{$D^+ \to K_{\mathrm{S}}^0(\mathrm{DD}) \pi^+$}, and \mbox{$D^+ \to \phi \pi^+$} channels. (Right) Distribution of the smeared $\mathrm{IP}(D^+)$ in the \mbox{$D^+ \to \phi \pi^+$} subsample associated with the first DD-type decay-time bin, obtained by smearing the $D^+$ decay vertex in \mbox{$D^+ \to \phi \pi^+$} candidates with a width of \mbox{$\sigma_{\mathrm{smear}} = 12.5\,\mathrm{mm}$}. The smeared $\mathrm{IP}(D^+)_{\mathrm{smeared}}$ distribution is compared with the target $\mathrm{IP}(D^+)$ distribution of the corresponding \mbox{$D^+ \to K_{\mathrm{S}}^0(\mathrm{DD}) \pi^+$} subsample.}
    \label{fig:Fig2}
\end{figure}
As shown in the left panel of \cref{fig:Fig2}, in the \DpTophipi and \DpToKSLLpi samples the prompt peak and the secondary tail at large \ipDp values can be clearly separated. 
In contrast, in the \DpToKSDDpi sample, the poorer \ipDp resolution makes the two categories largely indistinguishable.

For all subsamples corresponding to decay-time bins that include \DpToKSLLpi candidates (and the corresponding \DpTophipi candidates), a tight requirement of $\ipDp < 100\,\mum$ is applied. This selection is effective in suppressing secondary decays, thanks to the good vertex and momentum resolutions for \KSLL candidates. 
This selection is the same in all \KSLL decay-time bins, since for \DpToKSLLpi candidates the \ipDp distribution does not show a significant correlation with the neutral-kaon decay time.

For \DpToKSDDpi candidates, instead, it is not possible to efficiently separate secondary decays from promptly produced \Dp mesons using the \ipDp variable. The \KSDD decay vertex is reconstructed far from the PV and outside the \velo acceptance, resulting in an \ipDp resolution of about 100--500\mum, comparable to the typical impact parameter of \Dp mesons produced in $b$-hadron decays. Consequently, an \ipDp requirement has only a limited effect on the secondary-decay contamination in this sample. Nevertheless, it is effective in reducing combinatorial background from random track combinations not originating from a genuine \Dp decay vertex. For this reason, a looser requirement of $\ipDp < 350\,\mum$ is applied to \DpToKSDDpi candidates.
To reproduce an equivalent effect in the \DpTophipi subsamples associated with decay-time bins containing \DpToKSDDpi candidates, the \Dp decay vertex is smeared along the direction of the pion momentum. The smearing width $\sigma_{\rm smear}$ is tuned in bins of \KSDD decay time to reproduce the observed \ipDp distributions in data, with values ranging from approximately 7 to 12\mm. 
The resulting \ipDpsm distribution in a \DpTophipi subsample is shown in the right panel of \cref{fig:Fig2}.
After this procedure, the requirement $\ipDpsm< 350\,\mum$ is applied, ensuring a consistent treatment of the secondary contribution between the \DpTophipi and \DpToKSDDpi samples.

\begin{figure}
    \centering
    \includegraphics[page = 1, width=0.45\linewidth]{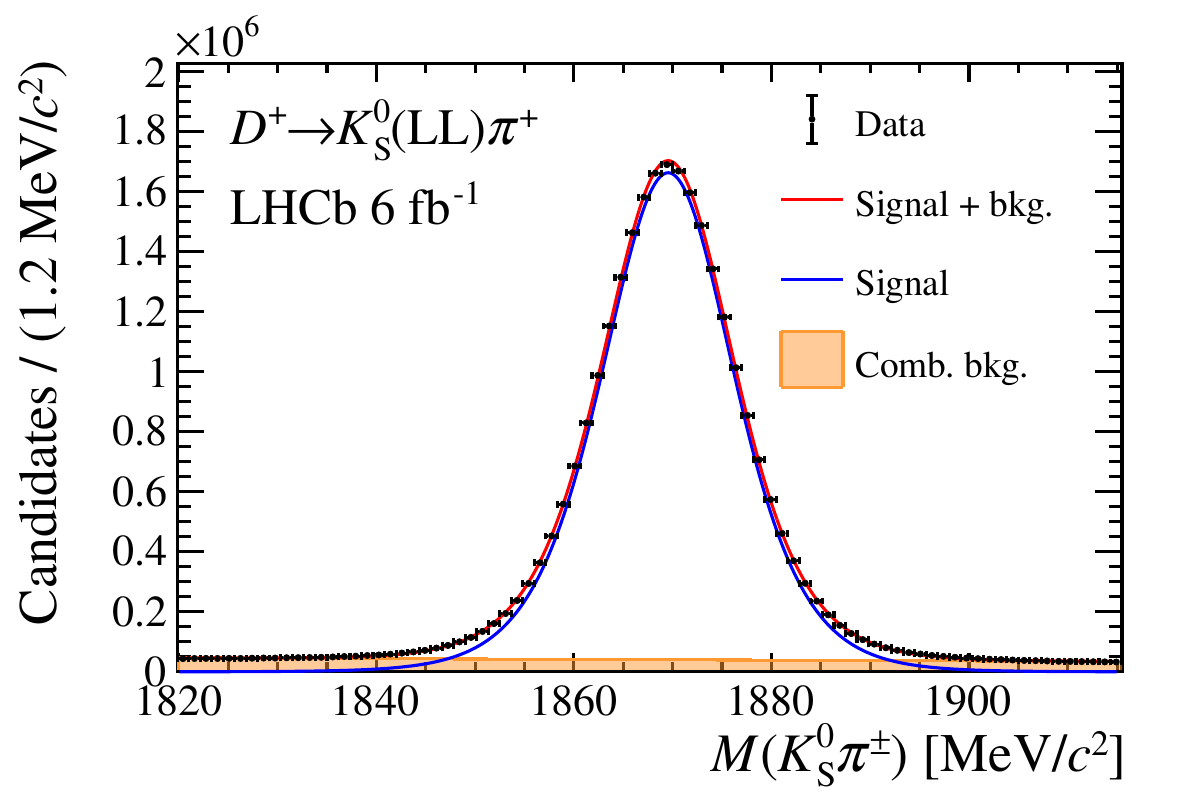}
    \includegraphics[page = 1, width=0.45\linewidth]{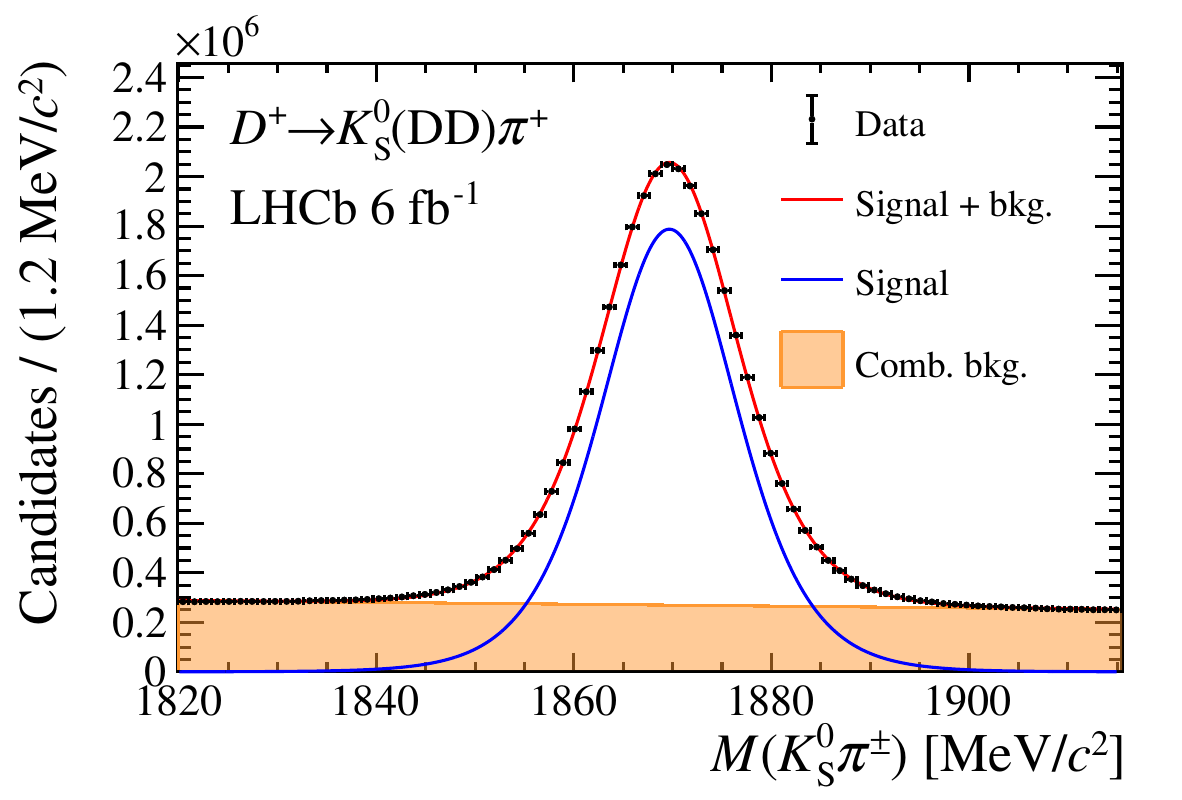}
    \includegraphics[page = 1, width=0.45\linewidth]{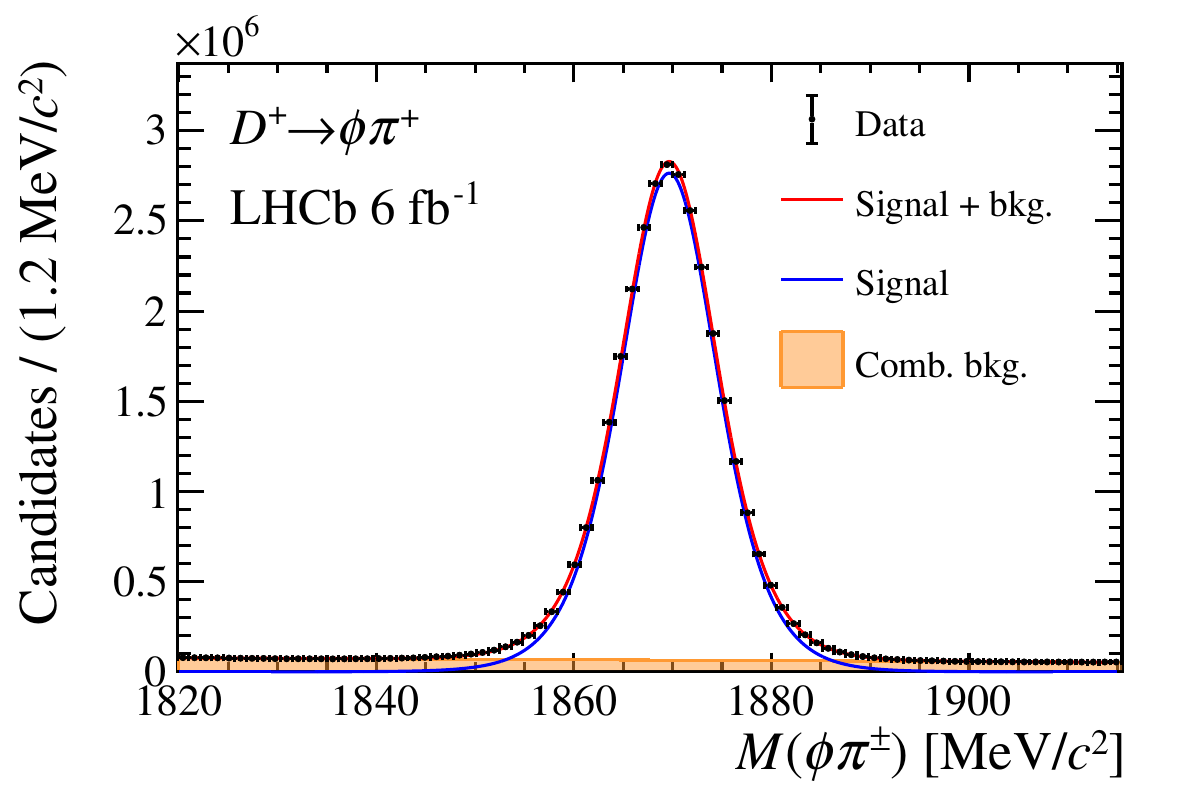}
    \caption{Invariant-mass distribution of (left) \mbox{$D^+ \to K_{\mathrm{S}}^0(\mathrm{LL}) \pi^+$}, (right) \mbox{$D^+ \to K_{\mathrm{S}}^0(\mathrm{DD}) \pi^+$}, and (bottom) \mbox{$D^+ \to \phi \pi^+$} decay candidates after the application of online and offline requirements.
    }
    \label{fig:Fig3}
\end{figure}

The invariant-mass distributions of \DpToKSpi and \DpTophipi candidates after all online and offline selections are shown in~\cref{fig:Fig3}. 
The results of fits to these distributions are 
also shown for illustrative purposes.
The probability density function (\PDF) employed to perform the fits is described in~\cref{sec:nuisance_asys}, where fits are performed on different subsamples to determine the asymmetry differences \DeltaAs.
The distribution of signal decay candidates has a standard deviation $\sigma$ of 8.6\mevcc for the \DpToKSLLpi decays, 9.0\mevcc for the \DpToKSDDpi decays, and 6.2\mevcc for the \DpTophipi decays. The background is dominated by genuine \KS and $\phi$ mesons associated with uncorrelated particles and displays, to a very good approximation, a linear behaviour. The presence of background contributions due to misidentified or partially reconstructed decays
under the \Dpm mass peaks is found to have a minor effect, and is included in the systematic uncertainties, as detailed in \cref{sec:systematics}.
Selecting candidates within a mass range of $\pm3\sigma$ around the known \Dp mass, signal purities of 94\% are achieved for the \DpToKSLLpi and \DpTophipi modes, and of 71\% for \DpToKSDDpi. The signal yields are 25.6 million, 28.4 million, and 31.5 million, respectively.

\section{Cancellation of nuisance asymmetries}\label{sec:nuisance_asys}

The \Dp production asymmetry in $pp$ collisions, $\Aprod(\Dp)$, originates from the fact that \cquarkbar quarks can hadronise by combining with valence quarks from the colliding protons, whereas this mechanism is not available to \cquark quarks. As a consequence, this asymmetry depends on the momentum of the \D mesons.
The detection asymmetry associated with the reconstruction of the trigger pion, $\Adet(\pip)$, instead arises from the geometrical acceptance of the \lhcb detector and from the different interaction cross-sections of particles and antiparticles with the detector material.
Similarly to $\Aprod(\Dp)$, this asymmetry depends on the momentum of the trigger pion.
Both $\Aprod(\Dp)$ and $\Adet(\pip)$ are largely suppressed by the requirements reported in \cref{sec:selection}, leaving only residual effects at the few-percent level due to tracking inefficiencies and detector misalignment. In addition, averaging the final results over the two magnet polarities further suppresses any residual asymmetries that may survive the selection criteria and the cancellation mechanism described in~\cref{eq:DeltaAKS0}.

More precisely, the subdivision into different subsamples is driven by the distribution of the \KS decay time. Six decay-time intervals are defined for the LL sample in the range $[0.0,\,0.4]\;\tau_\KS$, and seven decay-time intervals for the DD sample in the range $[0.1,\,3.0]\;\tau_\KS$. 
The edges of the decay-time bins are reported in Tab.~\ref{tab:Tab1}.
The kinematic distributions of the \Dp meson and of the trigger pion need to be equalised between the \DpToKSpi and \DpTophipi samples, separately for each neutral-kaon decay-time subsample~$s$. This procedure guarantees a cancellation of the $\Aprods(\Dp)$ and $\Adets(\pip)$ contributions to a very high level of precision in the measured \DeltaAs.
To ensure this, the \DpTophipi data sample is partitioned into 13 corresponding subsamples by assigning each candidate to a $\KS$ decay-time bin. For this purpose, three-dimensional histograms of kinematic variables (namely the $\pt$ and $\eta$ of the \Dp meson and the $\pt$ of the trigger pion) are constructed from the \DpToKSpi sample in each $\KS$ decay-time bin. These histograms determine the probability of assigning each \DpTophipi candidate to a decay-time bin, such that the resulting \DpTophipi subsamples 
reproduce the kinematic distributions of the corresponding \DpToKSpi decay-time bin. This procedure preserves the correlations between decay time and kinematics observed in the \DpToKSpi sample, thereby maximising the statistical sensitivity of the analysis and simplifying the subsequent weighting procedure~\cite{thesis_celestino}.

\begin{table}[t]
    \centering
     \caption{Effective signal yields of \mbox{$D^+ \to K_{\mathrm{S}}^0 \pi^+$} and \mbox{$D^+ \to \phi \pi^+$} candidates in the different decay-time bins after the kinematic weighting.
The decay-time bin boundaries are given in units of the $K_{\mathrm{S}}^0$ mean lifetime, $\tau_{K_{\mathrm{S}}^0}$.
Reported values refer to the full Run~2 dataset.}
    \label{tab:Tab1}
    \renewcommand{\arraystretch}{1.2} 
    \begin{tabular}{l|c|c}
    \hline
\hline

$t/\tau_\KS$ bin  & \DpToKSpi  & \DpTophipi \\ 

\hline
\hline

\KSLL  &  &   \\

[0.000, 0.048] &  2.7 M & 1.7 M \\ 

[0.048, 0.076] &  2.3 M & 1.3 M \\ 

[0.076, 0.104] &  1.7 M & 1.0 M \\ 

[0.104, 0.148] &  1.9 M & 1.0 M \\ 

[0.148, 0.216] &  1.4 M & 0.7 M \\ 

[0.216, 0.400] &  0.4 M & 0.2 M \\ 

\hline

\KSDD &  &   \\

[0.10, 0.36] &  3.3 M & 2.6 M \\ 

[0.36, 0.51] &  2.4 M & 1.7 M \\ 

[0.51, 0.66] &  2.2 M & 1.3 M \\ 

[0.66, 0.84] &  2.1 M & 1.2 M \\ 

[0.84, 1.05] &  1.9 M & 0.9 M \\ 

[1.05, 1.41] &  1.8 M & 0.7 M \\ 

[1.41, 3.00] &  1.2 M & 0.5 M \\
\hline
\hline
    \end{tabular}
\end{table}

After this splitting procedure, each \DpTophipi subsample is associated with the corresponding \DpToKSpi subsample.
For each pair of subsamples (\ie for each decay-time bin), the momentum distributions of both the \Dp meson and the trigger pion are equalised using a multidimensional weighting procedure.
The weighting is performed on \DpTophipi candidates iteratively, through a sequence of seven three-dimensional weighting steps, each designed to match a selected combination of momentum components of the \Dp and \pip mesons between the two subsamples. The procedure is terminated once all relevant kinematic distributions are in agreement; specifically, when the ratios of the background-subtracted normalised histograms of each variable between the \DpToKSpi and \DpTophipi subsamples are compatible with unity within statistical uncertainties in the majority of bins.
At the end of the procedure, small residual differences remain between the kinematic distributions of the \DpToKSpi and \DpTophipi samples, particularly near the boundaries of the momentum acceptance. These intrinsically arise from the fact that the procedure targets a highly correlated six-dimensional kinematic space, while operating through an iterative weighting of lower-dimensional, three-dimensional subspaces.
Any further refinement of the procedure is found to induce changes that are much smaller than the final statistical uncertainty. A dedicated systematic uncertainty is therefore assigned to account for these residual differences, as described in~\cref{sec:systematics}. The weighting procedure leads to a loss of the statistical power of the selected candidates. 
The effective signal yield of each decay mode can be estimated as \mbox{$N_{\rm eff} = (\sum_{i=1}^K w_i)^2/\sum_{i=1}^K w_i^2$}, where $K$ is the total number of
candidates and $w_i$ are the weights, which account for both the background subtraction and kinematic weights. 
The resulting effective yields are $14.8$ million for \DpTophipi candidates, $10.4$ million for \DpToKSLLpi candidates, and $14.9$ million for \DpToKSDDpi candidates. 
The effective signal yields of each \DpToKSpi and \DpTophipi subsample after the kinematic weighting are reported in \cref{tab:Tab1}.

Finally, after the splitting and kinematic weighting procedures, the raw asymmetries are determined independently in each \KS decay-time bin for the \DpToKSpi and \DpTophipi samples. They are extracted via a simultaneous binned \chisq fit to the invariant-mass distributions of \Dp and \Dm candidates, namely $M(\KS\pipm)$ and $M(\phi\pipm)$, in the \DpToKSpi and \DpTophipi channels, respectively.
The fits employ empirical PDFs to model the signal and combinatorial background contributions. All PDF parameters are determined independently for each decay-time interval and channel. The signal PDFs 
are described by a single \jsu function~\cite{Johnson:1949zj}. Within each decay-time bin, the signal shape is constrained to be identical for \Dp and \Dm candidates, except for a global shift of the peak position
to account for small residual inaccuracies in the momentum calibration for particles of opposite charge.
The combinatorial background
is modelled using a normalised first-order polynomial function, with the same functional form and shared parameters between \Dp and \Dm within each fit.
The average goodness of fit is moderate, with an average \chisqndf of 1.64 (1.68) and ndf = 79, across all fits in the \DpToKSpi (\DpTophipi) data samples, which are subdivided by decay-time bin, data-taking year, and magnet polarity. 
This behaviour is expected, as a deliberately simple model with only 11 free parameters is employed, chosen to ensure the robustness of the analysis and the numerical stability of the fits across all subsamples.

\begin{figure}[t]
    \centering    
    \includegraphics[page = 1, width=0.485\linewidth]{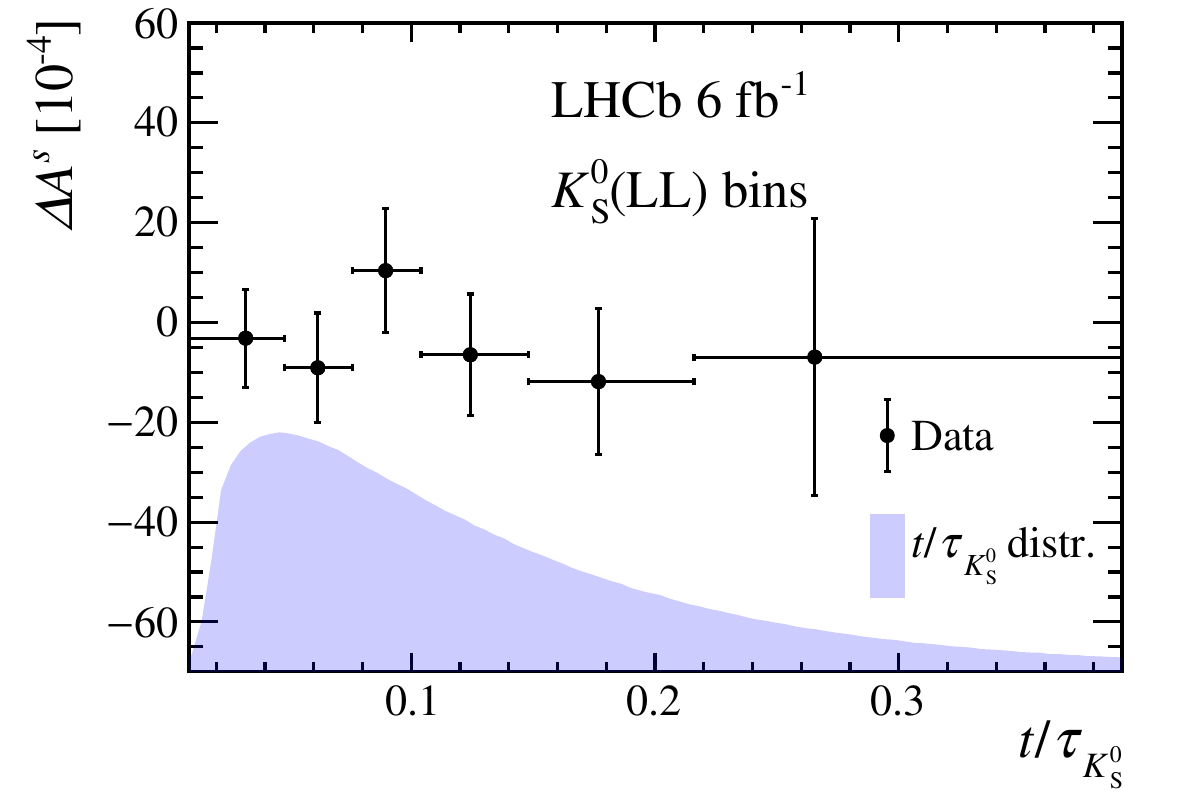}
    \includegraphics[page = 1, width = 0.485\textwidth]{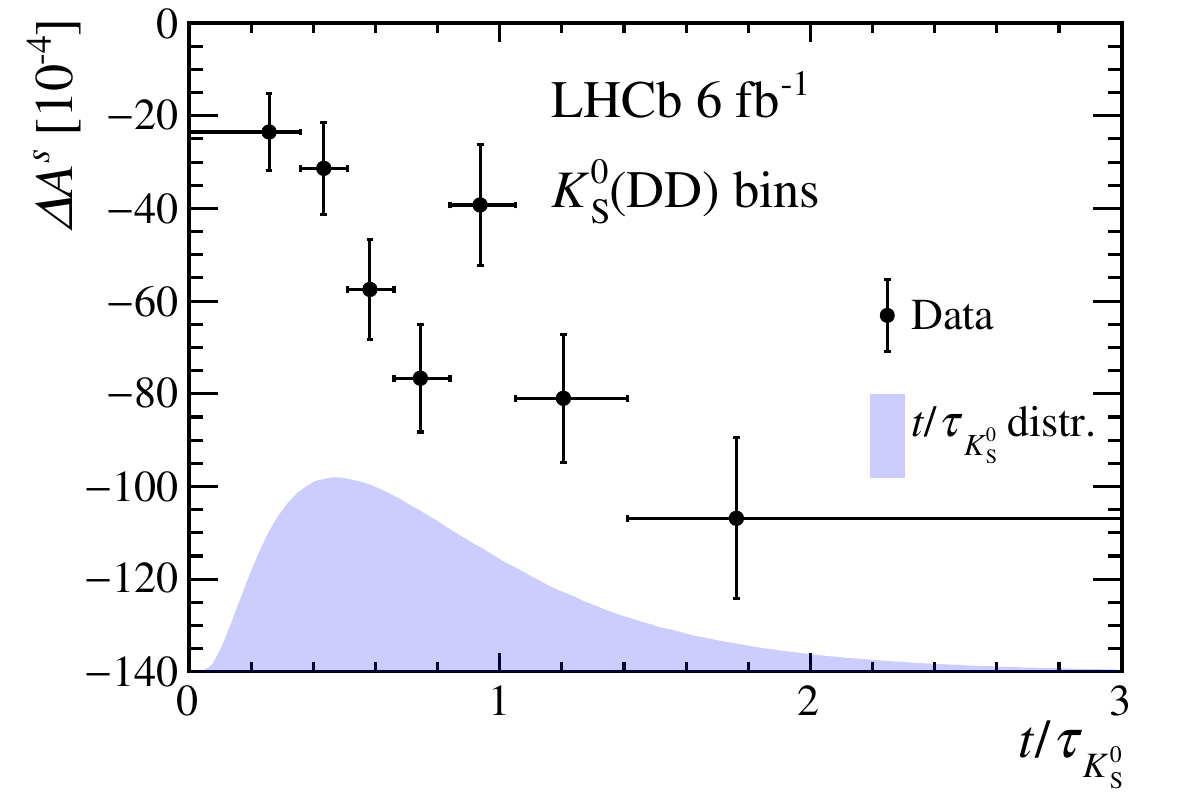}
    \caption{Measurements of $\Delta A^s$ in the different (left) $K_{\mathrm{S}}^0(\mathrm{LL}) $ and (right) $K_{\mathrm{S}}^0(\mathrm{DD}) $ subsamples, obtained as average on the Run~2 dataset. The horizontal bars have illustrative purpose and show the edges between the bins.
    The filled histogram shows the $t/\tau_{K_{\mathrm{S}}^0}$ distribution in arbitrary units.}
    \label{fig:Fig4}
\end{figure}

After determining the raw asymmetries for all \DpToKSpi and \DpTophipi subsamples, their differences \DeltaAs, as defined in \cref{eq:DeltaAKS0}, are measured. This is performed separately by data-taking year and magnet polarity. The results of each time bin show good consistency across different datasets, with a global \pvalue of 61\%. Since no dependence on data-taking conditions remain, the results from all Run~2 datasets are combined into averaged values of \DeltaAs in each decay-time bin, as shown in~\cref{fig:Fig4}. These observables are used to constrain the models of the time dependence of the neutral-kaon asymmetry in LHCb, to measure \ACPDPhiPi and to determine the CF-DCS interference parameters in the \DpToKSpi decay, \rpi and \deltapi, as presented in~\cref{sec:meas_Acp_d2phipi} and~\cref{sec:meas_rpi_deltapi_plugin}.

\section{Time-dependent neutral-kaon asymmetry}\label{sec:ADKS0Model}

The prediction of the time-dependent neutral-kaon asymmetry in each subsample,
\ADTilde, is obtained using a phenomenological model originally developed in Ref.~\cite{Stahl:2014xlz} for
previous \lhcb measurements~\cite{LHCB-PAPER-2014-013,LHCb-PAPER-2019-002,LHCb-PAPER-2022-024}.
The computation is performed on a candidate-by-candidate basis by propagating the
neutral-kaon state through the detector material, using the detailed
\geant-based description of the \lhcb geometry.

In \DpToKSpi decays, the neutral kaon is produced as a coherent superposition of the \Kz and \Kzb states, with coefficients determined by the CF and DCS amplitudes. These are parametrised by \rpi, \deltapi, and $\varphi$, as defined in \cref{eq:DCS-CF_ratio}. The weak phase $\varphi$, fixed by the CKM matrix elements, changes sign under charge conjugation and induces a small direct \CP asymmetry in the charm decay of order $8\times10^{-5}$.
Up to an overall normalisation, the initial state of the neutral kaon can be written as
\begin{equation}\label{eq:psiKplus0_KzKzb}
\begin{split}
 \ket{\psi_K^{(+)}(0)}
 &\propto \mathcal{A}(\Dp \to \Kzb \pip)\ket{\Kzb}
 + \mathcal{A}(\Dp \to \Kz \pip)\ket{\Kz} \\
 &\propto \ket{\Kzb}
 + \rpi \, e^{i(\deltapi+\varphi)} \ket{\Kz} \, ,
\end{split}
\end{equation}
while for the \CP-conjugate decay \DmToKSpi the corresponding initial state is
\begin{equation}\label{eq:psiKminus0_KzKzb}
\begin{split}
    \ket{\psi_K^{(-)}(0)}
    &\propto \mathcal{A}(\Dm \to \Kz \pim)\ket{\Kz}
    + \mathcal{A}(\Dm \to \Kzb \pim)\ket{\Kzb} \\
    &\propto \ket{\Kz}
    + \rpi \, e^{i(\deltapi-\varphi)} \ket{\Kzb} \, .
\end{split}
\end{equation}
The kaon trajectory is discretised into steps corresponding to homogeneous material layers.
At each step, the neutral-kaon state is updated to account for regeneration and \CP-violating effects using the formalism for time evolution in a homogeneous medium.
The evolution is obtained by solving Schrödinger’s equation with an effective Hamiltonian incorporating \Kz--\Kzb mixing, \CP violation, and coherent forward scattering on nuclei~\cite{Pais:1955sm,Good:1957zza,Fetscher:1996fa}.
In the $\ket{\KS}$--$\ket{\KL}$ basis, the state at decay time $t$ is written as
\begin{equation}
\ket{\psi_K^{(\pm)}(t)} = \alphaSpm(t)\ket{\KS} + \alphaLpm(t)\ket{\KL} \, ,
\end{equation}
where the coefficients $\alphaSLpm(t)$ depend on the initial conditions and the Hamiltonian parameters, as summarised in \cref{app:ADKS0Model}.
The evolution is governed by the difference of the eigenvalues of the Hamiltonian, 
which encodes mixing and decay in the neutral kaon system, and by the regeneration parameter $\Delta\chi$, determined by the difference of the interaction cross-sections of \Kzb and \Kz with nuclei, $\sigma_\Kzb-\sigma_\Kz$~\cite{Gsponer:1978dt,PhysRevLett.75.2070},
\begin{equation}\label{eq:Deltachi}
 \Delta\chi \equiv \cfrac{p \rho N_A }{ 2M_K A\sin(\arg\Delta f)}\ (\sigma_\Kzb-\sigma_\Kz) \ e^{i\arg\Delta f}
    \propto p^{(1-n)} \cdot \rho \cdot A^{(m-1)}   \, .
\end{equation}
Here, $N_A$ is the Avogadro number, $M_K$ is the neutral kaon mass, $p$ is  the neutral kaon momentum, $A$ is the atomic weight of the material, $\rho$ is the material density. The following experimental values, determined in Refs.~\cite{Gsponer:1978dt,PhysRevLett.75.2070}, are used: $\arg\Delta f=(-124.7\pm0.8)^\circ$, $n = 0.614\pm0.009$ and $m = 0.758\pm 0.003$.
This framework allows the coefficients $\alphaSL(t)$ to be updated at each material layer of the LHCb detector and enables the computation of the decay width of a neutral kaon produced in a \Dp or \Dm decay into the $\pip\pim$ \CP-even final state at decay time $t$. This width can be estimated from the projection of the state $\ket{\psi_K^{(\pm)}(t)}$ onto the \CP-even state $\ket{\Kone} \propto \ket{\Kz} - \ket{\Kzb}$, as
\begin{equation}
       \Gamma(\psi_K^{(\pm)}(t) \to \pip\pim)  \propto  \left|\alphaSpm(t) + \varepsilon \cdot \alphaLpm(t)\right|^2,
\end{equation}
where $\varepsilon$ is complex parameter, of the order of $10^{-3}$, that quantifies \CP violation in neutral kaon mixing. Direct \CP violation in the neutral-kaon decay, which is of order $10^{-6}$~\cite{Sozzi:1087897}, has been neglected.
The resulting per-candidate asymmetry is defined as
\begin{equation}\label{eq:asy_KSKL_strongphase}
    a(\KS, t) \equiv 
    \frac{
        \Gamma\!\left[\psi_K^{(+)}(t) \to \pip\pim\right]
        - \Gamma\!\left[\psi_K^{(-)}(t) \to \pip\pim\right]
    }{
        \Gamma\!\left[\psi_K^{(+)}(t) \to \pip\pim\right]
        + \Gamma\!\left[\psi_K^{(-)}(t) \to \pip\pim\right]
    } \, ,
\end{equation}
where $\psi_K^{(\pm)}(t)$ denotes the neutral-kaon state at time $t$, originating from \DpmToKSpi decays at time zero.
The asymmetry depends on the parameters governing neutral-kaon mixing, decay, and regeneration in the \lhcb detector, as well as on the parameters \rpi, \deltapi, and $\varphi$.
The only free parameters of the model are the strong parameters \rpi and \deltapi, which are allowed to take values in the ranges $[-1, \ 0]$ and $[-\pi, \ \pi]$, respectively.
All other parameters are fixed to their known values, as detailed in \cref{app:ADKS0Model}, and their uncertainties give a negligible contribution to the asymmetry, when compared to the statistical uncertainties.
The dominant source of systematic uncertainty arises from the imperfect knowledge of the \lhcb material distribution and is discussed in \cref{sec:systematics}.

\begin{figure}[t]
    \centering
     \includegraphics[ page = 1, width = 0.45\textwidth]{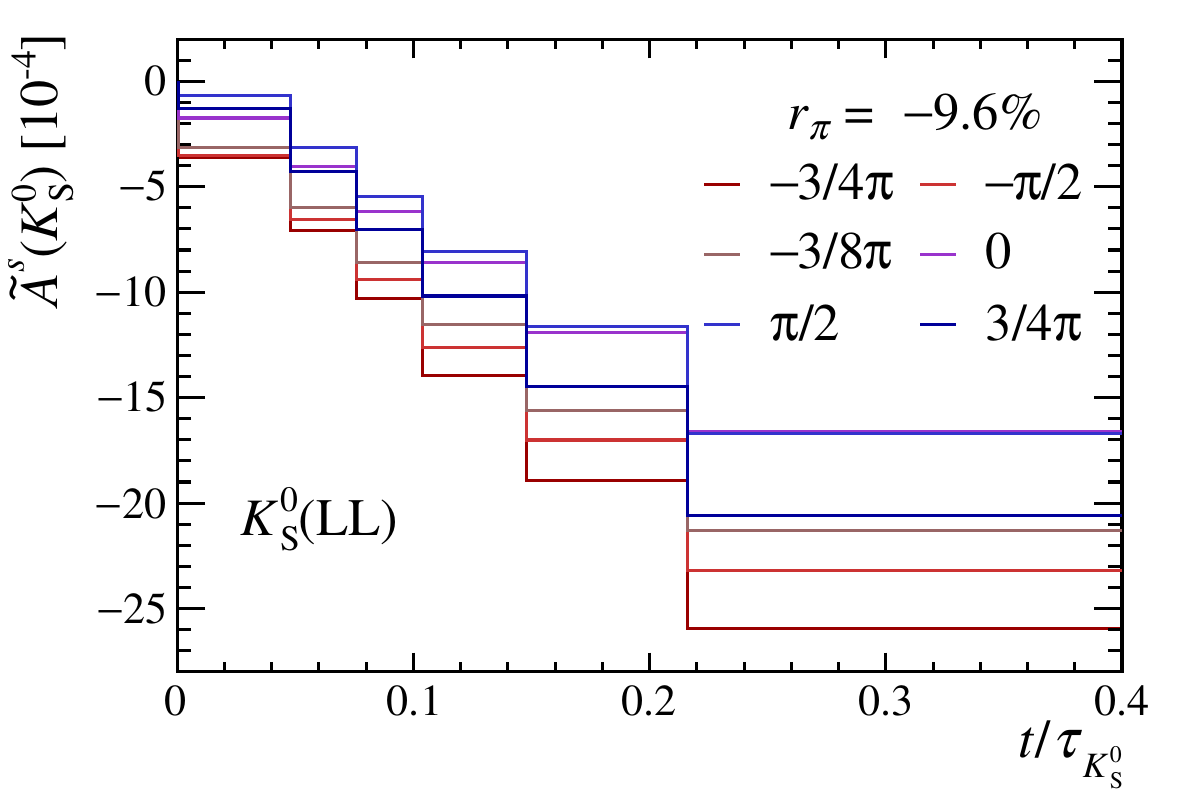}
     \includegraphics[ page = 1, width = 0.45\textwidth]{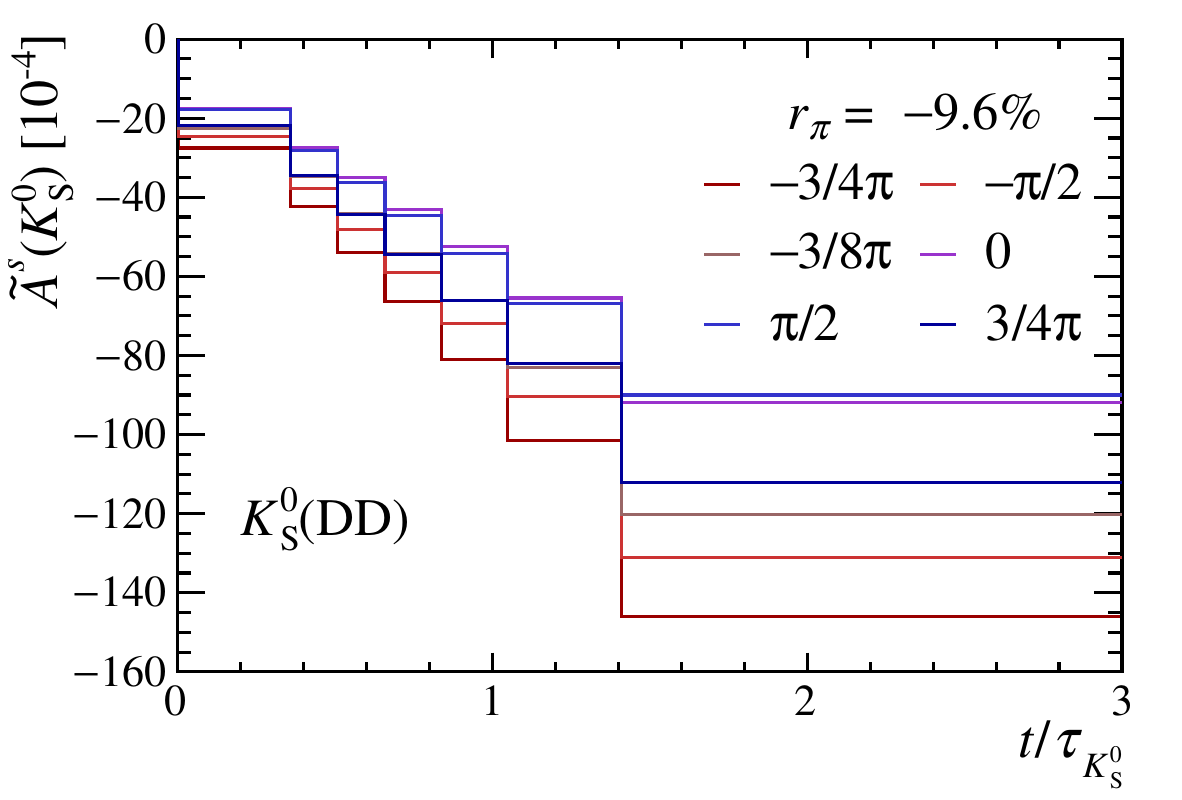}
      \includegraphics[ page = 1, width = 0.45\textwidth]{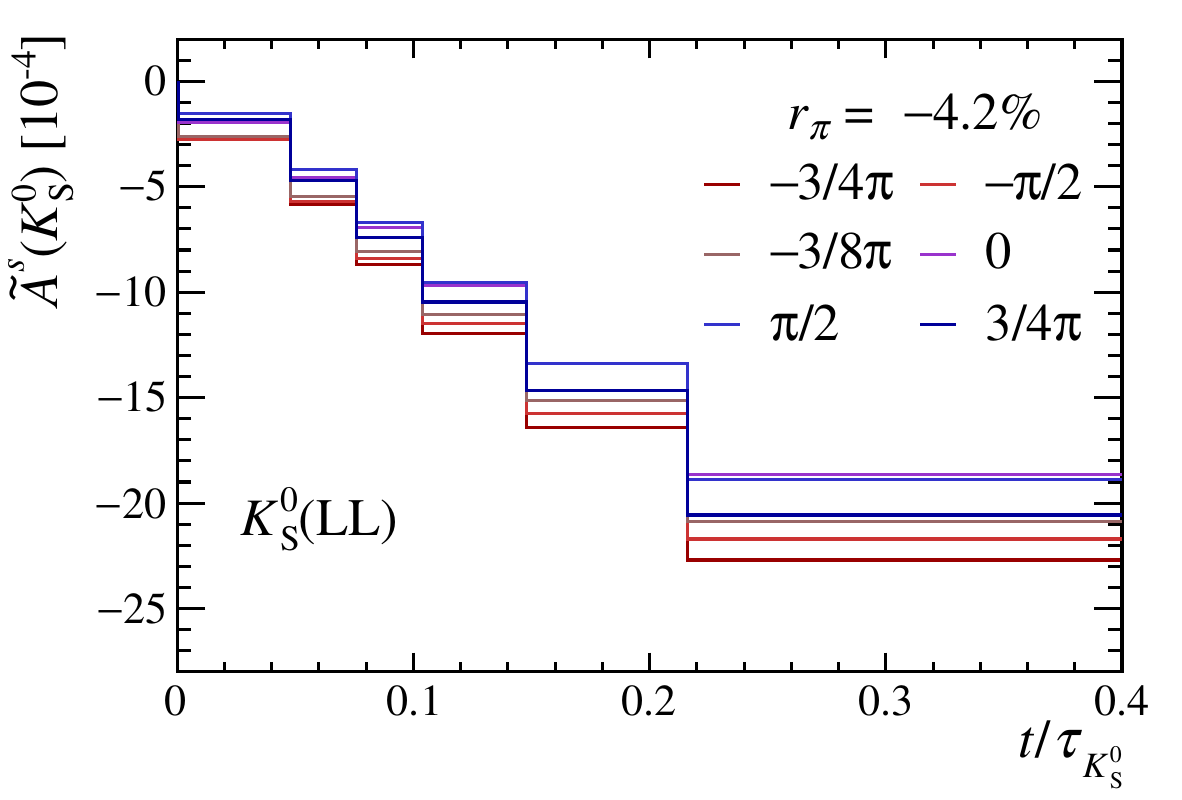}
     \includegraphics[ page = 1, width = 0.45\textwidth]{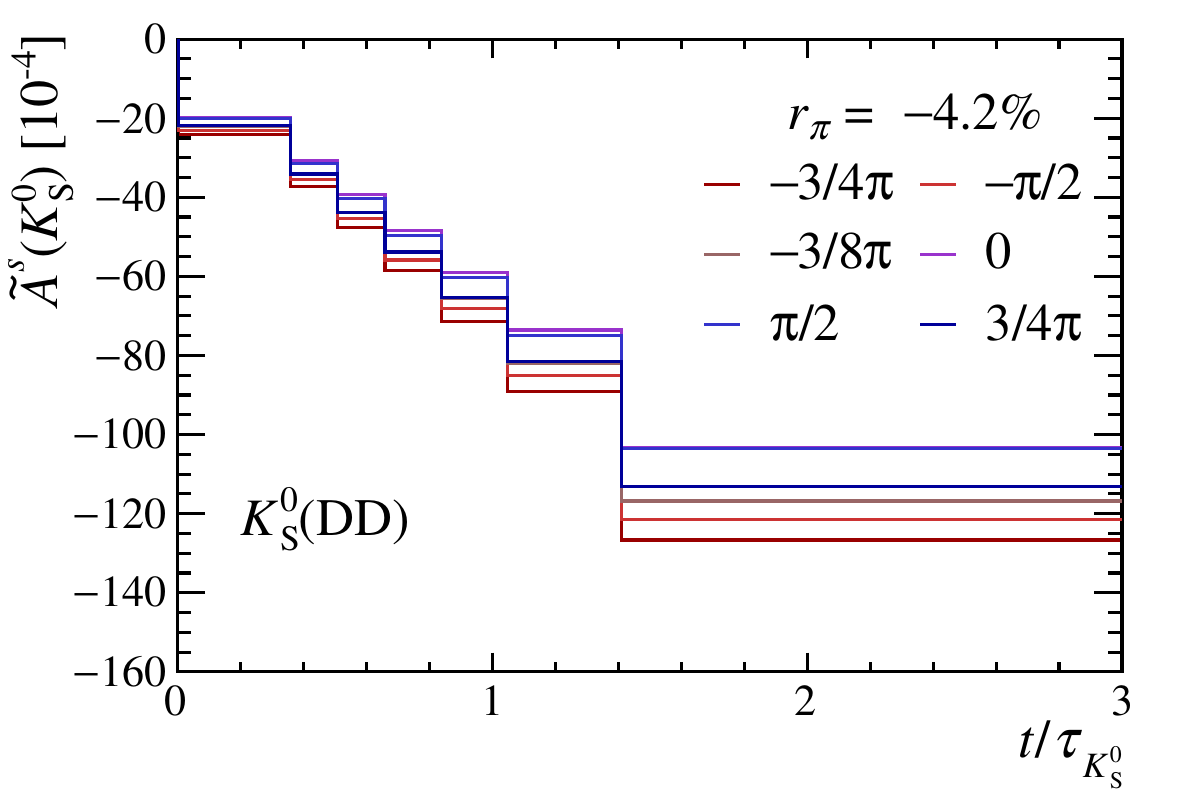}
    \caption{Set of templates for $\delta_\pi$ free to vary and $r_\pi$ fixed (top) at 
    $-9.6\%$ and (bottom) 
     $-4.2\%$
    in (left) $K_{\mathrm{S}}^0(\mathrm{LL}) $ and (right) $K_{\mathrm{S}}^0(\mathrm{DD}) $ bins. For large values of $r_\pi$, different strong phases result in very different templates, especially at large $K_{\mathrm{S}}^0$ decay-time values.}\label{fig:Fig5}
    \end{figure}

The per-candidate asymmetries are combined with appropriate weights, depending on the kinematics of each single \DpToKSpi candidate, to obtain the expected value of the neutral-kaon asymmetry \ADTilde, in each bin of decay time.
The predicted asymmetry values for different input values of the \rpideltapi pairs are shown in \cref{fig:Fig5}. As expected, the dependence on the relative strong phase \deltapi increases in magnitude for larger \rpi values. 
By using the theoretical predictions for $\rpi = -0.073$ and $\deltapi = -1.39 \ (+1.39)$ rad~\cite{Yu:2017oky,Wang:2017ksn}, the time-integrated asymmetry over all subsamples is about $-8.7\times 10^{-4} \ (-5.9 \times 10^{-4})$ in the \KSLL sample, and it increases to about $-33.6 \times 10^{-4}\ (-26.8 \times 10^{-4})$ when including also the \KSDD sample. 

With respect to the prediction obtained neglecting the DCS contribution (\ie $\rpi = 0$), as done in previous publications~\cite{LHCB-PAPER-2014-013,LHCb-PAPER-2019-002,LHCb-PAPER-2022-024}, the predicted asymmetry shifts by $1.1\times10^{-4}$ ($1.7\times10^{-4}$) in the \KSLL sample for negative (positive) strong phase. When the \KSDD sample is also included, the corresponding shifts become $2.3\times10^{-4}$ ($4.5\times10^{-4}$).

\section{Measurement of \texorpdfstring{\boldmath{\ACPDPhiPi}}{ACP(Dp 2 phi pi)}}\label{sec:meas_Acp_d2phipi}

After the kinematic weighting, the model for the difference of raw asymmetries \DeltaAs in each decay-time bin is given by
\begin{equation}
\DeltaAsTilde\qrpideltapi = \ADTilderdelta - q \, ,
\end{equation}
where the time-dependent asymmetry \ADTilde incorporates all the asymmetry contributions from the \DpToKSpi decay and from the \KS evolution and decay.
The best-fit value of the constant offset $q$, corresponding to \ACPDPhiPi, is determined using the profile-likelihood method.
A minimum $\chisq$ fit to the \DeltaAs values is performed, treating both \rpi and \deltapi as nuisance parameters,
and minimising the $\chisq(q)$ statistic, defined as 
\begin{equation}\label{eq:deltachisq_definition_q}
    \chisq(q) \equiv \min_{\rpi \,  , \, \deltapi} \ 
    \chisq\qrpideltapi
    =  \min_{\rpi \,  , \, \deltapi}
    \sum_{s=1}^{13} \frac{\left( \DeltaAs - \DeltaAsTilde\qrpideltapi \right)^2}{\sigma_{\DeltaAs}^2}\, ,
\end{equation}
where $\sigma_{\DeltaAs}$ is the statistical uncertainty on the \DeltaAs observable in each decay-time bin $s$.  
In the fit, the strong phase \deltapi is left free to vary across the full range $[-\pi,\,\pi]$, while a constraint is applied to the value of \rpi. The ratio \rpi is expected to be close to the ratio between the DCS (\decay{c}{d \bar{s} u}) and CF (\decay{c}{s \bar{d} u}) CKM matrix elements,
$\left| \Vcds \Vus /\Vcss \Vud\right| = (5.336 \pm 0.010)\times 10^{-2}$~\cite{CKMfitter2015}, up to small contributions from strong interactions.  
The magnitude of hadronic corrections to this ratio can be constrained using analogous decays of $\Dz$, \Dp, and $\Lc$ hadrons, in which the dominant tree-level amplitudes proceed through the same CF and DCS weak transitions as in \DpToKSpi decays. Representative examples are $\Lc \to p \Kp \pim$ and $\Dz \to K^*(892)^- \pip$, whose DCS-to-CF amplitude ratios are $(4.2\pm0.3)\times 10^{-2}$~\cite{PDG2024} and $(9.5\pm0.2)\times 10^{-2}$~\cite{BaBar:2018cka}, respectively. 
Similar values are observed in other charm-hadron decays, such as $\Dz\to \Km\pip$, with DCS-to-CF amplitude ratios lying in the range $[0.042, 0.096]$~\cite{PDG2024,BaBar:2018cka}.
The only exception is the decay $\Dp \to \Km\pip\pip\piz$, which exhibits a slightly higher ratio of $(13.9\pm0.6)\times 10^{-2}$. This is considered an outlier, as discussed in Ref.~\cite{BESIII:2020wnc}, likely due to large isospin-symmetry violations between $\Dp \to \Kp\pim\pip\piz$ and $\Dz \to \Kp\pim\pip\pim$, possibly caused by final-state interactions and differing resonance structures. Ignoring this decay, the ratio \rpi is constrained to lie in the range $[-0.096 , \ -0.042]$ 
following the approach of Refs.~\cite{Charles:2016qtt, Hocker:2001xe}, by adding the following term to the $\chisq\qrpideltapi$ statistic,
\begin{equation}\label{eq:GaussUniformConstraint}
    \chisq_{c}(\rpi) = \left\{
    \begin{array}{cl}
         0 &  \text{ if } r_\pi \in [-0.096 , \ -0.042] \\
         \left(\frac{\rpi-r_0}{0.8\cdot \sigma_0}\right)^2 - \left(\frac{1}{0.8}\right)^2 & \text{ otherwise} \\
    \end{array}
    \right. 
\end{equation}
where $r_0 = -0.069$ is the arithmetic mean of the interval extremes and $\sigma_0 = 0.027$ is their semi-difference. All considered DCS-to-CF ratios lie within one standard deviation of this interval, with the exception of $\Dp \to \Km\pip\pip\piz$, discussed above, which is covered at the two-standard-deviation level.

\begin{figure}
    \centering
    \includegraphics[page=1, width=0.45\linewidth]{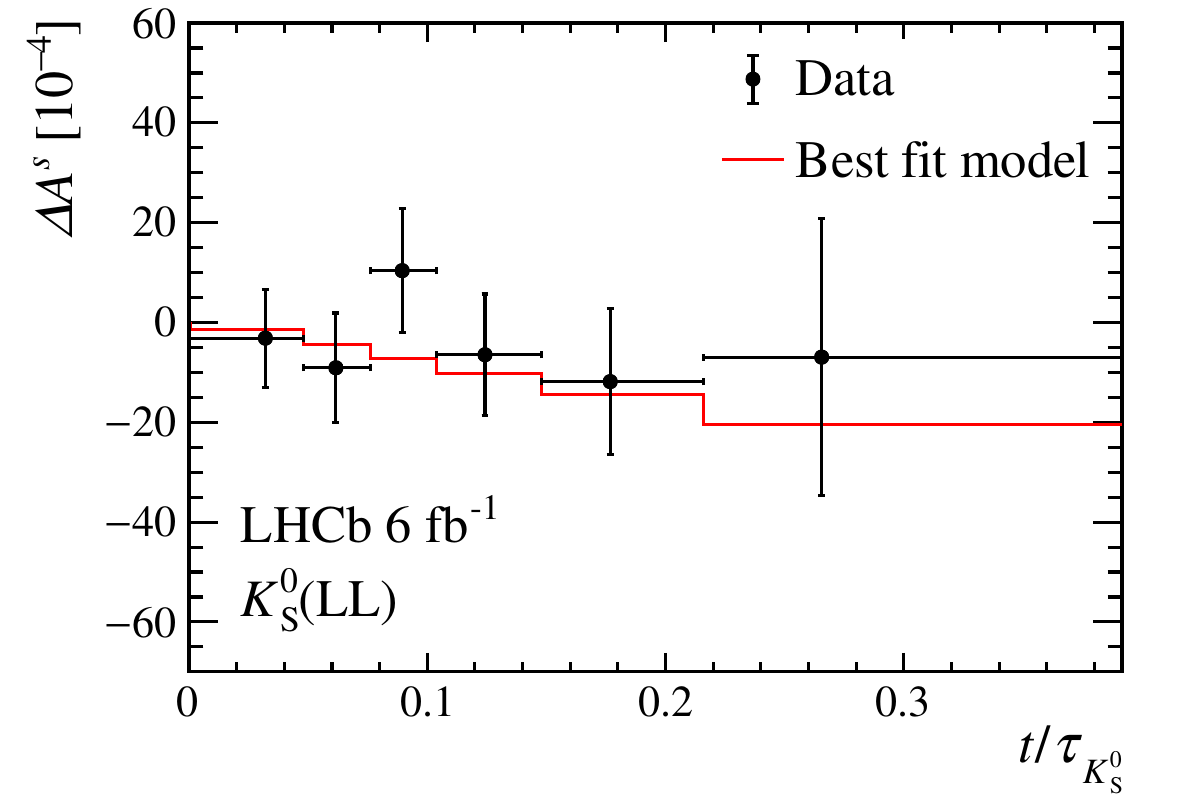}
    \includegraphics[page=1,width=0.45\linewidth]{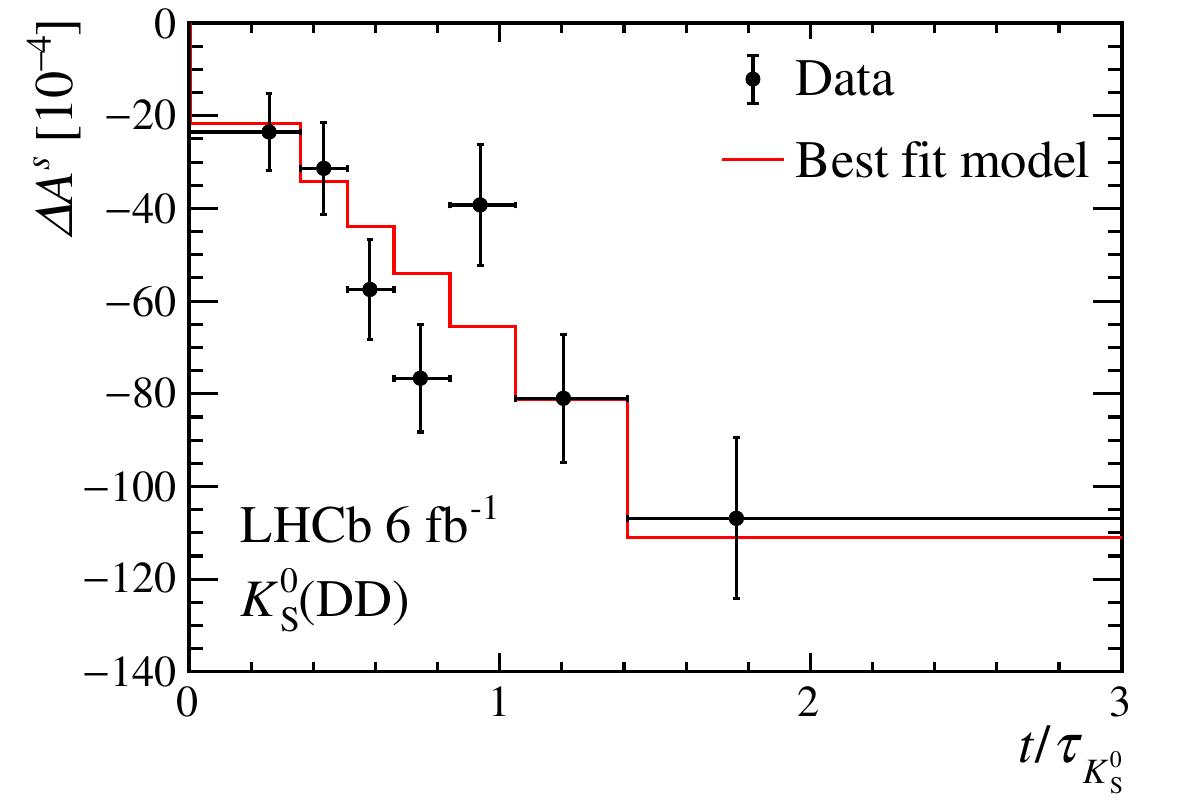}
    \caption{Measured difference of raw asymmetries, $\Delta A^s$ as a function of $K_{\mathrm{S}}^0$ decay time for (left) $K_{\mathrm{S}}^0(\mathrm{LL}) $ and (right) $K_{\mathrm{S}}^0(\mathrm{DD}) $ decay-time bins. 
    The fit projection is superimposed to the data points.}
    \label{fig:Fig6}
\end{figure}

The best estimate for the constant offset $q$ obtained from the minimisation corresponds to $\hat{q} = \ACPDPhiPi =(0.1\ ^{+4.9}_{-5.0})\times10^{-4} = (0.1\pm 4.9)\times 10^{-4}$, where the latter uncertainty is estimated as the average between the upper and lower uncertainties.
The best-fit values for the CF–DCS interference parameters are $\hat{r}_{\pi} = -0.095$ and $\hat{\delta}_{\pi} = -3\pi/4$.
The fit model agrees with data with $\chisq(\hat{q}) = 12.2$ for 11 degrees of freedom, corresponding to a \pvalue of 35\%. The projection of the fit is superimposed on the measured \DeltaAs points in~\cref{fig:Fig6}. 
The $\chisq(q)$ statistic has approximately a parabolic shape as a function of the offset $q$.
The statistical uncertainties on $\hat{q}$ are determined from the values of $q$ in which the $\chisq(q)$ increases by one unit with respect to its minimum.

The total statistical uncertainty on \ACPDPhiPi 
arises both from the finite size of the data samples and from the uncertainty associated with the nuisance parameters \rpi and \deltapi, which are free to vary in the fit. The contribution from the data sample size alone is estimated by repeating the fit with \rpi and \deltapi fixed to their best-fit values, yielding a reduced statistical uncertainty of $3.3\times 10^{-4}$.

The fit strategy is validated using pseudoexperiments, each consisting of a set of 13 \DeltaAs measurements, one for each \KS decay-time bin. Each \DeltaAs value is randomly drawn from a Gaussian PDF with a mean given by the prediction of a model for a tested \qrpideltapi parameter set, and a standard deviation equal to the uncertainty observed in data for the corresponding decay-time bin. 
For a given set of \textit{true} parameters \qrpideltapi, the time-dependent fit for the offset $q$ is repeated for each pseudoexperiment. The bias on the offset is then estimated as the difference between the average measured offset and the input value $q$. This bias is found to be compatible with zero for all tested values of $\rpi \in [-0.096, \, -0.042]$ and $\deltapi \in [-\pi,\,\pi]$.

As a cross check, the fit minimisation is repeated with \rpi allowed to vary freely over the full range $[-1,\,0]$. 
The resulting best estimate, $\hat{q}$, differs from the baseline fit by $+0.3\times 10^{-4}$, and its statistical uncertainty slightly increases to $5.2\times 10^{-4}$.

\subsection{Alternative determinations using external inputs on \texorpdfstring{\boldmath{\rpi} and \boldmath{\deltapi}}{rpi and deltapi}}

The baseline measurement of \ACPDPhiPi, presented in the previous section, is obtained with no constraint on \deltapi and with a loose constraint on \rpi based solely on measured amplitude ratios in similar decay channels. This section presents two alternative determinations of \ACPDPhiPi, obtained by applying different constraints on the nuisance parameters, determined by theoretical and experimental results.
These alternative determinations are provided as robustness checks of the stability of the result under different assumptions on the hadronic parameters, and are further interpreted in~\cref{sec:results}.

The first alternative result is obtained by fixing the parameters \rpi
and \deltapi to the values predicted by the FAT approach~\cite{Yu:2017oky,Wang:2017ksn}, namely $\rpi = -0.073 \pm 0.004$ and
$\deltapi = +1.39 \pm 0.05$~rad or $\deltapi = -1.39 \pm 0.05$~rad. Using the two possible values of the strong phase, the corresponding estimates are \mbox{$\ACPDPhiPi = (-2.2 \pm 3.3)\times 10^{-4}$} for \mbox{$\deltapi = -1.39$ rad}, and \mbox{$\ACPDPhiPi = (4.6 \pm 3.3)\times 10^{-4}$} for $\deltapi = +1.39$ rad. The two results differ by $6.8\times 10^{-4}$, which is approximately twice the statistical uncertainty.
The fit quality yields $\chisqndf = 13.0/12$ for the negative phase, and $\chisqndf = 13.8/12$ for the positive phase.

A second alternative result is obtained by imposing an experimental constraint on the parameters \rpi and \deltapi, derived from the measurement performed by the CLEO collaboration~\cite{CLEO:2007rhw}, of
\begin{equation}
  \BigRpi\equiv
  \frac{\Gamma(\Dp \to \KS \pip) - \Gamma(\Dp \to \KL \pip)}
       {\Gamma(\Dp \to \KS \pip) + \Gamma(\Dp \to \KL \pip)} = (2.2 \pm 1.6 \pm 1.8)\% \, .
\end{equation}
This observable can be expressed in terms of the amplitude ratio \rpi and the strong-phase difference \deltapi~\cite{Wang:2017ksn} as
\begin{equation}
\label{eq:Rpi_complete}
  \begin{split}
    \BigRpi
    &=
    -2\rpi
    \frac{
      \cos(\varphi + \deltapi)(1 - |\varepsilon|^2)
      + 2\,\sin(\varphi + \deltapi)\,\Im(\varepsilon)
    }
    {
      |1 - \varepsilon|^2
      + |1 + \varepsilon|^2 \rpi^2
    } \\
    &\simeq
    -2\,\rpi \cos(\varphi + \deltapi),
  \end{split}
\end{equation}
where $\varepsilon$ is the complex parameter quantifying \CP violation in the neutral-kaon system, with $|\varepsilon| \sim 10^{-3}$. The approximate expression in the second line is obtained by expanding to leading order in both $\varepsilon$ and \rpi.
The measurement of \ACPDPhiPi is repeated by applying a Gaussian constraint to the experimental value of \BigRpi~\cite{CLEO:2007rhw}. 
The resulting best-fit offset is $\ACPDPhiPi = (0.4 \pm 5.0) \times 10^{-4}$, with a goodness-of-fit of $\chisqndf = 12.6/11$.

\section{Systematic uncertainties}
\label{sec:systematics}

The systematic uncertainties are divided into two main categories. The first category includes all sources affecting the determination of the asymmetry difference in bins of decay time, \DeltaAs. The second category comprises uncertainties associated with the time-dependent model used to predict the neutral-kaon asymmetry, \ADTilde.

\subsection{Systematic uncertainties on \texorpdfstring{\boldmath{\DeltaAs}}{Delta As}}\label{sec:systematics_DeltaAs}

The systematic uncertainties affecting the determination of \DeltaAs arise from four main sources. The first is the modelling of the invariant-mass distributions used to extract the raw asymmetries, which includes contributions from partially reconstructed or misidentified background (\textit{mass model}). The second source is residual contamination from \Dp mesons originating from decays of bottom hadrons (\textit{secondary decays}). The third arises from imperfect agreement between the kinematic distributions of the \Dp meson and the trigger pion in the \DpToKSpi and \DpTophipi decay modes (\textit{kinematic weighting}). Finally, a nonzero detection asymmetry in the reconstruction of nonresonant $\Kp\Km$ pairs under the $\phi$ mass peak (\textit{nonresonant $\Kp\Km$}) is also considered.

All uncertainties affecting the \DeltaAs measurements are summarised in~\cref{tab:Tab2}. The dominant contributions originate from the mass modelling and residual kinematic differences, while the overall uncertainties remain largely dominated by statistical effects.

\subsubsection{Mass model}

A systematic uncertainty is assigned to account for possible imperfections in the empirical PDFs used in the mass modelling. To evaluate this uncertainty, the mass fits are repeated using a set of 15 alternative models, all providing an adequate description of the data, and the corresponding raw asymmetries entering the determination of the \DeltaAs values are extracted.
The alternative models employ more flexible signal parametrisations, either by allowing selected parameters to vary independently between \Dp and \Dm or by introducing additional Gaussian components, and a more complex combinatorial background description with the sum of two exponentials. Among these models, six also include background contributions from partially reconstructed and misidentified decays populating the mass region of interest.
All dominant decay modes potentially contributing in the relevant mass window have been investigated, including \decay{D^+_{(s)}}{\KS/\phi \pip \piz} decays with an unreconstructed \piz, semileptonic \decay{D^+_{(s)}}{\KS/\phi \ l^+ \nu} decays where the lepton ($l=\mu,e$) is misidentified as a pion, and \decay{\Lc}{\KS/\phi \ p} decays with a misidentified proton. The mass shapes of these contributions are studied using the fast phase-space simulation framework \textsc{RapidSim}~\cite{Cowan:2016tnm}, which parametrically accounts for the detector resolution and acceptance, as well as kinematic requirements on momenta and lifetime-related quantities.
Most of these contributions are found to be negligible, either because of small branching fractions or nearly 
uniform mass distributions, and 
can be effectively described by the combinatorial background model.
The only exceptions are \decay{\Dsp}{\phi \pip \piz} decays in the \DpTophipi channel and \decay{\Dp}{\KS l^+ \nu} decays in the \DpToKSpi channel, whose asymmetric mass shapes populate the lower-mass sideband. These contributions are included in the corresponding alternative fit models, with their shapes fixed to simulation. When their yields are left free in the fit, they are found to be compatible with zero in most cases and induce fit instabilities.
For this reason, the expected contamination from these backgrounds is upper-bounded using the known PDG branching fractions, the measured production cross-section of the parent meson, and the fraction of events populating the mass fit region, as estimated from the \textsc{RapidSim} shapes. In the alternative fit models, the yields of the partially reconstructed background components are fixed to three different values, each equal to or smaller than this upper bound, while the associated asymmetries are left free to vary.
Owing to the increased number of free parameters, the alternative models generally provide an improved description of the data compared to the baseline model. The fit model that best describes the \DpToKSpi and \DpTophipi mass shapes employs a signal parameterisation given by the sum of a \jsu and a Gaussian function, together with a double-exponential background. This model achieves an average \chisqndf of 1.01 (1.15), with ndf = 76, across all fits performed independently in the \DpToKSpi (\DpTophipi) data samples, which are further subdivided by decay-time bin, data-taking year, and magnet polarity. 
The systematic uncertainty associated with the mass modelling is evaluated independently in each decay-time bin as the RMS of the \DeltaAs values obtained from the ensemble of alternative fit models.

\subsubsection{Secondary decays}

Residual contamination from \Dp mesons originating from $b$-hadron decays can bias the measurement of the asymmetry differences if the relative secondary fraction differs between the \DpToKSpi and \DpTophipi samples. 

\begin{figure}
    \centering
    \includegraphics[page = 1, width = 0.45\textwidth]{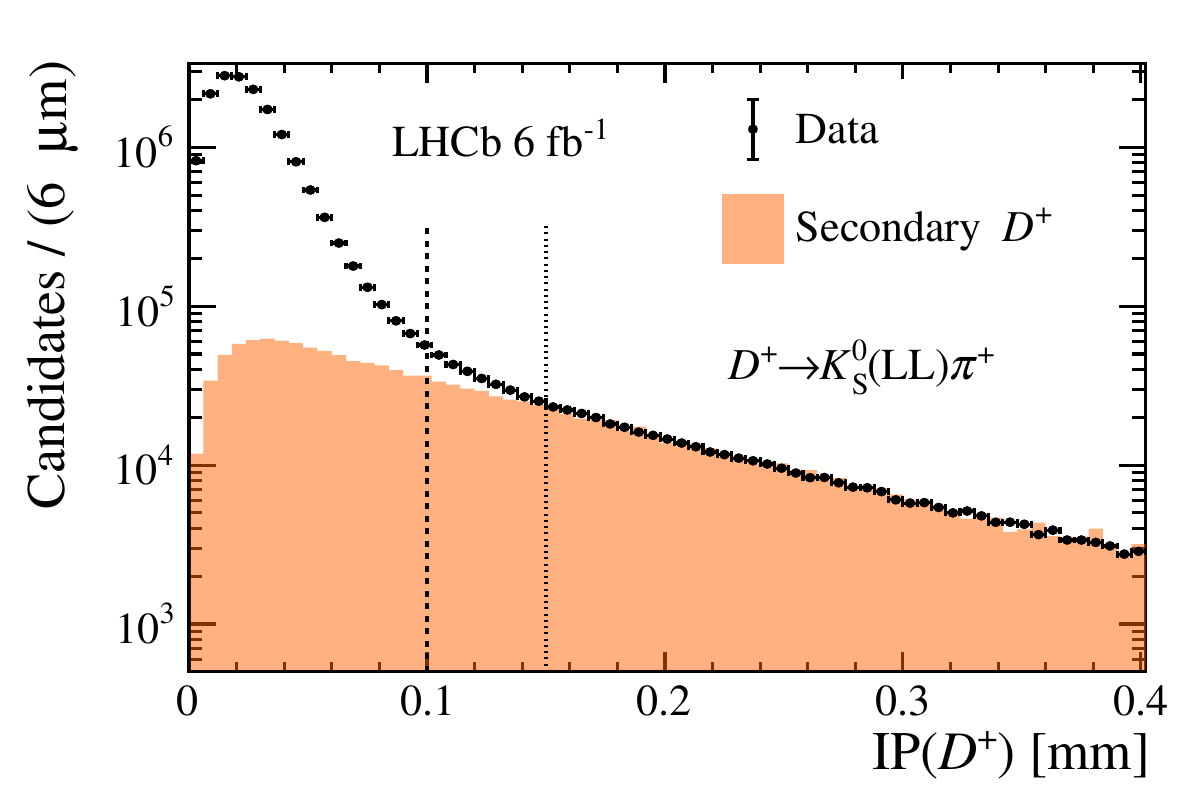} 
    \includegraphics[page = 1, width = 0.45\textwidth]{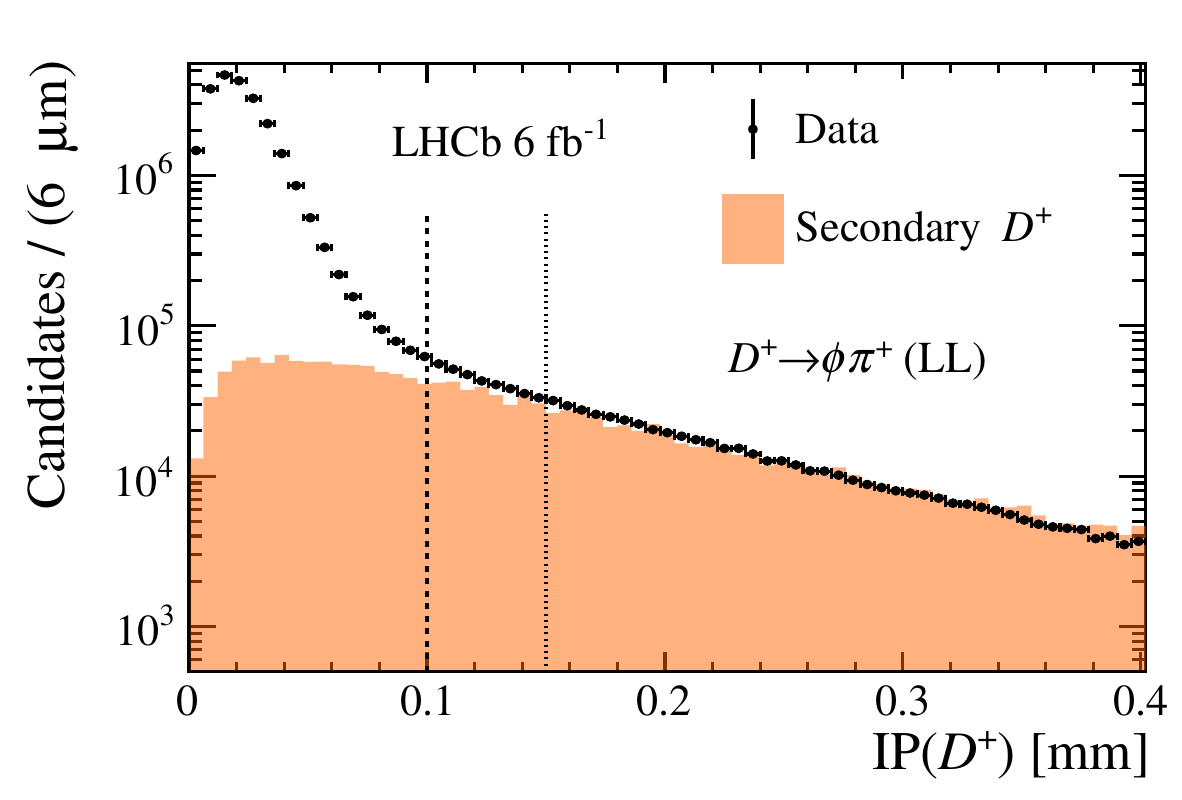} 
    \includegraphics[width = 0.45\textwidth]{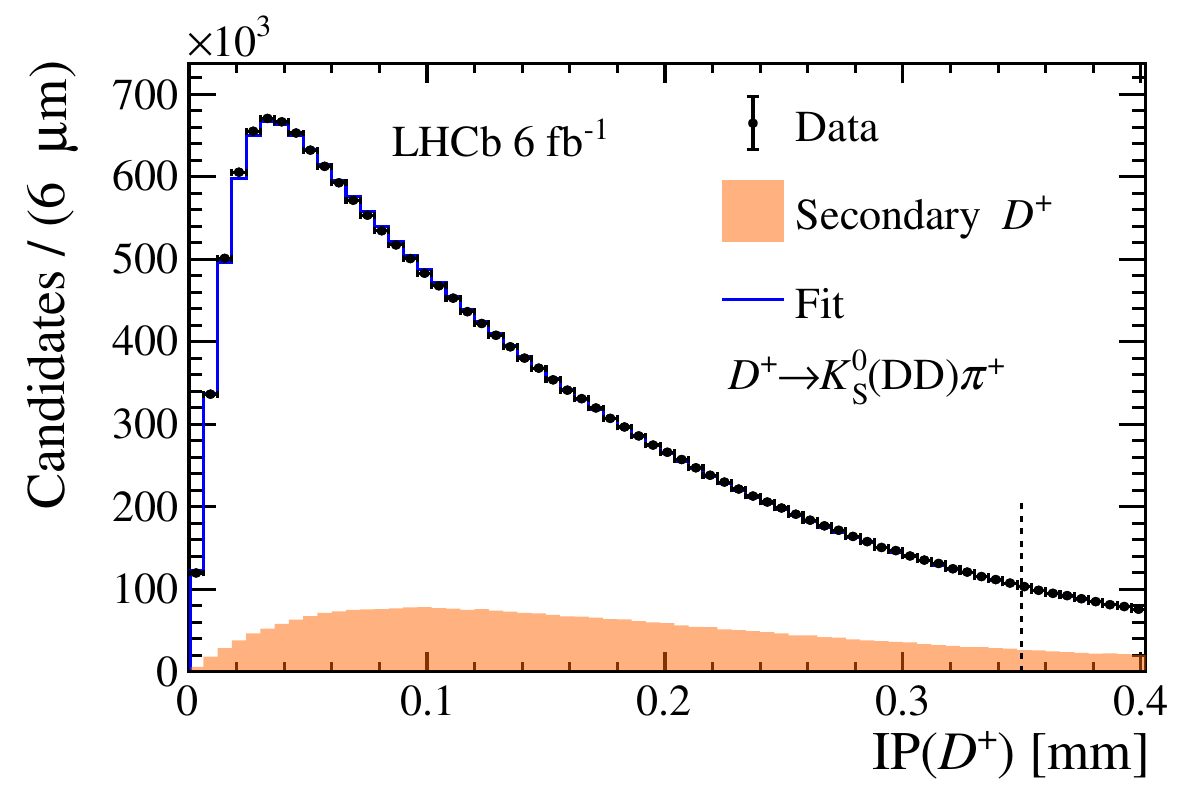}
    \includegraphics[page = 1, width = 0.45\textwidth]{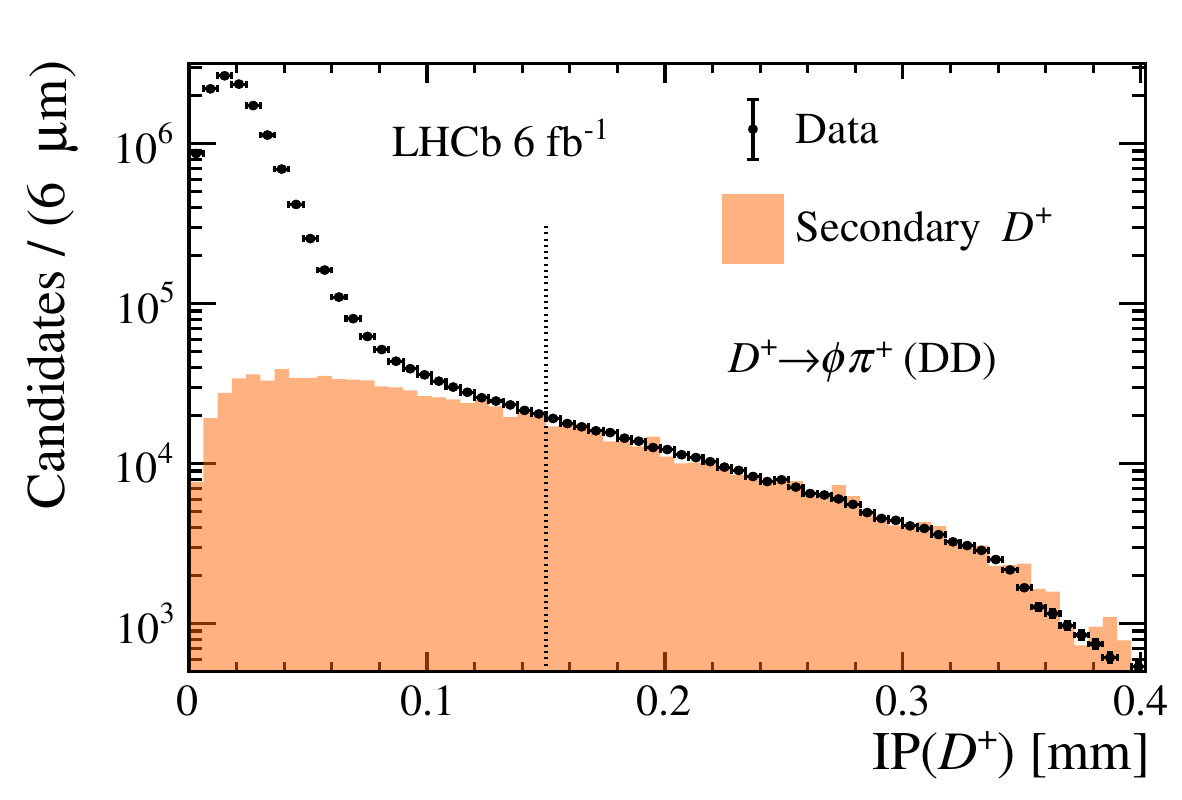} 

     \caption{Comparison between the $\textrm{IP}(D^+)$ distributions observed in data and in simulated secondary decays for (left) \mbox{$D^+ \to K_{\mathrm{S}}^0 \pi^+$} and (right) \mbox{$D^+ \to \phi \pi^+$} candidates, in the (top) LL and (bottom) DD subsamples. The dashed lines indicate the offline requirements on $\textrm{IP}(D^+)$, while the dotted lines mark the beginning of the secondary-enriched region used to normalise the simulated secondary component in the \mbox{$D^+ \to K_{\mathrm{S}}^0(\mathrm{LL}) \pi^+$} and \mbox{$D^+ \to \phi \pi^+$} samples. In the bottom-right plot, the impact of the $\textrm{IP}(D^+)_{\rm smeared} < 350\mum$ requirement is clearly visible.}
     \label{fig:Fig7}
\end{figure}

In the \KSLL subsamples, the requirement on the \Dp impact parameter \mbox{($\IP < 100\,\mum$)} suppresses this secondary component to a few-percent level. 
In each decay-time subsample, the residual bias is estimated as
\begin{equation}\label{eq:secondaries}
\Delta A_{\rm sec} =
(f_{\rm sec}^{\phi\pi} - f_{\rm sec}^{\KS\pi}) \cdot
\langle A_{\rm sec} - A_{\rm prompt} \rangle,
\end{equation}
where $f_{\rm sec}^{\phi\pi}$ and $f_{\rm sec}^{\KS\pi}$ denote the fractions of secondary decays in the \DpTophipi and \DpToKSpi samples, respectively, and $\langle A_{\rm sec} - A_{\rm prompt} \rangle$ is the weighted average of the difference between the observed asymmetries of secondary and promptly produced \Dp candidates in the \DpTophipi and \DpToKSpi samples.
These asymmetries $A_{\rm sec}$ and $A_{\rm prompt}$ are not expected to be equal because the production asymmetry of $b$-hadrons decaying to \Dpm mesons is different from the production asymmetry of prompt \Dpm mesons.
The bias is evaluated in each \KS decay-time bin using simulation of \DpTophipi and \DpToKSLLpi decays originating from a mixture of known $b$-hadron decay modes. The simulation provides the \ipDp distribution, which is then used to determine $\Delta A_{\rm sec}$. By normalising this distribution to data in a secondary-enriched region ($\ipDp > 150\,\mum$), the fractions $f_{\rm sec}^{\phi\pi}$ and $f_{\rm sec}^{\KS\pi}$ are determined with satisfactory accuracy, as shown in \cref{fig:Fig7}. The fraction of secondary decays in the $\ipDp < 100\,\mum$ region are found to be between $3.5\%$ and 6\% in the \DpToKSLLpi bins and between $2.5\%$ and 4\% in the \DpTophipi bins.  
The asymmetries $A_{\rm sec}$ and $A_{\rm prompt}$ are estimated, to good approximation, from the raw asymmetries measured in the secondary-enriched ($\ipDp > 150\,\mum$) and prompt-enriched ($\ipDp < 100\,\mum$) samples, respectively.

In the \KSDD subsamples, given the poorer \ipDp resolution, it is not possible to define a sufficiently pure secondary-enriched region.
The secondary fraction is therefore determined from a fit to the \ipDp distribution, using \lhcb simulation to model the PDFs of the prompt and secondary \DpToKSDDpi decay candidates. The \PDF of the secondary component is fixed from simulation, whereas the parameters of the \PDF describing the prompt component are allowed to vary by a few percent to account for residual mismodelling of the \ipDp resolution in simulation. This provides a satisfactory description of the data while preserving the distinctive features of the original simulated distribution, as shown in~\cref{fig:Fig7}.
The fraction of secondary decays in the $\ipDp < 350\,\mum$ region ranges between $10\%$ and $15\%$ across the \DpToKSDDpi bins. 
The fraction of secondary decays in \DpTophipi candidates assigned to the \KSDD bins is estimated using the procedure described above, by normalising the simulation to data in the secondary-enriched region, $\ipDp > 150\mum$. In this case, the secondary fraction is determined after applying the smearing procedure to the reconstructed decay vertex of the \DpTophipi candidates and considering the full \ipDp interval, as shown in~\cref{fig:Fig7}. These fractions range between 5\% and 8\% in the different subsamples into which the \DpTophipi candidates are divided. 
The \DpTophipi candidates are also used to estimate $A_{\rm sec}$ and $A_{\rm prompt}$ in each DD-type bin, using the same strategy discussed above.

The resulting bias is estimated using \cref{eq:secondaries} and found to be consistent with zero in all decay-time bins, for all \KSLL and \KSDD subsamples. The uncertainty on the bias, typically $\mathcal{O}(10^{-4})$, is assigned as a systematic uncertainty.

\subsubsection{Kinematic weighting}

The cancellation of nuisance asymmetries in the measurement of the \DeltaAs observables relies on the momentum distributions of the \Dp meson and the trigger pion being identical in the \DpToKSpi and \DpTophipi samples. 
This condition is achieved through the kinematic weighting procedure described in \cref{sec:nuisance_asys}.
Residual mismatches between the two samples can lead to an incomplete cancellation of production and detection asymmetries, potentially biasing the measured \DeltaAs values.
For a particle $X$ with a momentum-dependent production or detection asymmetry, $A_{\rm P/D}^X(\vec{p}^X)$, the resulting bias is estimated following Ref.~\cite{LHCb-PAPER-2022-024} as
\begin{equation}
    \Delta A_{\rm P/D}^X=
    \int d\vec{p}^X \,
    \bigl[ \mathcal{P}_a(\vec{p}^X) - \mathcal{P}_b(\vec{p}^X) \bigr]
    \, A_{\rm raw}(\vec{p}^X),
\end{equation}
where $\mathcal{P}_{a,b}(\vec{p}^X)$ are the normalised momentum distributions in the two samples. In each \KS decay-time bin, the systematic uncertainty is obtained by summing in quadrature the contributions from $\Delta A_{\rm P}^\Dp$ and $\Delta A_{\rm D}^\pip$.

\subsubsection{Nonresonant \texorpdfstring{\boldmath{\Kp\Km}}{K+K- } background}

The detection asymmetry of the $\Kp\Km$ pair from the decay \decay{\phi}{\Kp\Km} is assumed to be zero, as the $\phi$ meson is a \CP-invariant initial state. However, residual nonresonant $\Kp\Km$ pairs are present below the $\phi$ mass peak~\cite{CLEO:2008msk}, for which the positive and negative kaons can have different momentum distributions. This can induce a nonzero detection asymmetry, potentially biasing the measured \DeltaAs observables.
This bias is estimated using simulated samples to model the charged kaon detection asymmetry, following the approach of Ref.~\cite{LHCb-PAPER-2019-002}, as
\begin{equation}
    \Delta\Adet(\Kp\Km) =
    \int dp_{\rm SS} \, dp_{\rm OS} \,
    \bigl[ \Adet(\Kp,p_{\rm SS}) - \Adet(\Kp,p_{\rm OS}) \bigr]
    \, \mathcal{P}(p_{\rm SS},p_{\rm OS}),
\end{equation}
where $p_{\rm SS}$ and $p_{\rm OS}$ are the momenta of the same-sign and opposite-sign kaons with respect to the \Dp charge, and $\mathcal{P}(p_{\rm SS},p_{\rm OS})$ is their joint momentum distribution.

\begin{table}[t]
    \centering
     \caption{Uncertainties on $\Delta A^s$ in all the $K_{\mathrm{S}}^0$ decay-time subsamples, in units of $10^{-4}$. 
     The decay-time bin boundaries are given in units of the $K_{\mathrm{S}}^0$ mean lifetime, $\tau_{K_{\mathrm{S}}^0}$. The total systematic uncertainty, $\sigma_{\rm syst}$, is computed as the sum in quadrature of the systematic contributions listed in the first four columns; this is added in quadrature with the statistical uncertainty, $\sigma_{\rm stat}$, to obtain the total uncertainty, $\sigma_{\rm tot}$.
     }
    \label{tab:Tab2}
    \renewcommand{\arraystretch}{1.2} 
    \begin{tabular}{c|
    >{\centering\arraybackslash}p{1.45cm}|
    >{\centering\arraybackslash}p{1.85cm}|
    >{\centering\arraybackslash}p{1.85cm}|
    >{\centering\arraybackslash}p{2.25cm}||
    c|c|c}
    \hline
\hline

$t/\tau_\KS$ bin  & Mass model & Secondary decays & Kinematic weighting & Nonresonant $\Kp\Km$&  $\sigma_{\rm syst}$ & $\sigma_{\rm stat}$ &$\sigma_{\rm tot}$\\ 

\hline
\hline

\KSLL  &  & && & &   \\

[0.000, 0.048] &  1.6 & 1.0 & 1.3 & 0.47 & 2.3& 9.8 & 9.8 \\ 

[0.048, 0.076] &  3.8 & 0.8 & 1.8 & 0.43 & 4.3& 11 & 12 \\ 

[0.076, 0.104] &  1.2 & 0.8 & 1.4 & 0.30 & 2.0& 13 & 13 \\ 

[0.104, 0.148] &  3.8 & 0.7 & 1.9 & 0.37 & 4.3& 12 & 13 \\ 

[0.148, 0.216] &  4.3 & 0.9 & 3.4 & 0.18 & 5.6& 15 & 16 \\ 

[0.216, 0.400] &  10 & 1.2 & 8.7 & 0.15 & 13& 28 & 31 \\ 

\hline

\KSDD &  & && & &   \\

[0.10, 0.36] &  2.4 & 3.0 & 0.8 & 0.41 & 3.9& 8.3 & 9.2 \\ 

[0.36, 0.51] &  2.1 & 4.8 & 0.7 & 0.22 & 5.3& 9.9 & 11 \\ 

[0.51, 0.66] &  4.5 & 6.9 & 3.3 & 0.22 & 8.9& 11 & 14 \\ 

[0.66, 0.84] &  1.8 & 8.2 & 1.3 & 0.13 & 8.5& 12 & 14 \\ 

[0.84, 1.05] &  2.0 & 8.8 & 2.0 & 0.13 & 9.2& 13 & 16 \\ 

[1.05, 1.41] &  2.5 & 9.0 & 3.7 & 0.01 & 10& 14 & 17 \\ 

[1.41, 3.00] &  4.3 & 13 & 4.7 & 0.41 & 14& 17 & 22 \\ 
\hline
\hline
\end{tabular}   
\end{table}

\subsection{Model for neutral-kaon asymmetry}\label{sec:systematics_model}

The second class of systematic uncertainties includes effects related to the modelling of the neutral-kaon time-dependent asymmetry, and consequently the expected values \ADTilde for each decay-time subsample. These uncertainties mainly arise from the limited accuracy of the LHCb detector geometry description and from the modelling of the neutral-kaon evolution in matter. In particular, \KSLL candidates interact only with the material of the \velo detector, while a fraction of \KSDD candidates traverse additional material in the first RICH detector, placed between the \velo and the second tracking detector, enhancing the sensitivity to the detector material description and to regeneration effects.
The LHCb material map has an uncertainty on the total amount of material of the \velo layers of $\pm6\%$, as discussed in Ref.~\cite{LHCB-DP-2014-001}. This uncertainty represents the dominant contribution to the uncertainty on \ADTilde. To propagate this effect, the \ADTilde predictions are recomputed introducing, in addition to \rpi and \deltapi, a third free parameter corresponding to a relative variation of the material density, $\varepsilon_\rho$, allowed to vary within the interval $[-18,\,18]\%$. 
This alternative set of templates, with $\varepsilon_\rho$ different from zero, is employed in \cref{sec:systematics_Acp} to assign the systematic uncertainty due to the imperfect knowledge of the \lhcb material on \ACPDPhiPi.

A further source of systematic uncertainty arises from the modelling of neutral-kaon regeneration, which is described by the $\Delta\chi$ parameter and detailed in \cref{app:ADKS0Model}. In the model used to predict \ADTilde, the momentum dependence of $\Delta\chi$, given in \cref{eq:Deltachi}, is obtained from a fit to experimental data for kaons with momenta $p({\KS}) > 20\gevc$~\cite{Gsponer:1978dt}. At lower momenta, the regeneration amplitudes exhibit a different momentum dependence in materials heavier than carbon, as determined from dedicated measurements using various target materials~\cite{Kleinknecht:1994,Carithers:1977pj,Dydak:1976bv,Foeth:1970yz}.
To assess the associated systematic uncertainty, alternative \ADTilde predictions are computed by interpolating these experimental results and adopting the corresponding momentum dependence of $\Delta\chi$ for low-momentum \KS candidates traversing materials heavier than carbon. The resulting alternative templates, differing only in the assumed momentum dependence of $\Delta\chi$, are then used to assign the systematic uncertainty on \ACPDPhiPi arising from this effect.

The model for the neutral-kaon asymmetry depends on several additional parameters, which are fixed to their best-known values as detailed in \cref{app:ADKS0Model}. The impact of their uncertainties on the predicted asymmetries, $\ADTilde$, is found to be negligible compared to the systematic sources discussed in this section, and remains much smaller than the statistical uncertainties.

\subsection{Systematic uncertainties on \texorpdfstring{\boldmath{\ACPDPhiPi}}{ACP(D+ -> Phi Pi+)}}\label{sec:systematics_Acp}

\begin{table}
    \centering
    \caption{Systematic, statistical, and total uncertainties, in units of  $10^{-4}$, on the measurement of \mbox{$a_{C\!P}(D^+ \to \phi \pi^+)$}, using the 
    Run 2 dataset. The three columns are for the baseline analysis strategy (Hat constraint on $r_\pi$) and for two alternative strategies using external inputs based either on theoretical predictions for $r_\pi$ and $\delta_\pi$ (FAT predictions) or on experimental constraints on $r_\pi$ and $\delta_\pi$ (Constraint on $R_\pi$).
   }
    \label{tab:Tab3}
        \renewcommand{\arraystretch}{1.2} 
    \begin{tabular}{l|c|c|c}
        \hline
        \hline
        Source 
        & Hat constr. on \rpi
        & FAT predictions 
        & Constraint on \BigRpi \\
        \hline
        \hline
        Mass model                & $1.1$ & $0.68$ & $1.3$ \\
        Secondary decays              & $1.5$ & $1.3$ & $1.9$ \\
        Kinematic weighting          & $0.89$ & $0.54$ & $0.92$ \\
        Nonresonant $\Kp\Km$  & $0.20$ & $0.08$ & $0.16$ \\
        \hline
        $p(\KS)$ dependence 
                                 & $0.10$ & $0.60$ & $0.30$ \\
        Material density         & $1.2$ & $0.89$ & $1.1$ \\
        FAT \rpi and \deltapi & $-$ & $0.20$ & $-$ \\
        \hline
        Systematic uncertainty   & $2.4$ & $1.9$ & $2.7$ \\
        Statistical uncertainty  & $4.9$ & $3.3$ & $5.0$ \\
        Total uncertainty        & $5.5$ & $3.8$ & $5.7$ \\
        \hline
        \hline
    \end{tabular}
\end{table}

To propagate the systematic uncertainties associated with the determination of \DeltaAs to the measurement of \ACPDPhiPi, the fit for the offset is repeated for each systematic contribution, adding it in quadrature to the statistical uncertainty $\sigma_{\DeltaAs}$ of \cref{eq:deltachisq_definition_q}. 
Correlations between systematic uncertainties across different decay-time bins are found to be negligible.
The difference in quadrature between the resulting uncertainty on \ACPDPhiPi and that from the baseline fit is taken as the corresponding systematic uncertainty.

The uncertainty associated with the LHCb material map is evaluated by repeating the fit using three nuisance parameters: \rpi and \deltapi, constrained as described in \cref{sec:meas_Acp_d2phipi}, and the material-density variation $\varepsilon_\rho$, included with a Gaussian constraint with width
$6\%$. The systematic uncertainty from the material description is then obtained as the difference in quadrature between the total uncertainty from this fit and that obtained by fixing $\varepsilon_\rho $ to zero.
Finally, \ADTilde templates obtained using an alternative momentum dependence of the regeneration parameter $\Delta\chi$ are used to evaluate the corresponding systematic uncertainty. The difference with respect to the baseline result is assigned as the associated uncertainty.
In the baseline analysis strategy and in the result with the constraint on \BigRpi, the strong parameters \rpi and \deltapi are optimised directly on data, thus the uncertainty due to their determination is included in the statistical uncertainty. The size of their contribution to the uncertainty can be estimated as the difference in quadrature between the total statistical uncertainty and the statistical uncertainty obtained fixing them, which is $3.3\times 10^{-4}$. This contribution is $3.7\times 10^{-4} $ and $ 3.8\times 10^{-4}$ for the baseline result and for that with the constraint on \BigRpi, respectively. On the other hand, the results based on the FAT predictions are obtained by fixing both \rpi and \deltapi to their predicted values. A systematic uncertainty is then assigned as the maximum variation of \ACPDPhiPi when allowing these parameters to vary within their quoted uncertainties.

The total uncertainty on \ACPDPhiPi is obtained repeating the fit for the offset, adding all the systematic contributions in quadrature to the statistical uncertainty $\sigma_{\DeltaAs}$ of \cref{eq:deltachisq_definition_q} and leaving the $\varepsilon_\rho$ parameter free to vary with a Gaussian constraint.
The systematic uncertainties for the three measurement strategies presented in this analysis are summarised in~\cref{tab:Tab3}.

\section{Consistency checks}
\label{sec:checks}

\begin{figure}
    \centering
    \includegraphics[page=1, width=0.4\textwidth]{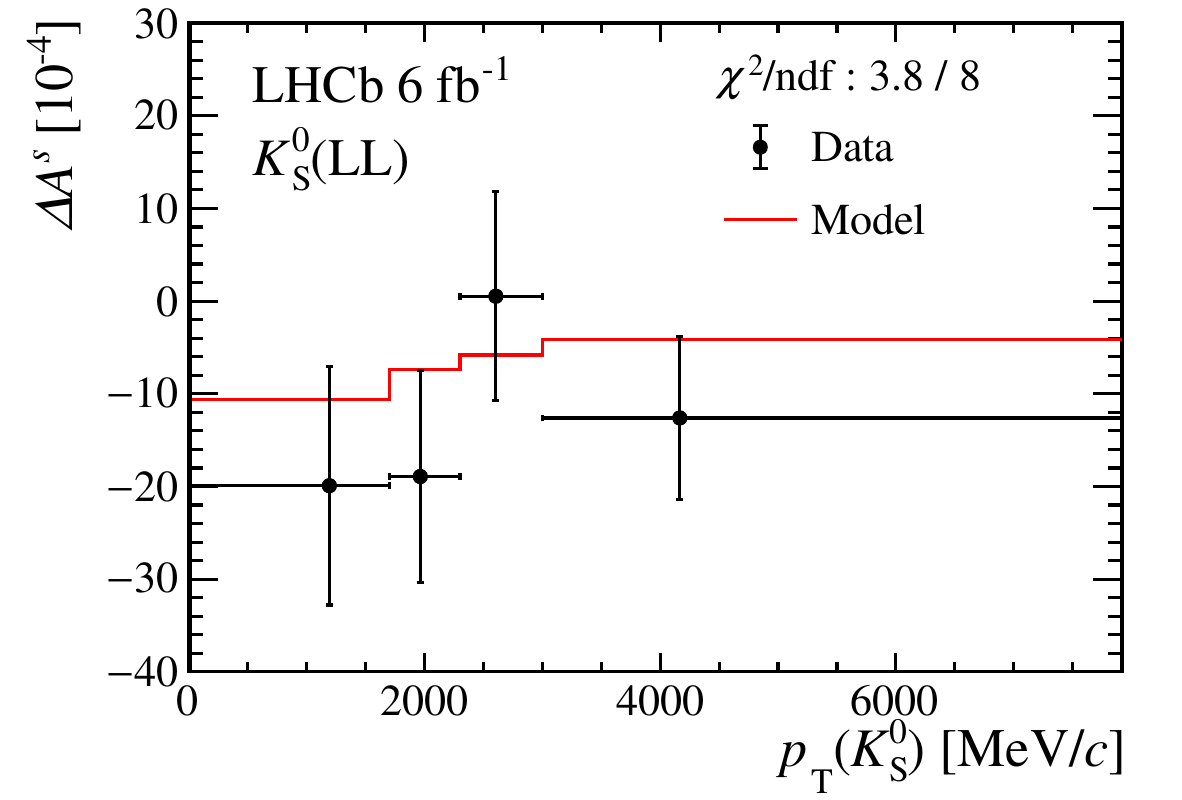}
    \includegraphics[page=1, width=0.4\textwidth]{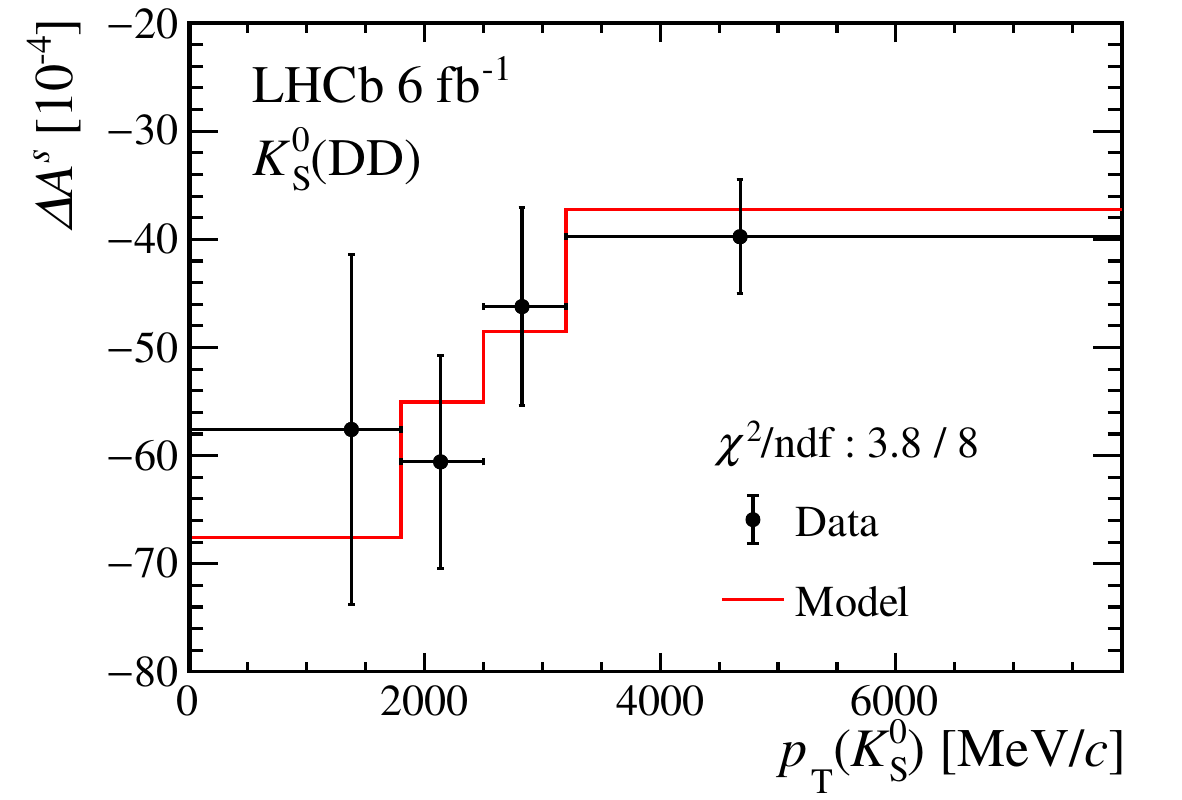}
    \includegraphics[page=1, width=0.4\textwidth]{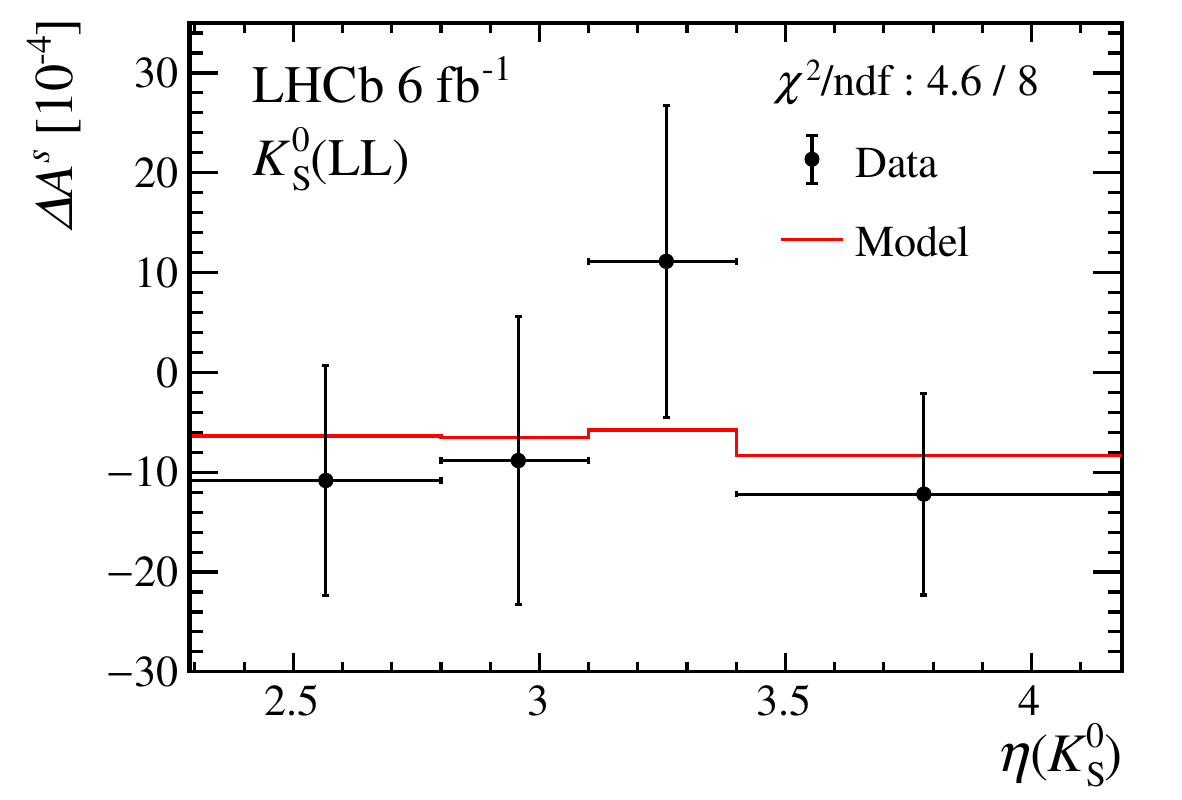}
    \includegraphics[page=1, width=0.4\textwidth]{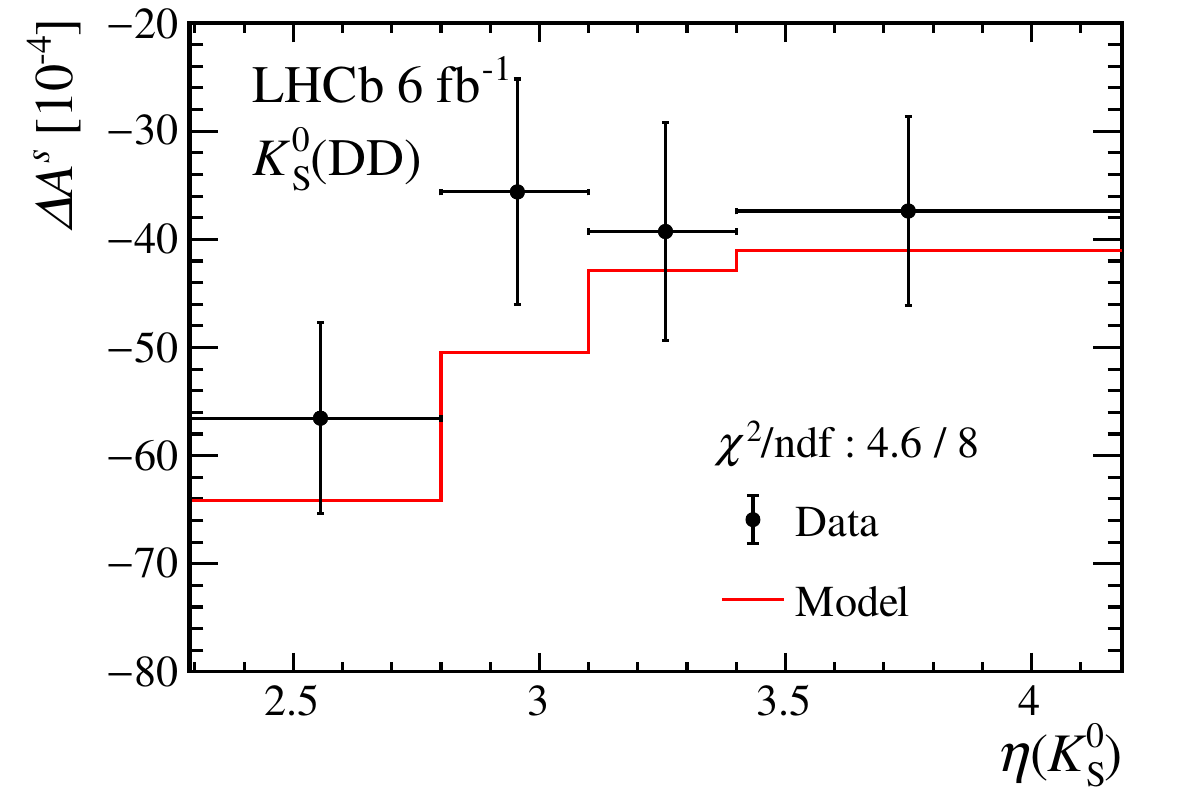}
    \caption{Comparison between the measured $\Delta A^s$ asymmetry values and the fit projection in bins of (top) $p_\mathrm{ T}(K_{\mathrm{S}}^0)$ and (bottom) $\eta(K_{\mathrm{S}}^0)$ using the (left) $K_{\mathrm{S}}^0(\mathrm{LL})$ and (right) $K_{\mathrm{S}}^0(\mathrm{DD})$ candidates.}
    \label{fig:Fig8}
\end{figure}

Several consistency checks are performed to test the robustness of the analysis. 
In the baseline analysis, the neutral-kaon asymmetries, \ADTilde, are modelled and fitted to data as a function of the \KS decay time. 
However, this quantity also depends on the trajectory of the \KS meson, which can traverse different amounts of detector material. Moreover, a different \KS momentum results in different interaction cross-sections. 
For this reason, the model fitted in~\cref{sec:meas_Acp_d2phipi} is projected on $\pt(\KS)$, $\eta(\KS)$, and the \KS azimuthal angle $\varphi(\KS)$. 
The splitting and weighting procedures of~\cref{sec:nuisance_asys} are repeated by dividing the samples into bins of these kinematic variables instead of bins of decay time. 
New \ADTilde asymmetry templates are produced by fixing the values of the relevant parameters \ACPDPhiPi, \rpi and \deltapi to the values obtained in the baseline fit of~\cref{sec:meas_Acp_d2phipi}. 
The results of these fit projections for the $\pt(\KS)$ and $\eta(\KS)$ variables are shown in~\cref{fig:Fig8}. All projections of the fitted model are in agreement with the asymmetries observed in the data, with a \pvalue of $88\%$, $79\%$, and $85\%$ for the $\pt(\KS)$, $\eta(\KS)$, and $\varphi(\KS)$ checks, respectively.

The measurement of \ACPDPhiPi is also repeated changing
the binning schemes of the \KS decay time and of the equalisation of the kinematic variables of the \Dp and \pip particles. All the results obtained with alternative binning schemes are compatible with the baseline result, with a significance ranging between 0.25 and 1.1 equivalent Gaussian standard deviations. 
Additional robustness tests are performed to check that the measured value of \ACPDPhiPi does not display unexpected dependencies on various observables, namely the pseudorapidity of the neutral kaon, the $z$ coordinate of the decay vertex of the \Dp candidate, and the 
IP of the trigger pion. No significant dependencies of \ACPDPhiPi on any of these variables are found. 

In addition, the analysis is repeated separately in six independent subsamples divided based on the year of collection of the data samples (2015+2016, 2017, and 2018) and the LHCb magnet polarity. The results are compatible, with a \pvalue of 8.2\%. The measurement of \ACPDPhiPi is repeated also separately for the two magnet polarities, and the compatibility between the two results is at the level of 0.2 Gaussian standard deviations.
Finally, \ACPDPhiPi is measured independently using only \KSLL and \KSDD candidates, and the results are in agreement with each other, with a compatibility at the level of 0.7 Gaussian standard deviations.

\section{Measurement of \texorpdfstring{\boldmath{\rpi}}{rpi} and \texorpdfstring{\boldmath{\deltapi}}{deltapi}}\label{sec:meas_rpi_deltapi_plugin}

Confidence intervals in the \rpideltapi plane are constructed by comparing the measured time-dependent asymmetry differences \DeltaAs with predictions generated for a grid of \rpideltapi values. 
Since Wilks' theorem~\cite{Wilks:1938dza} is not valid in the small-\rpi regime, in which the sensitivity on \deltapi becomes small, the \textit{Plugin} method (also known as the $\widehat{\mu}$ method)~\cite{Bodhisattva:2009uba,gammacombo} is adopted to calibrate the distribution of the test statistic with pseudoexperiments and to obtain frequentist confidence regions without relying on asymptotic \chisq approximations.
This frequentist approach allows the determination of confidence levels for the parameters of interest in the presence of nuisance parameters, here represented by the overall offset $q$ and the material density variation $\varepsilon_\rho$.
The determination of the \rpideltapi confidence intervals is performed by incorporating all the systematic uncertainties described in \cref{sec:systematics}.

\begin{figure}[t]
    \centering
    \includegraphics[page = 1, width = 0.45\textwidth]{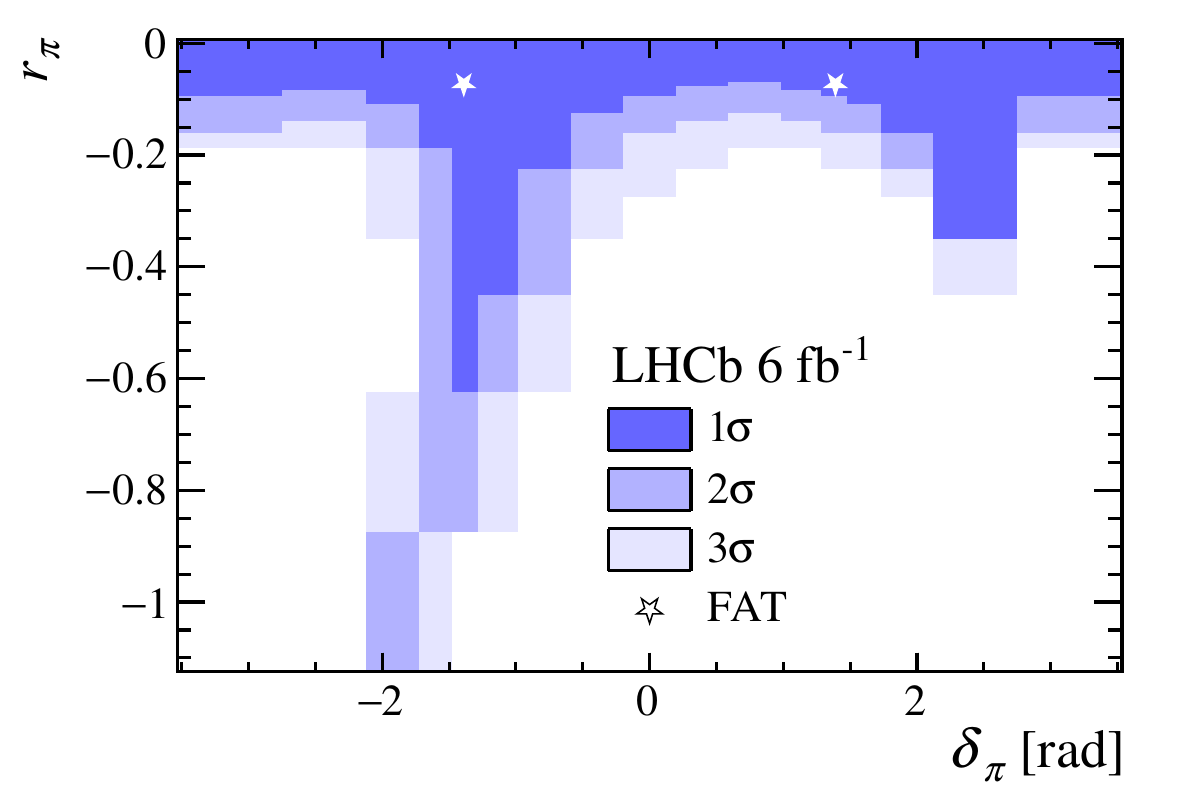}
    \includegraphics[page = 1, width = 0.45\textwidth]{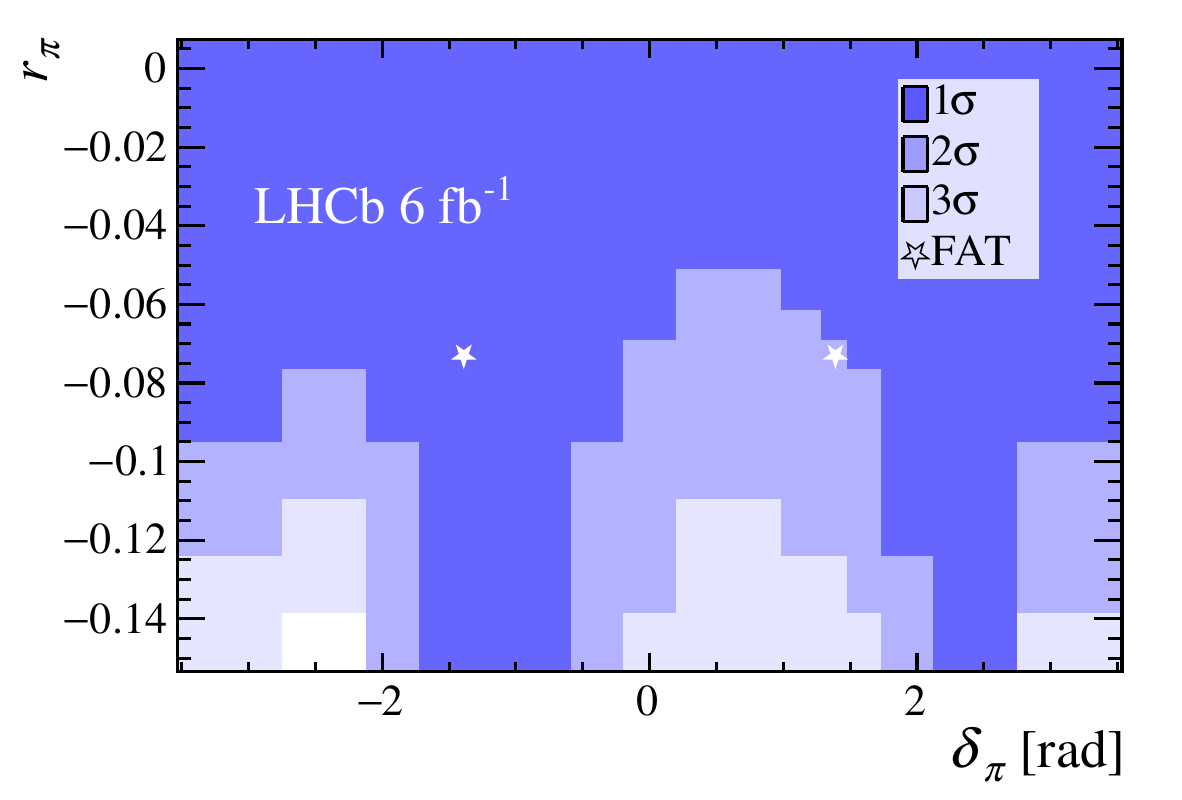}
    \caption{Confidence intervals applying different confidence-level thresholds to the \textit{p}-value in (left) the whole parameters space and (right) in the $r_\pi > -0.150$ region. White regions correspond to \textit{p}-values smaller than $0.3\%$.}
    \label{fig:Fig9}
\end{figure}

 The model used for the asymmetry is $\DeltaAsTilde = \ADTilderdeltarho-q$.
Several sets of predicted values of the asymmetry differences are generated on a four-dimensional grid spanning $\rpi \in [-1,0]$, $\deltapi \in [-\pi,\pi]$, $q \in [-124, 124]\times 10^{-4}$, and $\varepsilon_\rho\in[-18,\ 18]\%$. 
The material density variation $\varepsilon_\rho$ is constrained with a Gaussian prior of mean zero and width $6\%$.
The \DeltaAs data values are therefore compared with the predictions by profiling a similar test statistic as that of~\cref{eq:deltachisq_definition_q}, 
\begin{equation}
\Delta\chisq\rpideltapi =
\min_{q\, , \, \varepsilon_\rho}\chisq\qrpideltapiDeltarho -
\min_{q\, ,\,r'_\pi\, ,\,\delta'_\pi,\, \varepsilon_\rho'}\chisq(q\, ,\,r'_\pi\, ,\,\delta'_\pi,\, \varepsilon_\rho'),
\end{equation}
where $\chisq\qrpideltapiDeltarho$ is defined in analogy with \cref{eq:deltachisq_definition_q}. The uncertainties on the \DeltaAs data incorporate both the statistical and the systematic contributions, as detailed in \cref{tab:Tab2}.
For each \rpideltapi pair, 10\,000 pseudoexperiments are generated starting from the predicted values $\DeltaAsTilde$.  These predictions are obtained using the $(q\, ,  \varepsilon_\rho)$ values that minimise the $\chisq\qrpideltapiDeltarho$ function with \rpi and \deltapi fixed. The predicted values are then fluctuated according to  Gaussian distributions with widths given by the uncertainties $\sigma_{\DeltaAs}$ measured in data. 
Each pseudodataset is fitted using the same procedure as applied to data, and the resulting distribution of the test statistic $\Delta\chisq\rpideltapi$  is constructed. The \pvalue for a given \rpideltapi pair is then computed as the fraction of pseudoexperiments yielding a value of $\Delta\chisq\rpideltapi$ larger than that observed in data. Finally, confidence regions in the \rpideltapi plane, obtained using the full Run~2 dataset, are defined by the contours corresponding to
$p \geq 31.7\%$ ($1\sigma$),
$p \geq 4.6\%$ ($2\sigma$),
and $p \geq 0.3\%$ ($3\sigma$), and shown in~\cref{fig:Fig9}. \cref{app:pvalue_maps} reports the complete \pvalue map in the $\rpi>-0.150$ region.

The theoretical predictions from the FAT approach\cite{Yu:2017oky,Wang:2017ksn} are also shown in~\cref{fig:Fig9}. The prediction with a negative strong phase value, $\deltapi = -1.39$ rad, lies within the $1\sigma$ region, with a \pvalue of $48\%$. The scenario $\rpi=0$, corresponding to the absence of DCS charm transitions, is also compatible with the data, yielding a \pvalue of $76\%$. 
The solution with the opposite sign of the strong phase, $\deltapi = +1.39$ rad, is mildly disfavoured, with a \pvalue of $21\%$, placing it within the $2\sigma$ region. This behaviour is consistent with the findings of the Belle measurement of $\ACPdir(\DpToKSpi)$~\cite{Belle:2012ygt}, as discussed in Ref.~\cite{Yu:2017oky}.
Other theoretical determinations based on topological-amplitude and factorisation-inspired approaches are also available~\cite{Pakhlov:2021brf,Buccella:2019kpn}, including a very recent calculation~\cite{Lai:2026ysy}. Their predicted values, for the solution corresponding to a negative phase, fall within the $1\sigma$ confidence interval of our data-driven determination.

\section{Final results and conclusions}\label{sec:results}

A measurement of the direct \CP asymmetry in the \DpTophipi decay is presented using proton-proton collision data collected by the \lhcb experiment at a centre-of-mass energy of $13\tev$, corresponding to an integrated luminosity of $6\invfb$ recorded between 2015 and 2018. The result, obtained with a profile likelihood approach, is
\begin{equation*}
\begin{split}
   \ACPDPhiPi &=
   \left(0.1 \pm 4.9\stat \pm 2.4\syst\right)\times10^{-4} \, .\\
\end{split}
\end{equation*}
The result is compatible with \CP-symmetry at less than one Gaussian-equivalent standard deviation, 
and with the previously published \lhcb result~\cite{LHCb-PAPER-2019-002}, 
as well as the result reported in Ref.~\cite{LHCb-PAPER-2024-019} using $\Dp \to \Kp\Km\pip$ decays, based on a binned comparison between \Dp and \Dm phase-space distributions.
The measurement presented in this article supersedes the \lhcb result reported in Ref.~\cite{LHCb-PAPER-2019-002}, which used only a subset of the Run~2 dataset. 
Compared to that analysis, this work incorporates the 2018 data-taking period, corresponding to about $2\invfb$, and includes for the first time the \DpToKSpi candidates in which the neutral kaon decays downstream of the silicon vertex detector.
The inclusion of these candidates is essential for a reliable modelling of the neutral-kaon asymmetry and for a robust assessment of the associated systematic uncertainties. However, the online and offline requirements used to select candidates are significantly more stringent than those of Ref.~\cite{LHCb-PAPER-2019-002}, in order to precisely control production and detection asymmetries, as well as the neutral-kaon asymmetry. 
The resulting total signal yields of the samples used in this measurement are about a factor of 1.5 larger than those of the previous \lhcb measurement with a partial Run~2 dataset.
Furthermore, the result of this article can be combined with the previous LHCb determination based on independent data samples collected at centre-of-mass energies of $7$ and $8\tev$~\cite{LHCb-PAPER-2012-052}, under the assumption that statistical and systematic uncertainties are uncorrelated. The combination of Run 1 and Run 2 results is $\ACPDPhiPi =
   \left(-0.2 \pm 5.3\right)\times10^{-4}$.

A novel aspect of this analysis is the development of a time-dependent model of the neutral-kaon detection asymmetry that includes, for the first time, the interference between CF and DCS charm decay amplitudes. This allows constraining the hadronic parameters \rpi and \deltapi directly from data, without relying on external theoretical inputs or approximations such as $\rpi=0$, adopted in Refs.~\cite{LHCb-PAPER-2019-002,LHCb-PAPER-2022-024}. 
In addition to the baseline result, which does not rely on external inputs, apart from allowing \rpi to vary within a broad interval (chosen primarily on the basis of CKM factors and experimental measurements of decay processes governed by the same weak transitions), two alternative determinations of \ACPDPhiPi are reported. The first is model-dependent and is obtained by constraining \rpi and \deltapi to their theoretical predictions~\cite{Yu:2017oky,Wang:2017ksn}, yielding
\begin{equation*}
\begin{split}
   \ACPdir(\DpTophipi; \ \textrm{FAT}-) &=
   \left(-2.2 \pm 3.3\stat \pm 1.9 \syst\right)\times10^{-4}\, , \\
    \ACPdir(\DpTophipi; \ \textrm{FAT}+) &=
   \left(+4.6 \pm 3.3\stat \pm 1.9 \syst\right)\times10^{-4}\, , \\
\end{split}
\end{equation*}
where the two quoted central values correspond to the two predicted values of the relative strong phase \deltapi and the uncertainties are the same for both cases. The constrained parameters are $\rpi = -0.073 \pm 0.004$ and $\deltapi = -1.39\, (+1.39) \pm 0.05$~rad. The total uncertainty is $3.8\times 10^{-4}$, representing the most precise determinations of \ACPDPhiPi to date. This can be compared with the previously published \lhcb result~\cite{LHCb-PAPER-2019-002}, which achieves a total uncertainty of $5.1\times 10^{-4}$ under the assumption $\rpi=0$. 
The second alternative result is obtained by combining this measurement with the CLEO measurement of the \BigRpi observable, which is defined as the \KS/\KL asymmetry in $\Dp \to K^0_{\rm S/L}\pip$ decays~\cite{CLEO:2007rhw}, obtaining 
\begin{equation*}
\begin{split}
   \ACPdir(\DpTophipi; \ \BigRpi) &=
   \left(0.4 \pm 5.0 \stat \pm 2.7 \syst\right)\times10^{-4} \, .\\
\end{split}
\end{equation*}
This result has a slightly larger total uncertainty than the baseline result. Both alternatives are compatible with the main measurement.

This article also presents a measurement of the hadronic parameters \rpi and \deltapi governing CF-DCS interference in the \DpToKSpi decay, independent of any strong-interaction model. 
Confidence intervals in the \rpideltapi plane are obtained using a frequentist approach, accounting for both statistical and systematic uncertainties. Owing to the large Run~2 dataset collected by the \lhcb experiment, a wide region of the \rpideltapi parameter space is excluded at high confidence level.
The resulting confidence regions are compatible with the prediction based on the factorisation-assisted topological amplitudes approach~\cite{Yu:2017oky}, which yields $\rpi = -0.073 \pm 0.004$ and $\deltapi = -1.39 \pm 0.05$ rad and lies within the $1\sigma$ region. The alternative solution with positive strong phase, $\deltapi = +1.39$ rad, falls within the $2\sigma$ region, in agreement with the preference for a negative strong phase observed by the Belle collaboration in their measurement of $\ACPdir(\DpToKSpi)$~\cite{Belle:2012ygt}.

In conclusion, the methodology developed here represents a significant advance in the treatment of neutral-kaon detection asymmetries in high-precision flavour-physics analyses. The extension of the asymmetry model to include CF-DCS interference effects enables the determination of the associated hadronic parameters, with reduced reliance on external theoretical assumptions. The precision of the baseline result is currently limited by the statistical uncertainty on the interference parameters \rpi and \deltapi, rather than by detector-related systematic effects. This limitation is expected to be alleviated with the larger datasets collected during LHC Run~3 and beyond, which will allow tighter constraints on these parameters and a corresponding reduction of the uncertainty associated with the neutral-kaon asymmetry correction. The approach presented in this work is therefore well suited for future measurements with large datasets and can be directly applied to other charm and beauty decays involving neutral kaons, preparing for improved sensitivity to \CP-violating effects in the charm sector.

\section*{Acknowledgements}
%
%
\noindent We express our gratitude to our colleagues in the CERN
accelerator departments for the excellent performance of the LHC. We
thank the technical and administrative staff at the LHCb
institutes.
We acknowledge support from CERN and from the national agencies:
ARC (Australia);
CAPES, CNPq, FAPERJ and FINEP (Brazil); 
MOST and NSFC (China); 
CNRS/IN2P3 and CEA (France);  
BMFTR, DFG and MPG (Germany);
NKFIH (Hungary);              
INFN (Italy); 
NWO (Netherlands); 
MNiSW and NCN (Poland); 
MEC/IFA (Romania); 
MICIU and AEI (Spain);
SNSF and SER (Switzerland); 
NASU (Ukraine); 
STFC (United Kingdom); 
DOE NP and NSF (USA).
We acknowledge the computing resources that are provided by ARDC (Australia), 
CBPF (Brazil),
CERN, 
IHEP and LZU (China),
IN2P3 (France), 
KIT and DESY (Germany), 
INFN (Italy), 
SURF (Netherlands),
Polish WLCG (Poland),
IFIN-HH (Romania), 
PIC (Spain), CSCS (Switzerland), 
GridPP (United Kingdom),
and NSF (USA).  
We are indebted to the communities behind the multiple open-source
software packages on which we depend.
Individual groups or members have received support from
RTP (Australia), 
FWO Odysseus grant G0ASD25N (Belgium), 
Key Research Program of Frontier Sciences of CAS, CAS PIFI, CAS CCEPP (China); 
Minciencias (Colombia);
EPLANET, Marie Sk\l{}odowska-Curie Actions, ERC and NextGenerationEU (European Union);
A*MIDEX, ANR, IPhU and Labex P2IO, and R\'{e}gion Auvergne-Rh\^{o}ne-Alpes (France);
Alexander-von-Humboldt Foundation (Germany);
ICSC (Italy); 
Severo Ochoa and Mar\'ia de Maeztu Units of Excellence, GVA, XuntaGal, GENCAT, InTalent-Inditex and Prog.~Atracci\'on Talento CM (Spain);
the Leverhulme Trust, the Royal Society and UKRI (United Kingdom).

\newpage
\section*{Appendices}

\appendix

\section{Neutral kaon detection asymmetry in matter}
\label{app:ADKS0Model}

The time evolution of a neutral kaon state can be described using a phenomenological model, based on quantum mechanics, restricted to the two-dimensional space spanned by the \Kz and \Kzb states. 
The flavour eigenstates, $\ket{\Kz}$ and $\ket{\Kzb}$, are not eigenstates of the Hamiltonian, and this gives rise to \textit{mixing}, \ie\ if the neutral kaon is initially a $\Kz$ meson, there is a nonzero probability to observe after some time $t$ a $\Kzb$ meson.
The eigenstates of the Hamiltonian in vacuum are called $\ket{\KS}$ and $\ket{\KL}$, and in case of no \CP violation in the mixing, they correspond to the \CP-even and \CP-odd combinations of $\ket{\Kz}$ and $\ket{\Kzb}$.
This paper adopts the sign convention $\CP\ket{\Kz} = - \ket{\Kzb}$.
If there is \CP violation in the mixing, the eigenstates of the Hamiltonian differ from the \CP eigenstates by a correction quantified by the small parameter $\varepsilon$:
\begin{equation}\label{eq_KSKL_KzKzb}
\left\{
    \begin{array}{c}
        \ket{\KS} = \cfrac{1}{\sqrt{2(1+|\varepsilon|^2)}} \left((1+\varepsilon)\ket{\Kz} - (1-\varepsilon)\ket{\Kzb}\right),  \\
          \\
           \ket{\KL} = \cfrac{1}{\sqrt{2(1+|\varepsilon|^2)}} \left((1+\varepsilon)\ket{\Kz} + (1-\varepsilon)\ket{\Kzb}\right).\\
    \end{array}
    \right. 
   \end{equation}
Violation of \CP symmetry in mixing is one of the sources of neutral kaon asymmetry, and the experimental results for the modulus and phase of $\varepsilon$ are reported in Table~\ref{tab:Tab4}.  
The eigenvalues $ \lambda_{\rm S,L}$ are expressed as 
\begin{equation}
    \lambda_{\rm S,L} \equiv  \ m_{\rm S,L}-\frac{i}{2}\Gamma_{\rm S,L},  
\end{equation}
where the real parameters $m_{\rm S,L} \text{ and } \Gamma_{\rm S,L}$ are related to the mass and decay width of the neutral kaon, $M_K$ and $\Gamma_K$ respectively,
\begin{equation}
    M_K = \frac{m_{\rm S}+m_{\rm L}}{2} \qquad \text{and} \qquad \Gamma_K = \frac{\Gamma_{\rm S}+\Gamma_{\rm L}}{2}.
\end{equation}

If the neutral kaons evolve in matter, it is necessary to also include a contribution to the time evolution that takes into account the strong interaction between neutral kaons and nuclei. Due to the different quark compositions of \Kz and \Kzb mesons, their interaction cross-sections with nuclei are different.

The description of the time evolution of kaons in matter is based on  Refs.~\cite{Good:1957zza,Fetscher:1996fa}. 
To describe the coherent interaction of neutral kaons with matter, it is necessary to add another contribution to the Hamiltonian in vacuum. This new term can be parametrised in the $\ket{\Kz}-\ket{\Kzb}$ basis as
\begin{equation}
    \mathbf{H}_{\rm mat} = \left(\begin{array}{cc}
            \chi & 0 \\
             0 & \overline{\chi}
        \end{array}
        \right),
\end{equation}
where $\chi$ and $\overline{\chi}$ are the amplitudes of the interaction of a $\Kz$ or a $\Kzb$ with the material. It is possible to express the $\chi$ and $\overline{\chi}$ coefficients as a function of the neutral-kaon momentum $p$ and of the material properties, namely the nuclei density $n$ and the total interaction cross-section of a $\Kz$ or a $\Kzb$ with nuclei, denoted as $\sigma_{\Kz}$ and $\sigma_{\Kzb}$, respectively. 
Using the optical theorem, the matrix elements $\chi$ and $\overline{\chi}$ can be expressed in terms of the forward scattering amplitudes of the interaction of a $\Kz$ or a $\Kzb$ with nuclei, $f(\theta=0)$ and $\overline{f}(\theta=0)$. These amplitudes add coherently, since the elastic interaction with the neutral kaon leaves the nucleus unchanged and, working under Born's approximation, they can be written as
\begin{equation}
    \chi \equiv -\cfrac{2\pi n }{M_K} \ f(0) \qquad \text{and} \qquad
    \overline{\chi} \equiv -\cfrac{2\pi n }{M_K} \ \overline{f}(0) \, ,
\end{equation}
where $n$ can be expressed in terms of the mass density $\rho$ and atomic weight $A$ as $n = (\rho \ N_A)/{A}$, with $N_A$ the Avogadro number.
The time evolution of the \Kz--\Kzb system in matter depends only on the difference $\Delta f \equiv \overline{f}(0) - f(0)$, which can be determined as follows:
\begin{itemize}
    
    \item the imaginary part of $\Delta f$ is related, via the optical theorem, to the difference between the total interaction cross-section of a $\Kz$ and a $\Kzb$ with matter, including both the absorption and forward-scattering contributions,
\begin{equation}
    \label{eq:delta-sigma}
 \Delta\sigma \equiv \sigma_{\Kzb} - \sigma_{\Kz} = \frac{4\pi\hbar}{p}\ \Im\left(\Delta f \right).
\end{equation}
The difference between the cross-sections, $\Delta\sigma$, can be estimated with the following empirical scaling law derived in Ref.~\cite{Gsponer:1978dt}, based on measurements in C, Al, Cu, Sn, and Pb nuclei for neutral kaon momenta between 20 and 140\gevc,
\begin{equation}
    \label{eq:delta-sigma-numerical}
 \Delta\sigma(p,A) = 23.2\text{ mb} \cdot \frac{ A^{(0.758 \pm 0.003)}}{p^{(0.614\pm 0.009)}} \, ,  
\end{equation}
where $A$ is expressed in g/mol and $p$ in \gevc;
\item the phase of $\Delta f$ is determined using the phase-power relation
\begin{equation}
 \arg(\Delta f) = -\pi/2\cdot(1+\alpha) \, ,   
\end{equation}
where $\alpha$ is the exponent of the dependence of $\Delta \sigma$ on the neutral kaon momentum, $\Delta\sigma \propto p^{\alpha-1}$~\cite{Roehrig:1977db}. Using the value of $\alpha-1 = -0.614 \pm 0.009 $ from Ref.~\cite{Gsponer:1978dt} and the estimate of its uncertainty from Ref.~\cite{PhysRevLett.75.2070}, 
the phase is $\arg(\Delta f) = (-124.7 \pm 0.8)^\circ$. 
\end{itemize}
Using \cref{eq:delta-sigma}, $\Delta f$ can be expressed as
\begin{equation}
\Delta f(p,A) = \cfrac{p \ \Delta\sigma(p,A)}{4\pi\sin{(\arg{\Delta f})}} \  e^{i \arg{\Delta f}}
\propto p^{0.386} \cdot A^{0.758},
\end{equation}
where \cref{eq:delta-sigma-numerical} was used in the last step.
Finally, the quantity of interest for the time evolution of neutral kaons in matter, $\Delta\chi$, can be expressed as
\begin{equation}
 \Delta\chi = \cfrac{2\pi \rho N_A }{ M_K A}\ \Delta f 
    \propto p^{0.386} \cdot \rho \cdot A^{-0.242}   \, .
\end{equation}
The time evolution of the neutral kaon system can be found by solving Schr\"odinger's equation. In the $\ket{\KS}-\ket{\KL}$ basis, the state of the system at time $t$ can be written generally as    
\begin{equation}
    \ket{\psi_K(t)} = \alphaS(t)\ket{\KS}+\alphaL(t)\ket{\KL},
\end{equation}
where $\alphaS(t)$ and $\alphaL(t)$ depend on the coefficients of the initial state $\alphaSL(0)$ and the Hamiltonian parameters as described in Ref.~\cite{Fetscher:1996fa},
\begin{align}
\label{eq_alpha_S}
    \alphaS(t) & = e^{-i\Sigma t}\ \left(\alphaS(0) \left( \cos\Omega t  + i \cfrac{  \Delta\lambda}{{2\Omega}}\sin\Omega t \right)  -i \alphaL(0) \cfrac{\Delta\chi}{2\Omega} \sin\Omega t \right)\, ,\\
\label{eq_alpha_L}
    \alphaL(t) & = e^{-i\Sigma t}\ \left( \alphaL(0) \left( \cos\Omega t  - i \cfrac{  \Delta\lambda}{2\Omega} \sin\Omega t \right) - i\alphaS(0)\cfrac{\Delta\chi}{2\Omega} \sin\Omega t\right)\, ,
\end{align}
where the following parameters are used: 
\begin{itemize}

    \item $\Delta \lambda \equiv \lambda_{\rm L}-\lambda_{\rm S} = \Delta m - \cfrac{i}{2}\Delta\Gamma$, which depends on the eigenvalues of the vacuum Hamiltonian term;
    \item $\Delta\chi \equiv \chi - \overline{\chi}$, which depends on the material properties and kaon momentum; 
    \item $\Omega = \cfrac{1}{2} \ \sqrt{\Delta\lambda^2 + \Delta\chi^2}$, which determines the time dependence of $\alphaS$ and $\alphaL$;
    \item $\Sigma = \cfrac{1}{2} \ \left(\lambda_{\rm S}+\lambda_{\rm L} + \chi+ \overline{\chi} \right)$ describes the absorption of the strangeness components as well as the decay of the mass eigenstates. In particular, the decay width of the neutral kaon in any final state is proportional to
    \begin{equation}
        \left| e^{-i\Sigma t}\right|^2=\exp\left(-\left(\cfrac{\Gamma_{\rm S}+\Gamma_{\rm L}}{2} + \Im(\chi+ \overline{\chi})\right)t\right) \equiv \exp\left(-\left(\Gamma + \Gamma_{\rm abs}\right)t\right) \, .
    \end{equation}
\end{itemize}
The numerical values used for the parameters $\Delta m, \Delta\Gamma$, $\varepsilon$, and the phase $\arg(\Delta f)$ are summarised in Table~\ref{tab:Tab4}.

The difference $\Delta \chi$ is responsible for \KS \textit{regeneration} in matter, because according to \cref{eq_alpha_S,eq_alpha_L} it is possible to have $\alphaS(t) \neq 0$ even when $\alphaS(0) = 0$.
In particular, the size of regeneration can be quantified by the regeneration parameter $r$, defined similarly to Ref.~\cite{Ko:2010mk} as
\begin{equation}\label{eq:regeneration_r}
  r= \cfrac{\Delta\chi}{2\Omega} \, .
\end{equation}
This parameter is typically of the order of percent and, in the case of a pure initial \KL beam, it exactly quantifies the regenerated \KS component: $|\alphaS(t)|^2\propto |r|^2 \sin^2(\Omega t)$.

\begin{table}[t]
    \renewcommand{\arraystretch}{1.3} 

    \centering
        \caption{
        Measurements of the parameters used to estimate the neutral-kaon asymmetry in matter. To assess the contribution to the systematic uncertainties due to these external inputs, also the relative uncertainty is reported.}
    \label{tab:Tab4}
    \begin{tabular}{c|c|c|c}
         \hline
         \hline

        Parameter &  Measured value & Relative uncertainty & Ref.\\
         \hline
         \hline
         $M_K$ & $(497.611 \pm 0.013)\mevcc$ & $2.6 \times 10^{-5}$ &~\cite{PDG2024}  \\
         $\Delta m$ & $(0.5293 \pm 0.0009) \ \hbar/\text{s}$  & $1.7 \times 10^{-4}$ &~\cite{PDG2024}  \\
          $\tau_{\rm S}$ & $(0.8954 \pm 0.0004) \times 10^{-10} \sec$ & $4.5 \times 10^{-4}$&~\cite{PDG2024}   \\
         $\tau_{\rm L}$ & $(5.116 \pm 0.021) \times 10^{-8} \sec$ & $4.1 \times 10^{-3}$&~\cite{PDG2024}  \\
         $|\varepsilon|$ & $(2.228 \pm 0.011)\times 10^{-3}$& $4.9 \times 10^{-3}$&~\cite{pdg_CPV_K0_rew} \\
         $\arg(\varepsilon)$ & $(43.5 \pm 0.5)^\circ$& $1.1 \times 10^{-2}$&~\cite{pdg_CPV_K0_rew} \\
         $ \arg(\Delta f)$ & $(-124.7 \pm 0.8)^\circ$ & $6.4 \times 10^{-3}$&~\cite{Gsponer:1978dt,PhysRevLett.75.2070}\\
         \hline
         \hline

    \end{tabular}

\end{table}

Neutral kaon asymmetry can be estimated from the width of the decay of the state ${\psi_K(t)} \to \pi^+\pi^-$. Since the final state is \CP even, this decay amplitude can be estimated by projecting the state $\ket{\psi_K(t)}$ onto the \CP-even eigenstate $\ket{\Kone}$. Employing the following relations between the $\KS-\KL$ and the $\Kone-\Ktwo$ basis 
 \begin{equation}\label{eq_K1K2_KSKL}
\left\{
     \begin{array}{rc}
        \ket{\KS} = & \cfrac{\ket{\Kone} + \varepsilon \ket{\Ktwo}}{\sqrt{1+|\varepsilon|^2}},  \\
          \\
            \ket{\KL} = & \cfrac{\ket{\Ktwo} + \varepsilon \ket{\Kone}}{\sqrt{1+|\varepsilon|^2}};  \\
           
    \end{array}
\right.  
\end{equation}
this width can be written as
\begin{equation}\label{eq_gamma_KS0}
\begin{split}
       \Gamma(\psi_K(t) \to \pip\pim)  &\propto \left|\alphaS(t)\langle\Kone|\KS\rangle +  \alphaL(t)\langle\Kone|\KL\rangle\right|^2 \\
       &\propto \left|\alphaS(t) + \varepsilon \cdot \alphaL(t)\right|^2,
\end{split}
\end{equation}
which can be estimated, using \cref{eq_alpha_S,eq_alpha_L}, from the initial coefficients $\alphaSL(0)$, which depend on the decay in which the neutral kaon is produced. 
Using Eqs.~\eqref{eq:psiKplus0_KzKzb} and~\eqref{eq:psiKminus0_KzKzb}, the coefficients $\alphaSLpm(0)$ for the $\DpmToKSpi$ decays are given by
\begin{equation}\label{eq_alphaS+0_strongphase}
    \alphaSpm(0) = \sqrt{\cfrac{1+|\varepsilon|^2}{2(1+\rpi^2)}} \cdot \left(\cfrac{\mp 1}{1\mp\varepsilon} \pm \cfrac{\rpi e^{i(\deltapi\pm\varphi)}}{1\pm\varepsilon}\right)\,,
\end{equation}
\begin{equation}\label{eq_alphaL+0_strongphase}
    \alphaLpm(0) = \sqrt{\cfrac{1+|\varepsilon|^2}{2(1+\rpi^2)}} \cdot \left(\cfrac{1}{1\mp\varepsilon} + \cfrac{\rpi e^{i(\deltapi\pm\varphi) }}{1\pm\varepsilon}\right)\,.
\end{equation}

\subsection{Incoherent \texorpdfstring{\boldmath{\KS}}{KS} regeneration}

The regeneration of \KS mesons may occur either
coherently, through forward elastic scattering off all nuclei in the
medium, or incoherently, via diffractive scattering on individual
nuclei. Coherent regeneration interferes with \Kz--\Kzb mixing, and as discussed above is treated within the time evolution of the kaon system. Incoherent regeneration, instead, requires a different approach, being a pure probabilistic phenomenon. 

For a single
nucleus, the differential cross-section describing incoherent production of \KS mesons from a pure beam of \KL mesons through diffractive scattering can be expressed in terms of the difference between forward scattering amplitudes as~\cite{Kleinknecht:1973ny}
\begin{equation}
  \frac{d\sigma_{\rm diff}}{d\Omega}(\theta)
  \;=\;
  \frac{|f(\theta)-\bar{f}(\theta)|^2}{4}.
\end{equation}
The angular dependence of the amplitudes $f(\theta)$ and $\bar{f}(\theta)$
follows a Bessel-type behaviour and is described in detail in
Refs.~\cite{Baldini:1996ss,PhysRev.124.1223} using the eikonal
approximation. For scattering angles below a few degrees, the amplitudes
can be approximated as constant and equal to their forward values,
$f(0)$ and $\bar{f}(0)$. This approximation is justified by the stringent
kinematic and pointing constraints imposed by the \Dp meson reconstruction,
which limit the angular deviation of selected \KS candidates. The solid
angle can therefore be approximated as
$\Delta\Omega \simeq \pi(\delta\theta)^2$.
The allowed angle $\delta\theta$ can be interpreted as the uncertainty on the reconstructed direction of the neutral kaon. Deviations larger than this value would significantly distort the reconstructed decay geometry, leading to the rejection of the event. The value of $\delta\theta$ is estimated directly from data as the average angle between the reconstructed neutral-kaon flight direction and its momentum vector. This yields $\delta\theta < 0.25$ mrad for \KSLL candidates and $\delta\theta < 0.6$ mrad for \KSDD candidates.
This estimate is validated using \lhcb simulation by comparing the reconstructed \KS direction with the true one. The average angle between the two is measured to be consistent with the value found in data.

Under these assumptions, given an initial \KL meson travelling through matter, the probability of producing \KS mesons incoherently in the proper time interval $\deriv t$ is
\begin{equation}
 R_{\rm inc}
  \deriv t \equiv 
  \beta\gamma c\,n\,
  \frac{|\Delta f(0)|^2}{4}\,
  \Delta\Omega \ \deriv t\propto |r|^2,
\end{equation}
where the proportionality to the regeneration parameter $r$, defined in \cref{eq:regeneration_r}, is made explicit.
Once produced, incoherent \KS mesons propagate in matter with an
effective width
\begin{equation}
  \Gamma_{\rm eff}
  =
  \frac{\Gamma_{\rm S}+\Gamma_{\rm L}}{2}
  + \Gamma_{\rm abs},
\end{equation}
which accounts for both weak decays and nuclear absorption, with
$\Gamma_{\rm abs} \equiv \Im(\chi+\overline{\chi})$. For an initial pure
\KL state, the rate of incoherent $\KS\to\pi\pi$ decays occurring within a
uniform material layer, $\Gamma_{\rm inc}^{(\KL)}(t)$, is
\begin{equation}
  \Gamma_{\rm inc}^{(\KL)}(t)
  =
  \int_0^t \deriv t'\,
  R_{\rm inc}\,
  \Gamma_{\rm S}^{\pi\pi}\,
  e^{-\Gamma_{\rm eff}(t-t')} \nonumber =
  R_{\rm inc}\,
  \Gamma_{\rm S}^{\pi\pi}
  \frac{1 - e^{-\Gamma_{\rm eff}t}}{\Gamma_{\rm eff}}
  \, ,
  \label{eq:Gammainc_singlelayer}
\end{equation}
with $\Gamma_{\rm S}^{\pi\pi}$ the decay rate of a \KS into the $\pip\pim$ final state.
This expression is derived under the assumption that the material layer is sufficiently thin that the \KL component remains unchanged.
Thus, for a general initial state, this expression must be weighted by the
\KL component at production, $|\alphaL(0)|^2$.
In the full detector description, the neutral-kaon trajectory is divided
into discrete material layers. The probability of incoherent \KS
production in each layer is computed iteratively, taking into account
the survival probability of previously produced \KS mesons. The total
probability to produce an incoherent \KS along the full path is
\begin{equation}
  P_{\rm inc, tot}^\KS(t)
  =
  \sum_i
  |\alphaL(t_{i-1})|^2\,
  R_{{\rm inc},i}\,
  \frac{1 - e^{-\Gamma_{{\rm eff},i}\Delta t_i}}{\Gamma_{{\rm eff},i}}\,
  e^{-\sum_{k>i}\Gamma_{{\rm eff},k}\Delta t_k}.
\end{equation}
This probability is nearly identical for an initial \Kz or \Kzb state, differing
only through the small difference in the corresponding
$\alphaL(0)$ coefficients.
The effective rate of incoherent $\KS\to\pip\pim$ decays is
therefore
\begin{equation}
  \Gamma_{\rm inc}^{\KS \to \pip\pim}(t)
  =
  P_{\rm inc, tot}^\KS(t)\,
  \Gamma_{\rm S}^{\pi\pi}.
\end{equation}
This contribution is added to the coherent width defined in \cref{eq_gamma_KS0} when evaluating the neutral-kaon detection asymmetry.
The relative size of the incoherent and coherent components can be understood from their dependence on the regeneration parameter $r$, which is typically at the percent level. 
In the presence of an initial \KS component $(\alphaS(0)\neq 0)$, the coherent width contains terms of $\mathcal{O}(1)$, $\mathcal{O}(|r|)$ and $\mathcal{O}(|r|^2)$, whereas the incoherent contribution enters only at $\mathcal{O}(|r|^2)$ and is therefore strongly suppressed. 
The only situation in which the coherent and incoherent contributions become comparable is that of a pure \KL beam $(\alphaS(0)=0)$. 
In this case, both rates are proportional to $|r|^2$ and are of the same order of magnitude, with the incoherent contribution dominating in specific kinematic regimes~\cite{Baldini:1996ss,PhysRev.124.1223}. 
Numerically, the impact of incoherent regeneration is found to be negligible, at the level of $\mathcal{O}(10^{-8})$ for the combined \KSLL and \KSDD samples, compared with the dominant coherent contribution of order $10^{-4}$--$10^{-3}$.
As an additional check, this estimate is repeated using for $\delta\theta$ the maximum angle between the neutral kaon momentum and flight distance measured in data, corresponding to 1\mrad in \KSLL candidates and 3\mrad in \KSDD candidates. The corresponding contribution to the neutral kaon asymmetry due to incoherent regeneration is still found to be negligible, at the level of $\mathcal{O}(10^{-7})$ for the combined \KSLL and \KSDD samples.

\clearpage
\newpage
\section{Map of the \textit{p}-value for the strong parameters \texorpdfstring{\boldmath{\rpideltapi}}{(rpi,deltapi)}}\label{app:pvalue_maps}

\begin{figure}[ht!]
    \centering
     \includegraphics[width = \textwidth]{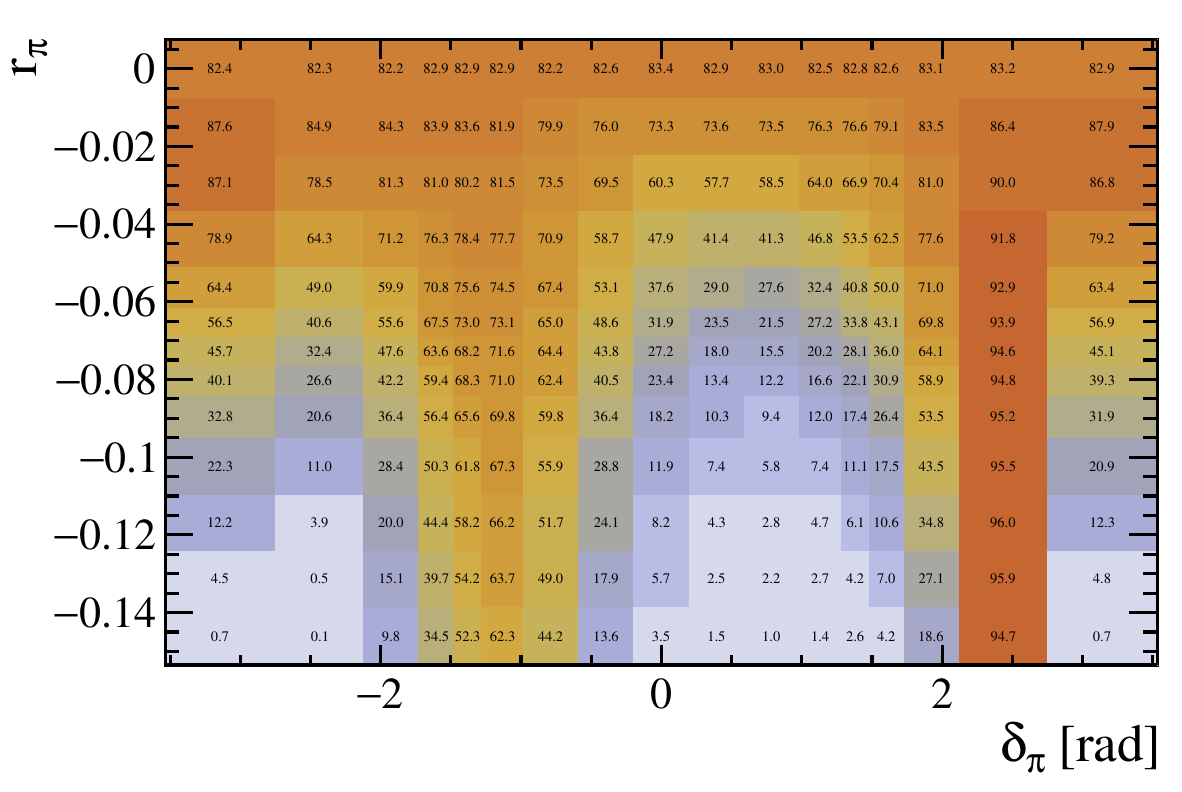}
    \caption{Map of the \textit{p}-value, in units of $10^{-2}$, as a function of the strong parameters $r_\pi$ and $\delta_\pi$ in the $r_\pi > -0.150$ region, considering both statistical and systematic uncertainties on the $\Delta A^s$ measurements, including density variations on the LHCb material map. }
    \label{fig:Fig10}
\end{figure}

\begin{table}[htbp]
\centering

\setlength{\tabcolsep}{2.2pt}
\renewcommand{\arraystretch}{1.2}
\caption{Map of the \textit{p}-values, in units of $10^{-2}$, as a function of the strong parameters \rpi and \deltapi in the $\rpi > -0.150$ region, considering both statistical and systematic uncertainties on the \DeltaAs measurements, including density variations on the LHCb material map.}
\label{tab:Tab5}
\resizebox{\textwidth}{!}{%
\begin{tabular}{c||c|c|c|c|c|c|c|c|c|c|c|c|c|c|c|c|c}
\hline
\hline
& & & & & & & & & & & & & & & & &\\
$r_\pi \backslash \delta_\pi$ &
$-\pi$ &
$-\cfrac{3\pi}{4}$ &
$-\cfrac{3\pi}{5}$ &
$-\cfrac{\pi}{2}$ &
$-1.39$ &
$-\cfrac{3\pi}{8}$ &
$-\cfrac{\pi}{4}$ &
$-\cfrac{\pi}{8}$ &
$0$ &
$\cfrac{\pi}{8}$ &
$\cfrac{\pi}{4}$ &
$\cfrac{3\pi}{8}$ &
$1.39$ &
$\cfrac{\pi}{2}$ &
$\cfrac{3\pi}{5}$ &
$\cfrac{3\pi}{4}$ &
$\pi$ \\
& & & & & & & & & & & & & & & & &\\
\hline
\hline
$0$      & 82.4 & 82.3 & 82.2 & 82.9 & 82.9 & 82.9 & 82.2 & 82.6 & 83.4 & 82.9 & 83.0 & 82.5 & 82.8 & 82.6 & 83.1 & 83.2 & 82.9 \\
$-0.015$ & 87.6 & 84.9 & 84.3 & 83.9 & 83.6 & 81.9 & 79.9 & 76.0 & 73.3 & 73.6 & 73.5 & 76.3 & 76.6 & 79.1 & 83.5 & 86.4 & 87.9 \\
$-0.029$ & 87.1 & 78.5 & 81.3 & 81.0 & 80.2 & 81.5 & 73.5 & 69.5 & 60.3 & 57.7 & 58.5 & 64.0 & 66.9 & 70.4 & 81.0 & 90.0 & 86.8 \\
$-0.044$ & 78.9 & 64.3 & 71.2 & 76.3 & 78.4 & 77.7 & 70.9 & 58.7 & 47.9 & 41.4 & 41.3 & 46.8 & 53.5 & 62.5 & 77.6 & 91.8 & 79.2 \\
$-0.058$ & 64.4 & 49.0 & 59.9 & 70.8 & 75.6 & 74.5 & 67.4 & 53.1 & 37.6 & 29.0 & 27.6 & 32.4 & 40.8 & 50.0 & 71.0 & 92.9 & 63.4 \\
$-0.065$ & 56.5 & 40.6 & 55.6 & 67.5 & 73.0 & 73.1 & 65.0 & 48.6 & 31.9 & 23.5 & 21.5 & 27.2 & 33.8 & 43.1 & 69.8 & 93.9 & 56.9 \\
$-0.073$ & 45.7 & 32.4 & 47.6 & 63.6 & 68.2 & 71.6 & 64.4 & 43.8 & 27.2 & 18.0 & 15.5 & 20.2 & 28.1 & 36.0 & 64.1 & 94.6 & 45.1 \\
$-0.080$ & 40.1 & 26.6 & 42.2 & 59.4 & 68.3 & 71.0 & 62.4 & 40.5 & 23.4 & 13.4 & 12.2 & 16.6 & 22.1 & 30.9 & 58.9 & 94.8 & 39.3 \\
$-0.088$ & 32.8 & 20.6 & 36.4 & 56.4 & 65.6 & 69.8 & 59.8 & 36.4 & 18.2 & 10.3 & 9.4  & 12.0 & 17.4 & 26.4 & 53.5 & 95.2 & 31.9 \\
$-0.102$ & 22.3 & 11.0 & 28.4 & 50.3 & 61.8 & 67.3 & 55.9 & 28.8 & 11.9 & 7.4  & 5.8  & 7.4  & 11.1 & 17.5 & 43.5 & 95.5 & 20.9 \\
$-0.117$ & 12.2 & 3.9  & 20.0 & 44.4 & 58.2 & 66.2 & 51.7 & 24.1 & 8.2  & 4.3  & 2.8  & 4.7  & 6.1  & 10.6 & 34.8 & 96.0 & 12.3 \\
$-0.131$ & 4.5  & 0.5  & 15.1 & 39.7 & 54.2 & 63.7 & 49.0 & 17.9 & 5.7  & 2.5  & 2.2  & 2.7  & 4.2  & 7.0  & 27.1 & 95.9 & 4.8  \\
$-0.146$ & 0.7  & 0.1  & 9.8  & 34.5 & 52.3 & 62.3 & 44.2 & 13.6 & 3.5  & 1.5  & 1.0  & 1.4  & 2.6  & 4.2  & 18.6 & 94.7 & 0.7  \\
\hline
\hline
\end{tabular}
}
\end{table}
\clearpage
\newpage


\addcontentsline{toc}{section}{References}
\bibliographystyle{LHCb}
\bibliography{refs/main,refs/standard,refs/LHCb-PAPER,refs/LHCb-CONF,refs/LHCb-DP,refs/LHCb-TDR}

\newpage
\centerline
{\large\bf LHCb collaboration}
\begin
{flushleft}
\small
R.~Aaij$^{38}$\lhcborcid{0000-0003-0533-1952},
M.~Abdelfatah$^{69}$,
A.S.W.~Abdelmotteleb$^{57}$\lhcborcid{0000-0001-7905-0542},
C.~Abellan~Beteta$^{51}$\lhcborcid{0009-0009-0869-6798},
F.~Abudin\'en$^{59}$\lhcborcid{0000-0002-6737-3528},
T.~Ackernley$^{61}$\lhcborcid{0000-0002-5951-3498},
A.A.~Adefisoye$^{69}$\lhcborcid{0000-0003-2448-1550},
B.~Adeva$^{47}$\lhcborcid{0000-0001-9756-3712},
M.~Adinolfi$^{55}$\lhcborcid{0000-0002-1326-1264},
P.~Adlarson$^{87,42}$\lhcborcid{0000-0001-6280-3851}
% ...

{\footnotesize \it

% ...
$^{38}$Nikhef National Institute for Subatomic Physics, Amsterdam, Netherlands\\
% ...
$^{42}$National Center for Nuclear Research (NCBJ), Warsaw, Poland\\
% ...
$^{47}$Instituto Galego de F{\'\i}sica de Altas Enerx{\'\i}as (IGFAE), Universidade de Santiago de Compostela, Santiago de Compostela, Spain\\
% ...
$^{51}$Physik-Institut, Universit{\"a}t Z{\"u}rich, Z{\"u}rich, Switzerland\\
% ...
$^{55}$H.H. Wills Physics Laboratory, University of Bristol, Bristol, United Kingdom\\
% ...
$^{57}$Department of Physics, University of Warwick, Coventry, United Kingdom\\
% ...
$^{59}$School of Physics and Astronomy, University of Edinburgh, Edinburgh, United Kingdom\\
$^{60}$School of Physics and Astronomy, University of Glasgow, Glasgow, United Kingdom\\
$^{61}$Oliver Lodge Laboratory, University of Liverpool, Liverpool, United Kingdom\\
% ...
$^{69}$Syracuse University, Syracuse, NY, United States\\
% ...
$^{87}$Department of Physics and Astronomy, Uppsala University, Uppsala, Sweden, associated to $^{60}$\\
% ...
}
\end{flushleft}




\end{document}